\documentclass{emulateapj}
\usepackage{apjfonts}
\usepackage{lscape}

\newcommand \microjy{$\mu$Jy}
\newcommand \plagas{power-law galaxies}
\newcommand \aah{Alonso-Herrero et al. (2006)}
\newcommand \spitzer{\textit{Spitzer}}
\newcommand \chandra{\textit{Chandra}}

\slugcomment{Accepted for publication in The Astrophysical Journal}
\shorttitle{Power-law AGN Candidates in the CDF-N}
\shortauthors{DONLEY ET~AL}

\begin{document}

\title{\spitzer\ Power-law AGN Candidates in the \chandra\ Deep Field-North}

\author{
J. L. Donley, \altaffilmark{1} G. H. Rieke, \altaffilmark{1}
P. G. P\'{e}rez-Gonz\'{a}lez, \altaffilmark{1,2} J. R. Rigby,
\altaffilmark{1} A. Alonso-Herrero \altaffilmark{3}}

\altaffiltext{1}{Steward Observatory, University of Arizona, 933 
North Cherry Avenue, Tucson, AZ 85721; jdonley@as.arizona.edu}
\altaffiltext{2}{Departamento de Astrof\'{\i}sica y CC. de la Atm\'osfera, Facultad de
CC. F\'{\i}sicas, Universidad Complutense de Madrid, 28040 Madrid,
Spain}
\altaffiltext{3}{Departamento de Astrof\'{\i}sica Molecular e Infrarroja, 
Instituto de Estructura de la Materia, CSIC, E-28006 Madrid, Spain}


\begin{abstract}

We define a sample of 62 galaxies in the \chandra\ Deep Field-North
whose \spitzer\ IRAC SEDs exhibit the characteristic power-law
emission expected of luminous AGN. We study the multiwavelength
properties of this sample, and compare the AGN selected in this way to
those selected via other \spitzer\ color-color criteria. Only 55\% of
the \plagas\ are detected in the X-ray catalog at exposures of $>
0.5$~Ms, although a search for faint emission results in the detection
of 85\% of the \plagas\ at the $\ge 2.5 \sigma$ detection level.  Most
of the remaining galaxies are likely to host AGN that are heavily
obscured in the X-ray.  Because the power-law selection requires the
AGN to be energetically dominant in the near- and mid-infrared, the
\plagas\ comprise a significant fraction of the \spitzer-detected AGN
population at high luminosities and redshifts.  The high 24 \micron\
detection fraction also points to a luminous population. The \plagas\
comprise a subset of color-selected AGN candidates.  A comparison with
various mid-infrared color selection criteria demonstrates that while
the color-selected samples contain a larger fraction of the X-ray
luminous AGN, there is evidence that these selection techniques also
suffer from a higher degree of contamination by star-forming galaxies
in the deepest exposures.  Considering only those \plagas\ detected in
the X-ray catalog, we derive an obscured fraction of 68\% (2:1).
Including all of the \plagas\ suggests an obscured fraction of $<
81$\% (4:1).

\end{abstract}

\keywords{galaxies: active --- infrared: galaxies --- X-rays: galaxies}


\section{Introduction}

Detecting complete samples of AGN, both locally and in the deep
cosmological fields, has been a major ongoing goal.  Hard X-ray
selection is a powerful way to detect both relatively uncontaminated
and complete samples of AGN.  Deep X-ray surveys with \chandra\ and
\textit{XMM-Newton} have now resolved 70-90\% of the cosmic X-ray
background (CXRB) at 2-8 keV into discreet sources (Mushotzky et
al. 2000; Giacconi et al. 2002; Alexander et al. 2003; Bauer et
al. 2004; Worsley et al. 2004, 2005), detecting the majority of the
X-ray unobscured AGN population in the deep X-ray fields.  At high
column densities, however, the dust and gas surrounding the central
engine (in combination with that located in the host galaxy) are
capable of hiding virtually all accessible AGN tracers. Therefore,
while the overall resolved fraction of the CXRB is high, it drops with
increasing energy to 60\% at 6-8~keV and to 50\% at $>$ 8 keV (Worsley
et al. 2004, 2005).

Population synthesis models of the CXRB and of X-ray luminosity
functions therefore predict a significant population of heavily
obscured AGN not detected in the deepest X-ray fields (i.e., the 1~Ms
\chandra\ Deep Field--South (CDF-S) and the 2~Ms \chandra\ Deep
Field--North (CDF-N)).  Predictions of their properties are in rough,
but not complete, agreement. Treister et al. (2004) predict an X-ray
incompleteness of 25\% at $N_{\rm H} = 10^{23}$~cm$^{-2}$ and of 70\%
at $N_{\rm H} = 10^{24}$~cm$^{-2}$. Ballantyne et al. (2006) estimate
that the deep X-ray surveys miss $\sim 50$\% of obscured AGN with
log~$L_{\rm x}$(ergs~s$^{-1}$)$> 44$ at all $z$, and that most of the
missing objects are Compton-thick (log~$N_{\rm H}$(cm$^{-2}$)$ > 24$).
Worsley et al. (2005) find that the unresolved sources are likely to
lie at redshifts of $z=0.5-1.5$, have column densities in excess of
$10^{23}$~cm$^{-2}$, and have intrinsic X-ray luminosities of
$<5\times 10^{43}$~ergs~s$^{-1}$.

Numerous attempts have been made to detect this population of heavily
obscured AGN, many of which have focused on the mid-infrared (MIR)
emission where the obscured radiation is re-emitted (Ivison et
al. 2004; Lacy et al. 2004; Hatziminaoglou et al. 2005; Stern et
al. 2005; Alonso-Herrero et al. 2006; Polletta et al. 2006) or on
combinations of MIR and multi-wavelength data (Johansson et al. 2004;
Yan et al. 2004, 2005; Donley et al. 2005; Franceschini et al. 2005;
Houck et al. 2005; Leipski et al. 2005; Mart\'{\i}nez-Sansigre et
al. 2005, 2006; Richards et al. 2006; Weedman et al. 2006). In the
MIR, luminous AGN can often be distinguished by their characteristic
power-law emission, which extends from the infrared to the ultraviolet
(e.g. Neugebauer 1979; Elvis et al. 1994).  This emission is not
necessarily due to a single source, but can arise from the combination
of non-thermal nuclear emission and thermal emission from various
nuclear dust components (e.g. Rieke \& Lebofsky 1981).  In contrast,
star-forming galaxies are characterized by dust and stellar emission
features redward and blueward of a local minimum at $\sim 5$ \micron.

We focus here on AGN with red power-law SEDs in the \spitzer\ 3.6-8.0
\micron\ bands. \aah\ selected a sample of 92 such sources in the CDF-S, 
70\% of which are hyper-luminous infrared galaxies (HyperLIRGS,
log~L$_{\rm IR}$(L$_\sun$)$ > 13$) or ultra-luminous infrared galaxies
(ULIRGS, log~L$_{\rm IR}$(L$_\sun$)$ > 12$). Nearly half (47\%) of
their power-law sample were not detected in X-rays at exposures of up
to 1~Ms. We use a selection similar to that of \aah\ to identify
\plagas\ in the 2~Ms CDF-N. Because the central regions of the \chandra\
Deep Fields are photon-limited, not background-limited, the 2~Ms CDF-N
is twice as sensitive as the 1~Ms CDF-S at the \chandra\ aimpoint
(e.g. Alexander et al. 2003).

We use a combination of spectroscopic and photometric redshifts to
estimate distances and luminosities for this class of AGN, placing
them in the context of the overall AGN population. The observations
and data reduction are outlined in \S2 and the selection criteria are
discussed in \S3.  The photometric redshift code is discussed in \S4.
Making use of the deepest available (2~Ms) X-ray data, we search for
faint X-ray emission from sources missed in the X-ray catalogs. The
X-ray, MIR, radio, and optical properties of the sample, including
detection fractions, X-ray luminosities, radio classifications, and
optical--MIR SEDs, are discussed in \S5.  In \S6, we compare the
power-law selection to other MIR color selection techniques, and
discuss the completeness and reliability of the power-law and
color-selection criteria. Finally, in \S7, we use the X-ray data to
calculate the obscuring columns of the \plagas\ and to estimate the
obscured fraction of the power-law sample. Throughout the paper, we
assume a cosmology of ($\Omega_{\rm m}$,$\Omega_{\rm
\Lambda},H_0$)=(0.3, 0.7, 72~km~s$^{-1}$~Mpc$^{-1}$).


\section{Observations and Data Reduction}

\spitzer\ was used to obtain MIPS and IRAC MIR observations of
the CDF-N, with 1400~s and 500~s exposures, respectively.  We use
these MIPS GTO images in place of the GOODS \spitzer\ data because the
former cover the full CDF-N, whereas the latter cover a field
approximately two times smaller.  The MIPS image was processed using
the MIPS GTO data analysis tool (Gordon et al. 2004, 2005); the IRAC
data analysis is described in Huang et al. (2004).  We used the {\sc
iraf} task \textit{allstar} to select sources at 24 \micron, and
SExtractor (Bertin \& Arnouts 1996) to select sources in the IRAC
bands. The 80\% completeness limit for the 24 \micron\ data is
83~\microjy\ (Papovich et al. 2004). The IRAC detection limits of our
survey (see \S3) are 1.8, 2.8, 14.5, and 18.0
\microjy, at 3.6, 4.5, 5.8, and 8.0 \micron, respectively.  Optical 
and NIR photometry were measured from the GOODS dataset
(\textit{bviz}, Giavalisco et al. 2004), as well as from the data of
Capak et al. (2004) (\textit{UBVRIz'HK'}).  The ground-based
photometry was aperture-matched for consistency, using the aperture of
the most sensitive band in which the source was detected.  For cases
in which multiple ground-based sources in a 1.5\arcsec\ radius caused
the aperture-matched flux density to differ from that of the nearest
source by a factor of $\ge$ 1.5, we replaced the aperture-matched flux
densities with the cataloged flux densities of the nearest source in
all bands.  In addition, if the ground-based photometry of sources
with multiple GOODS counterparts (in a 1.5\arcsec\ radius) differed
from the GOODS photometry by a factor of 2 or more in any of the
bands, we treated the ground-based photometry as upper limits. For
further details, see P\'{e}rez-Gonz\'{a}lez et al. (2005).


\section{Sample selection}

We selected as \plagas\ sources that were detected with $S/N \ge 6$ in
each of the four IRAC bands and whose IRAC spectra are well-fit by a
line of spectral index $\alpha \le -0.5$, where $f_{\nu} \propto
\nu^{\alpha}$.  In a similar study, \aah\ chose a limit of 
$\alpha \le -0.5$ based on the mean spectral slope of
optically-selected quasars ($\alpha \sim -1$, Elvis et al. 1994;
Neugebauer et al. 1979) and the optical spectral indices of Sloan
Digital Sky Survey (SDSS) QSOs ($\alpha = 0.5$ to -2, Ivezi{\'c} et
al. 2002). For comparison, broad-line AGN (BLAGN) from the AGN and
Galaxy Evolution Survey (AGES, C. S. Kochanek et al., in prep) have
3.6-8 \micron\ slopes of $\alpha=-1.07\pm0.53$ (Stern et al. 2005),
indicating that our cut lies within $\sim 1\sigma$ of the mean.  To
ensure a good linear fit, we applied a cut in the chi-squared
probability of $P_{\chi} \ge 0.1$. $P_{\chi}$ is the probability that
a fit to a power-law distribution would yield a value greater than or
equal to the observed chi-squared; a probability of 0.5 corresponds to
a reduced chi-squared of 1. $P_{\chi}$ tends to either lie close to
0.5, or is very small (see Bevington \& Robinson 2003). Because we are
primarily interested in the X-ray and radio properties of the
\plagas, we also restricted the sample to those galaxies with X-ray
exposures of $\ge 0.5$~Ms (Alexander et al. 2003); this selection also
ensures deep 1.4~GHz radio coverage (Richards 2000) and results in a
total survey area of 350~arcmin$^{2}$. In addition, we required the
IRAC SED to rise monotonically in $f_{\nu}$ to prevent contamination
from star-forming galaxies with possible stellar features in the IRAC
bands (resulting in the rejection of 10 sources).  Using these
criteria, we identified 79 \plagas\ in the CDF-N.

We removed 10 of the selected galaxies due to blended or problematic
IRAC photometry and we removed 1 galaxy (CDFN~22363) because of its
stellar-dominated spectrum. CDFN~22363 has a shallow slope of only
$-0.56 \pm 0.20$ and hence might have been scattered into the
power-law sample by noise (it has the second flattest slope in the
sample).  It is the only source in the power-law sample with a
spectroscopic redshift that is not detected in X-rays, and it lacks a
24 \micron\ counterpart (see \S5.2).  Six additional sources were
removed because their optical-MIR SEDs exhibited possible stellar
bumps.  The final sample of 62 power-law galaxies is listed in Table
1; radio through X-ray SEDs are shown in Figure 1.  We indicate in
Figure 1 the 27 sources that lie in the GOODS ACS region. In addition,
we flag all sources for which we have either replaced the ground-based
aperture-matched flux densities with the cataloged flux densities of
the nearest source or for which we treat the ground-based photometry
as upper limits (see \S2).

\begin{figure*}[t]
\epsscale{0.82}
\plotone{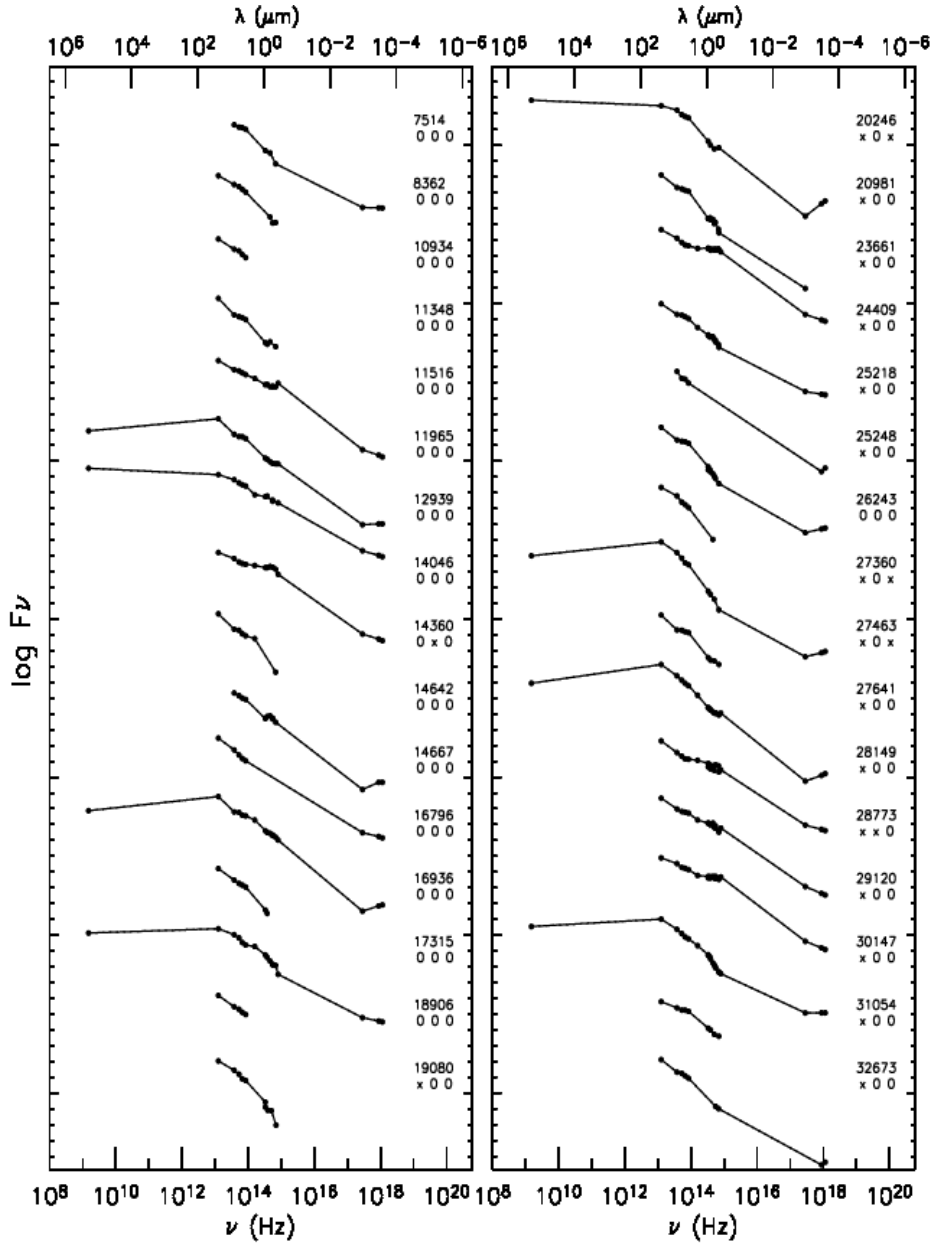}

\caption{X-ray through radio observed-frame SEDs of the \plagas.  The 
flags printed under the source ID indicate by x whether (1) the source
lies in the GOODS ACS field, (2) the aperture-matched flux densities
have been replaced by the cataloged flux densities of the nearest
source and (3) the ground-based photometry have been treated as upper
limits (see \S2).}
\end{figure*}

\begin{figure*}
\figurenum{1}
\epsscale{0.82}
\plotone{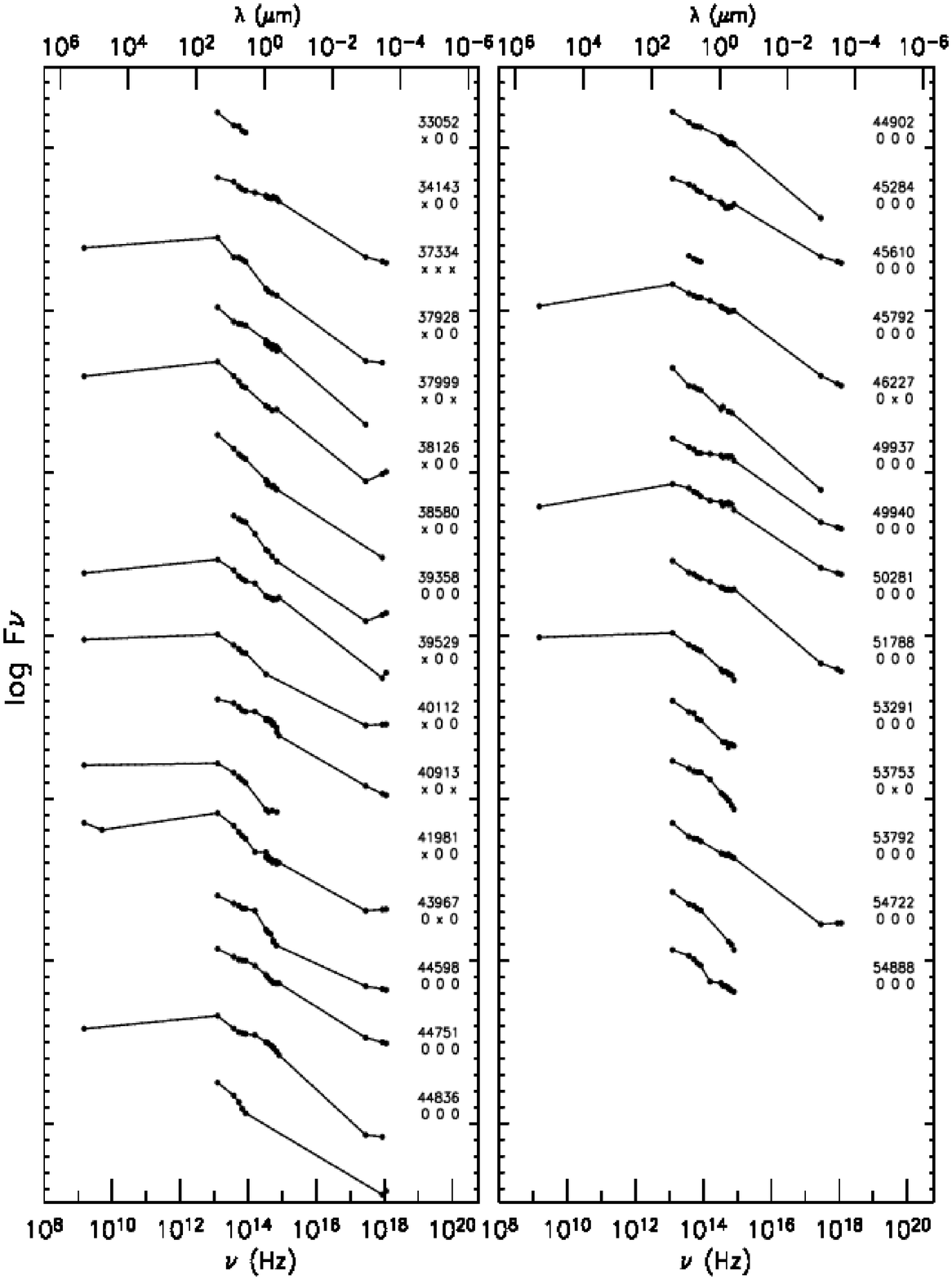}
\caption{Continued.}
\end{figure*}

The $S/N$ cut of 6 in each of the IRAC bands imposes the following
detection limits on the IRAC 3.6, 4.5, 5.8, and 8.0 \micron\ bands:
1.8, 2.8, 14.5, and 18.0 \microjy.  This large range of flux limits
imposes complicated selection effects on the power-law sample.  At
$\alpha < -0.67$, the sample is flux-limited by the 5.8 \micron\ band.
That is, all sources whose 5.8 \micron\ flux density exceeds the
detection limit of 14.5 \microjy\ are detectable in all of the IRAC
channels.  At $\alpha > -0.67$, however, the limiting flux shifts to
the 8.0 \micron\ band.  Because there is little change in the 5.8
\micron\ detection limit for red power-law sources with $-0.67 < \alpha <
-0.5$ ($f_{\rm lim}$ increases from 14.5 \microjy\ at $-3 < \alpha <
-0.67$ to 15.3 \microjy\ at $\alpha = -0.5$), the power-law sample is
essentially flux-limited in the 5.8 \micron\ band.  Comparisons
between the number of red ($\alpha \le -0.5$) and blue ($\alpha >
-0.5$) sources in the full IRAC sample, however, will be complicated
by these effects, as the 5.8 \micron\ limiting flux increases with
increasing spectral slope, reducing the number of blue sources that
meet the requirements of the power-law sample.

To minimize the chances of selecting non-active galaxies via the
power-law selection, we compared the spiral galaxy templates from
Devriendt et al. (1999) to the optical--MIR SEDs of the sources.  None
of the sources in the final sample are well-fit by a spiral
template. In addition to contamination from local spiral galaxies,
\aah\ found that cool ULIRGS at $z>2$ and with $\alpha > -1$ could
appear as \plagas. Of the 62 \plagas\ in the sample, nine have $\alpha
>-1$ and potentially lie at $z>2$.  Seven of these sources, however,
have X-ray counterparts, and all but one (a source with no redshift
estimate) have X-ray luminosities of log~$L_{\rm x}$(ergs~s$^{-1}$)$>
43$, and are therefore unlikely to be starburst-powered ULIRGS.  We
further explore the possibility of contamination due to star-forming
galaxies in \S6.2.2, using several updated SEDs, described in detail
in Appendix A.

We have also considered whether the contributions of 3.3 \micron\
($z>0.5$) and 6.2 \micron\ ($z<0.5$) aromatic emission features to the
flux in the IRAC 8.0 \micron\ band could create the appearance of a
red power-law continuum. Observations of starbursts detect these
aromatic features to equivalent widths of $\sim 0.1$ \micron\ and
$\sim 0.7$ \micron, respectively (Risaliti et al. 2006; B. R. Brandl
et al., in preparation).  Due to the large bandpass of the 8.0
\micron\ IRAC channel, however, the 3.3 and 6.2 \micron\ emission
features should have at most a 3\% or 20\% effect, respectively, on
the total flux. (The models of Devriendt et al. (1999) predict a
slightly larger (20\%) effect due to the 3.3 \micron\ feature (Stern
et al. 2005)).  Given the small impact of these features and the small
number of \plagas\ with $z<0.5$ for which the 8 \micron\ feature could
have an effect, we expect little contamination in the sample.

As a final check on the selection criteria, we plot in Figure 2 the
ratio of X-ray to optical emission, a well-known AGN diagnostic
(e.g. Maccacaro et al. 1988; Barger et al. 2003; Hornschemeier et
al. 2003). Sources not detected in R-band are assigned a lower limit
of 26~mag, the approximate completeness limit of the R-band catalog of
Capak et al. (2004).  As was found by \aah, all of the \plagas\ lie
within the region typically populated by AGN or transition objects,
given the current X-ray and optical limits.  Four of the X-ray sources
appear to be optically-faint with respect to their X-ray fluxes
(e.g. Rigby et al. 2005).

\begin{figure}
\centering
\epsscale{1.1}
\plotone{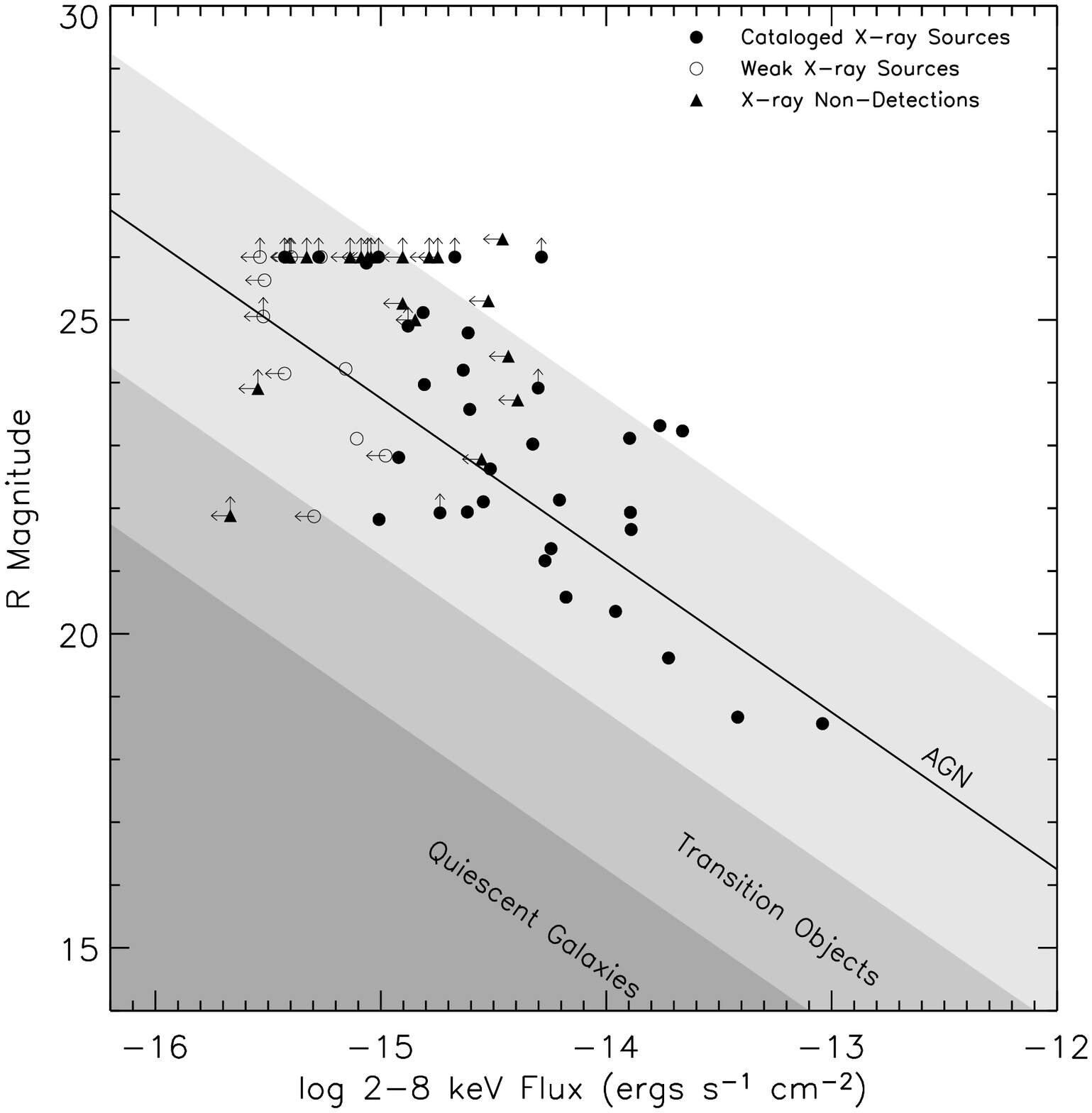}
\caption{Relationship between the observed R-band magnitude and hard
 (2--8 keV) hard X-ray fluxes of \plagas\ detected in the X-ray
 (filled circles), weakly-detected in the X-ray (open circles), and
 non-detected in the X-ray (triangles). For non-detected sources, we
 plot the 5$\sigma$ upper limits calculated as described in \S5.1. The
 lines and shading represent the regions populated by AGN ($f_{\rm
 X}/f_{\rm R} < |1|$), AGN and starbursts ($-2 <$ log$f_{\rm X}/f_{\rm
 R} < -1$), and quiescent galaxies, starbursts, and LLAGN (log $f_{\rm
 X}/f_{\rm R} < -2$) (see Barger et al. 2003; Hornschemeier et
 al. 2003).}
\end{figure}

\newpage
\section{Redshifts}

Twenty one of the \plagas\ have secure spectroscopic redshifts (Barger
et al. 2001, 2002, 2003; Hornschemeier et al. 2001; Dawson et
al. 2003; Cowie et al. 2004; Swinbank et al. 2004; Wirth et al. 2004
[Team Keck Redshift Survey]; Chapman et al. 2005; the Sloan Digital
Sky Survey [SDSS]) and 2 additional sources have photometric redshifts
from Barger et al. (2003). The spectroscopic redshifts range from
$z=0.29$~to~$2.92$, with a mean of $z=1.77$ and a median of
$z=2.02$. Of these 21 sources, 15 are classified as broad-line AGN, 4
have low $S/N$ narrow emission lines, and 2 are AGN whose line width is not
classified in the available literature.

We supplement the available spectroscopic redshifts with photometric
redshifts. Photometric redshifts were estimated with an improved
version of the method described in P\'erez-Gonz\'alez et
al. (2005). Our technique is based on the construction of a complete
set of SED templates composed from the galaxies with highly reliable
spectroscopic redshifts (about 1500 sources selected in the CDF-N and
CDF-S), which are later used to fit the SEDs of all the sample and
estimate a photometric redshift. The technique described in
P\'erez-Gonz\'alez et al. (2005) has been improved by significantly
increasing the resolution of the templates. This is achieved by
fitting the SEDs from the UV to the FIR to stellar population
synthesis and dust emission models. With this improvement, by
comparing with all the available spectroscopic redshifts, our
photometric redshifts have a value of $\Delta(z)/(1+z)$$<$0.1 for 88\%
of the galaxies in the CDF-N and CDF-S, and $\Delta(z)/(1+z)$$<$0.2 for
96\% of these sources. The average (median) value of $\Delta(z)/(1+z)$
is 0.05 (0.03). Further details about the photometric redshift
technique can be found in P. G. P\'erez-Gonz\'alez et al. (2007, in
preparation).

Using this technique, we estimated redshifts for an additional 20
sources, bringing the total number of \plagas\ with redshifts or
redshift estimates to 43 (69\%). We plot in Figure 3 a comparison
between the photometric and spectroscopic redshifts of what we will
refer to as the 'comparison sample'.  This sample consists of the 1420
IRAC sources that meet the detection cut of $S/N > 6$ in each of the
IRAC bands and that have X-ray exposures $\ge 0.5$~Ms; the 11 sources
in the original power-law sample removed due to bad or blended
photometry were excluded.  The mean (median) spectroscopic redshift of
the comparison sample is $z=0.74\ (0.64)$. Nearly half of the sources
in the comparison sample have IRAC SEDs that cannot be fit by a
power-law. Of the remaining galaxies, 6\% and 40\% can be fit by red
($\alpha \le -0.5$) or blue ($\alpha \ge -0.5$) power-laws,
respectively.  Only 50\% of the X-ray sources in the main CDF-N
catalog (Alexander et al. 2003) meet the exposure time and IRAC $S/N$
cuts used to define the comparison sample.

\begin{figure*}[t]
\epsscale{0.9}
\plottwo{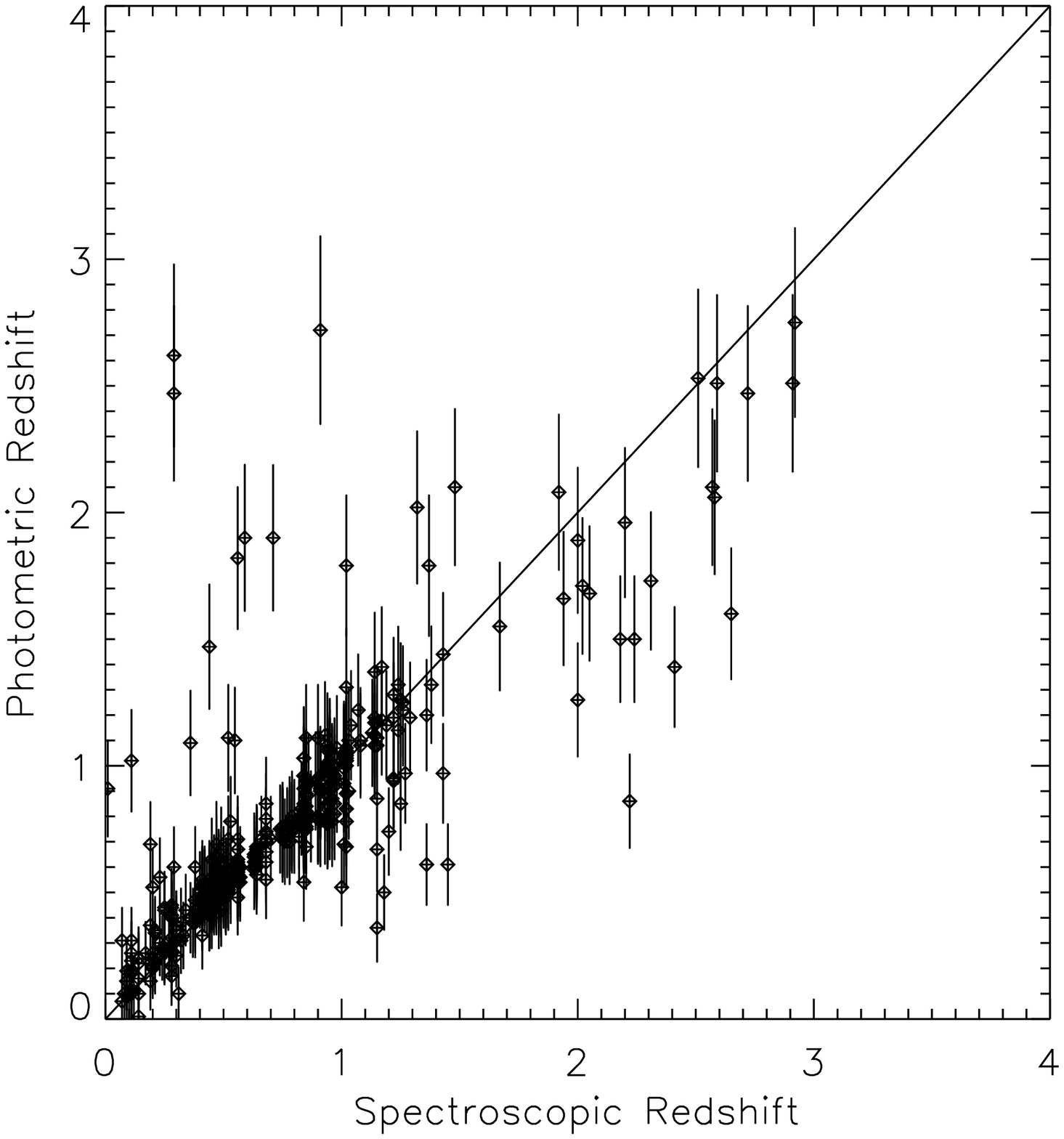}{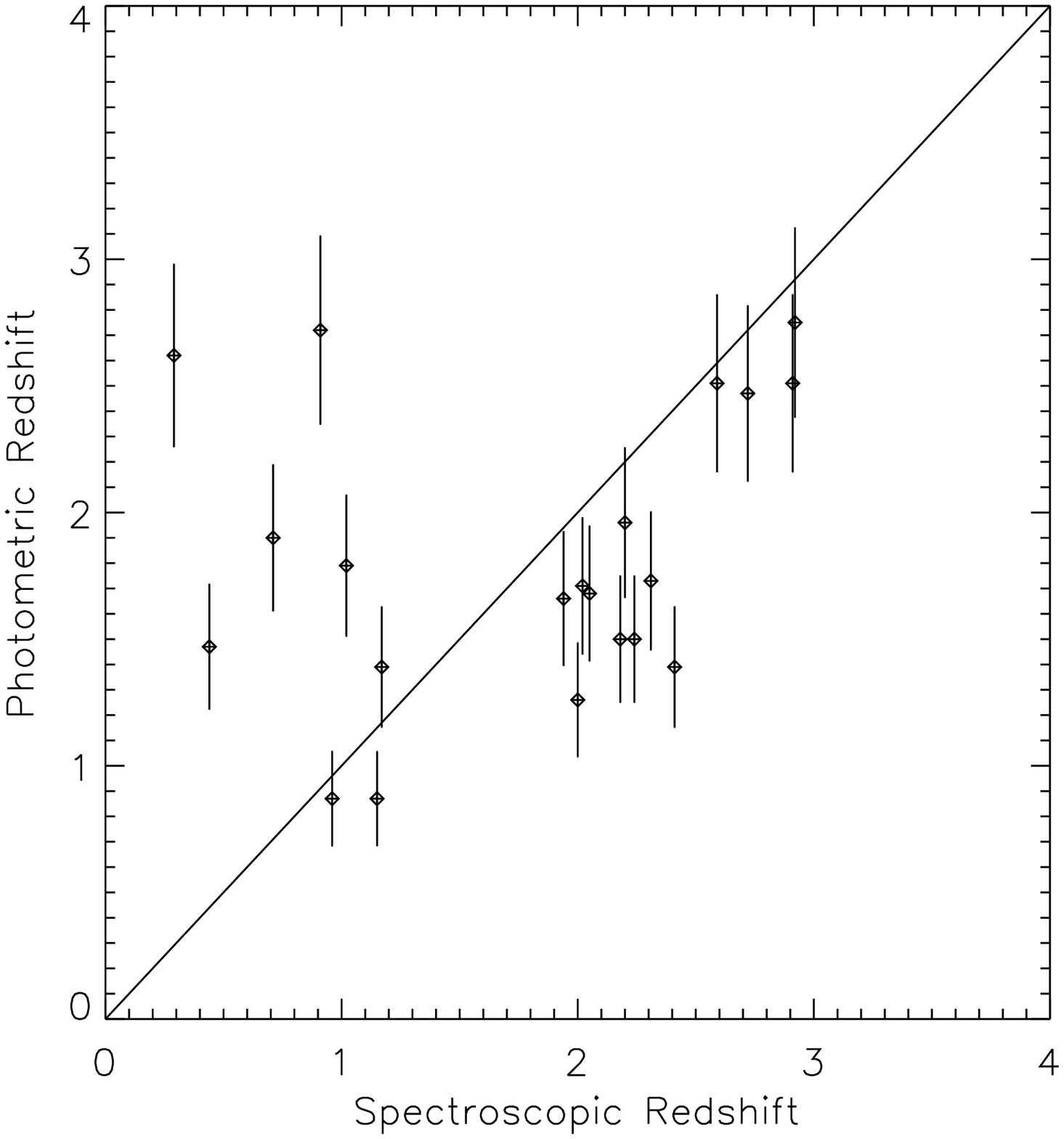}
\caption{Comparison between secure spectroscopic redshifts and photometric 
redshifts for the comparison sample (left) and the power-law galaxy
sample (right). We overplot on each panel a line of slope 1.}

\begin{center}
$\begin{array}{cccc}
\includegraphics[angle=0,scale=0.38]{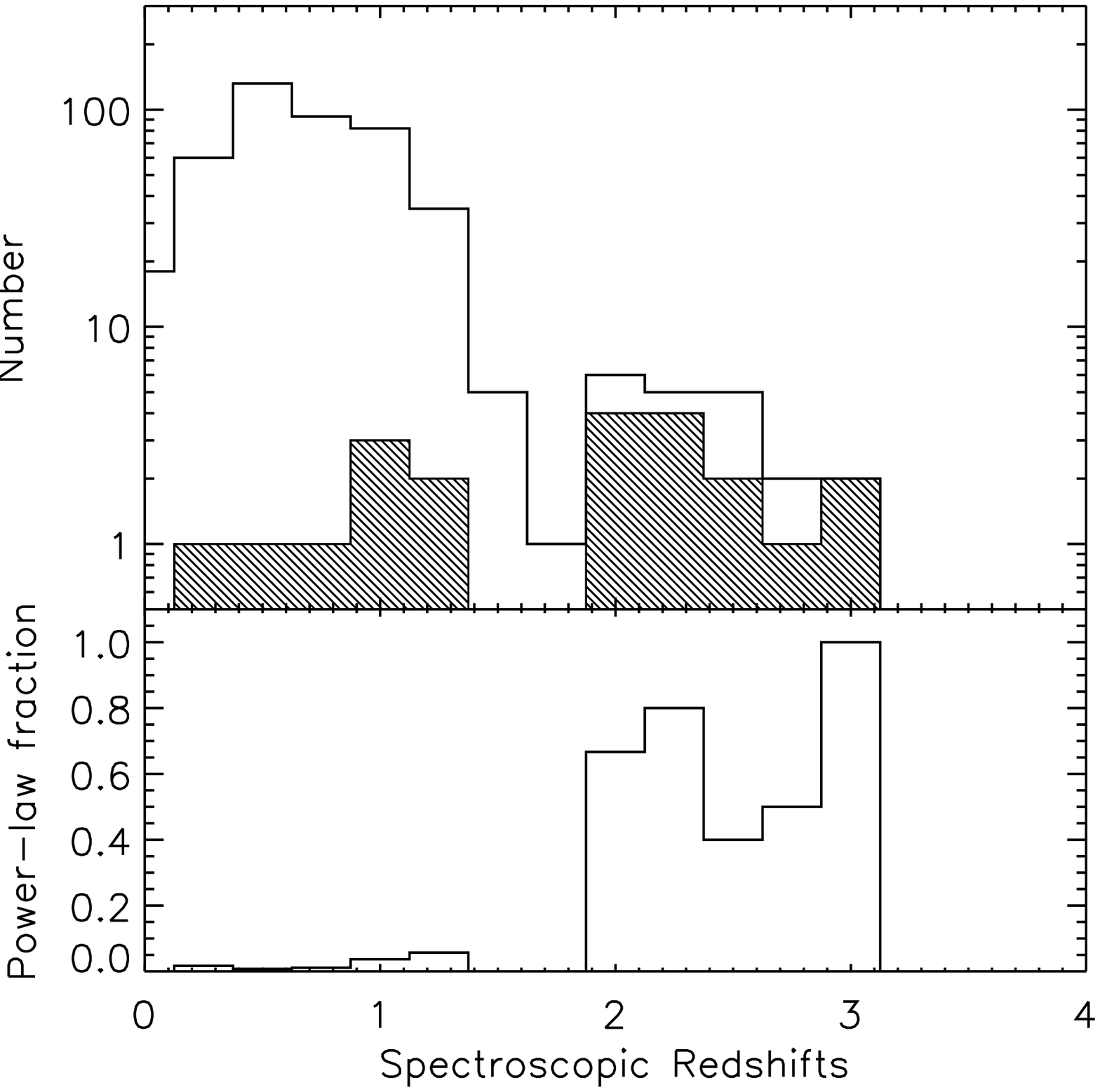} &
\includegraphics[angle=0,scale=0.38]{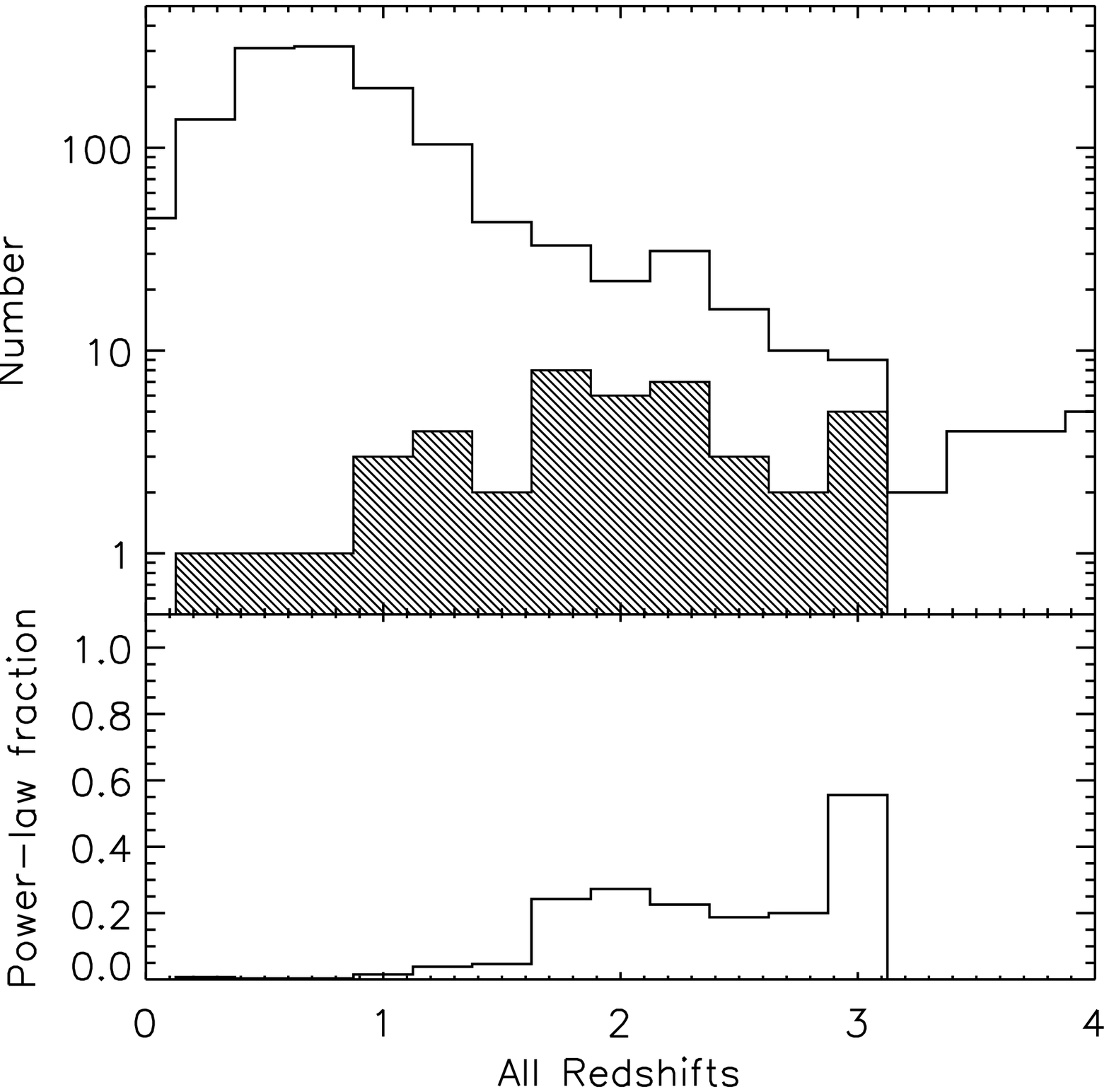} \\
\includegraphics[angle=0,scale=0.38]{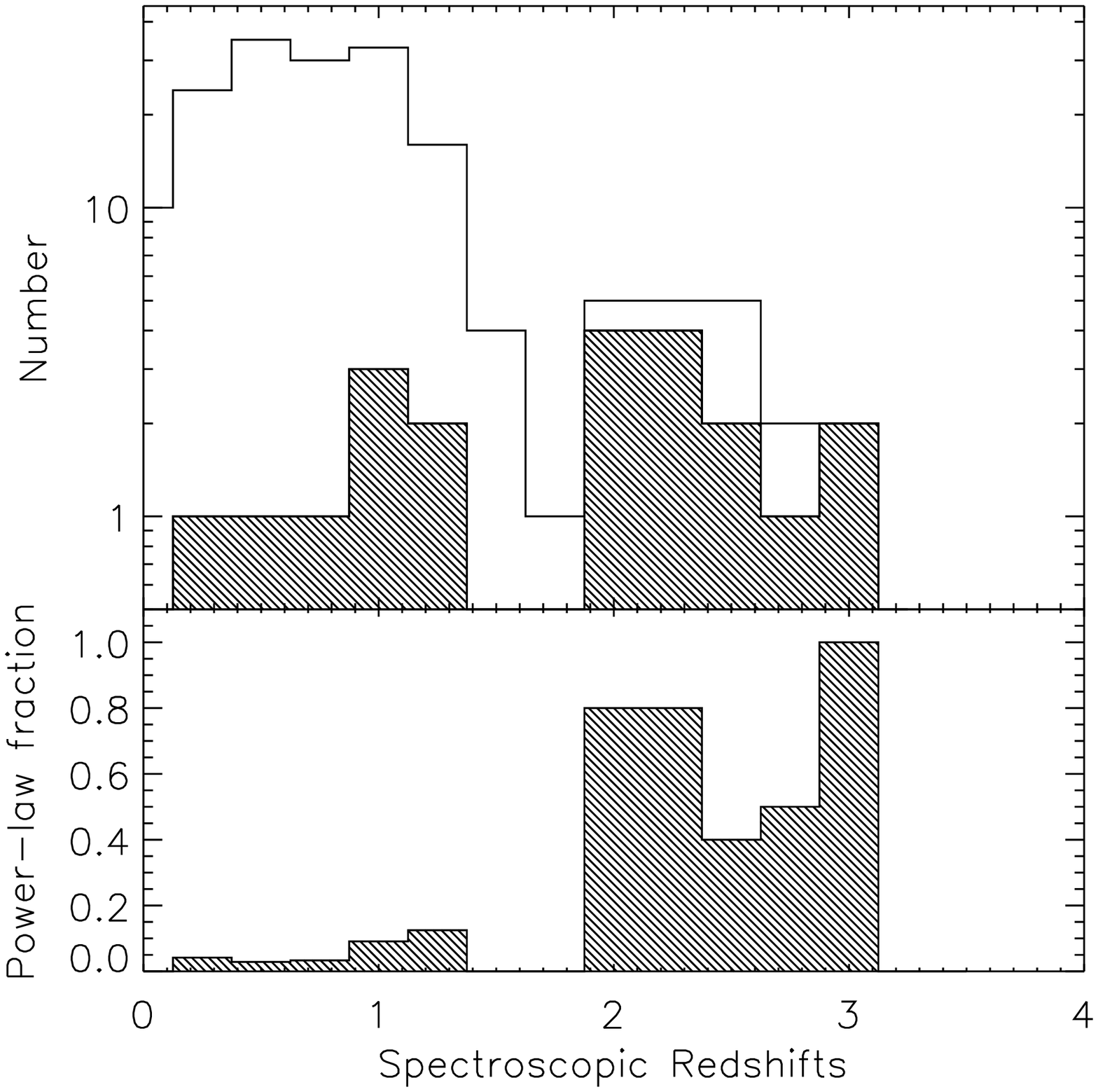} &
\includegraphics[angle=0,scale=0.38]{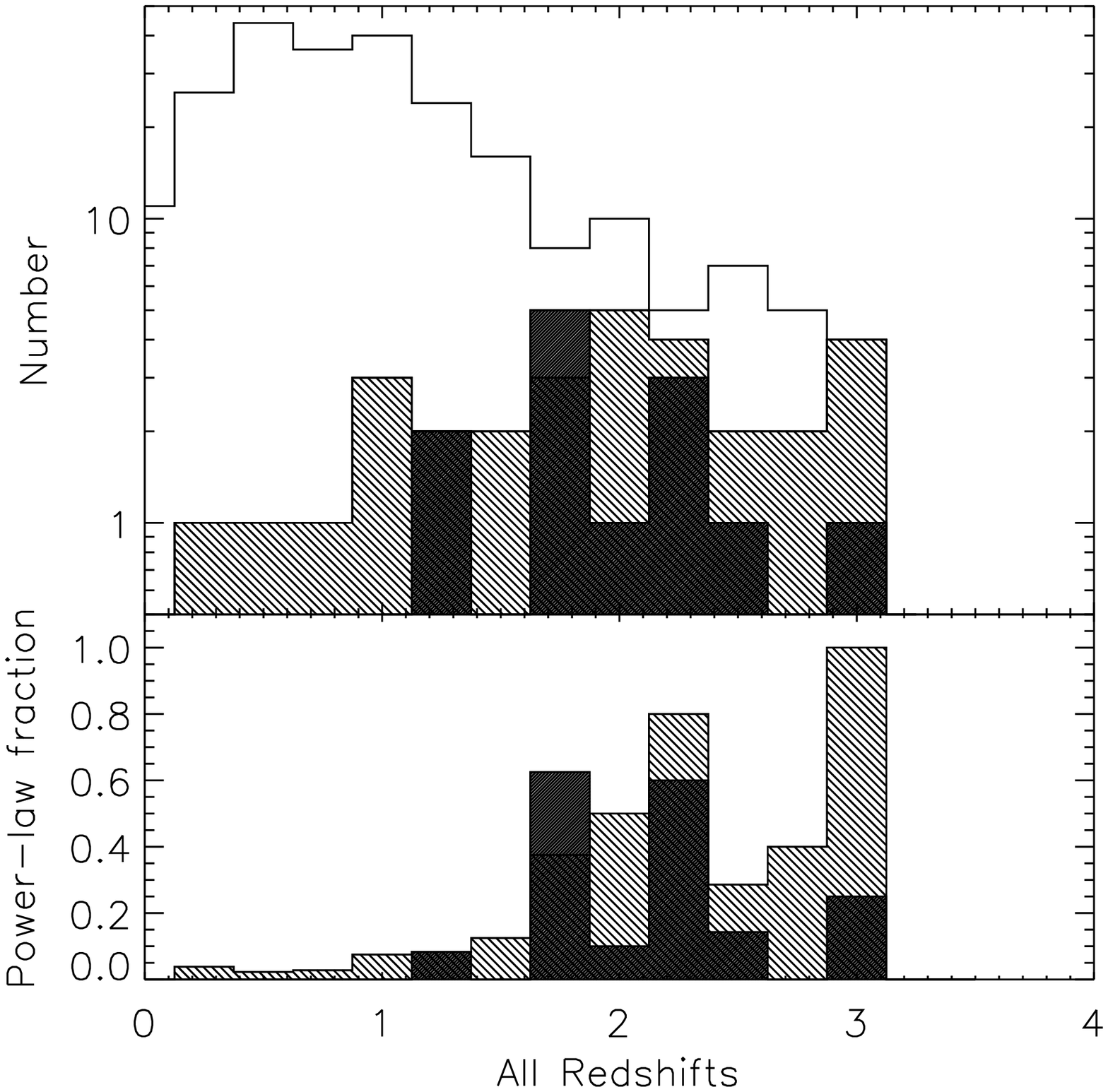} \\
\end{array}$
\end{center}
\caption{Top: Redshift distributions of the comparison
 sample (unshaded) and the power-law galaxy sample (shaded).  Bottom:
 Redshift distributions of the X-ray--detected members of the
 comparison sample (unshaded) and the X-ray--detected (lightly shaded)
 and non-detected (heavily shaded) members of the power-law galaxy
 sample. The distributions based only on spectroscopic redshifts are
 shown on the left; those based on both photometric and spectroscopic
 redshifts are shown on the right. The lower panels give the power-law
 fraction, as a function of redshift.}
\end{figure*}

We also plot in Figure 3 the redshift comparison for those galaxies
that meet the power-law criterion.  While the power-law galaxies tend
to be the farthest outliers, the agreement between the spectroscopic
and photometric redshifts is reasonable, given the difficulty in
assigning redshifts to sources with power-law dominated SEDs. Notably,
several \plagas\ at low spectroscopic redshifts have been assigned
high redshifts by our photometric code.  We take this into
consideration during the analysis, and rely on spectroscopic redshifts
alone wherever possible.  Nonetheless, the success rate in the
photometric redshifts (defined by agreement to 2 standard deviations)
implies that our total suite of redshifts is $\ge$ 80\% correct.

As shown in Figure 4, the \plagas\ tend to lie at significantly higher
redshift than both the average source from the comparison sample and
X-ray--detected members of the comparison sample, a trend that will be
investigated further in \S5.1.2.  The \plagas\ are comparable in
number to the X-ray--detected AGN in the comparison sample at $z>2$,
although as will be discussed in \S5.1, only half of the \plagas\ are
cataloged X-ray sources. \aah\ similarly found that the \plagas\ have
a relatively flat redshift distribution (with the exception of a spike
at $z=1.4$), with most sources lying at $z>1$.

\begin{figure*}[t]
\begin{center}
$\begin{array}{cccc}
\includegraphics[angle=0,scale=0.35]{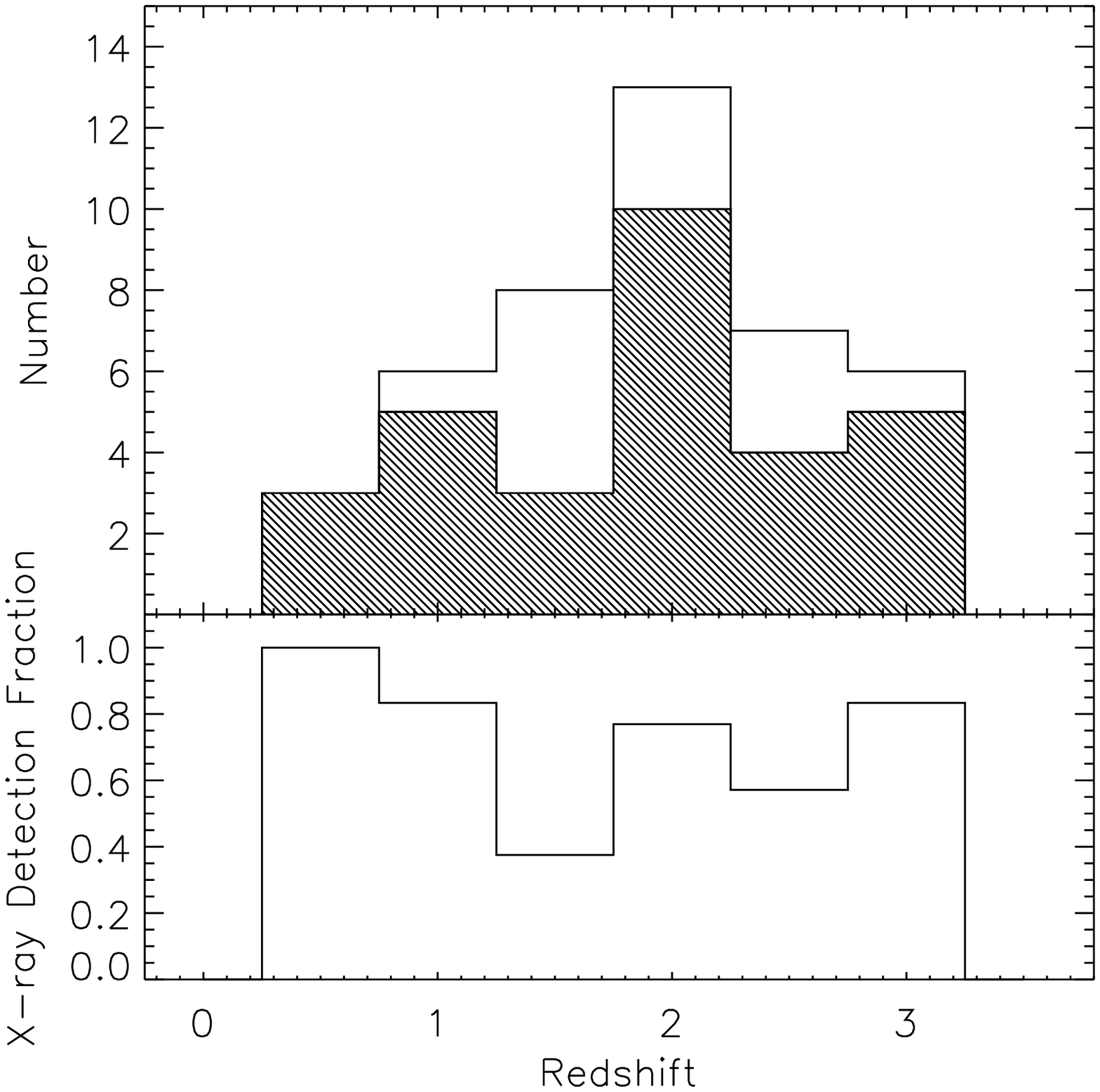} &
\includegraphics[angle=0,scale=0.35]{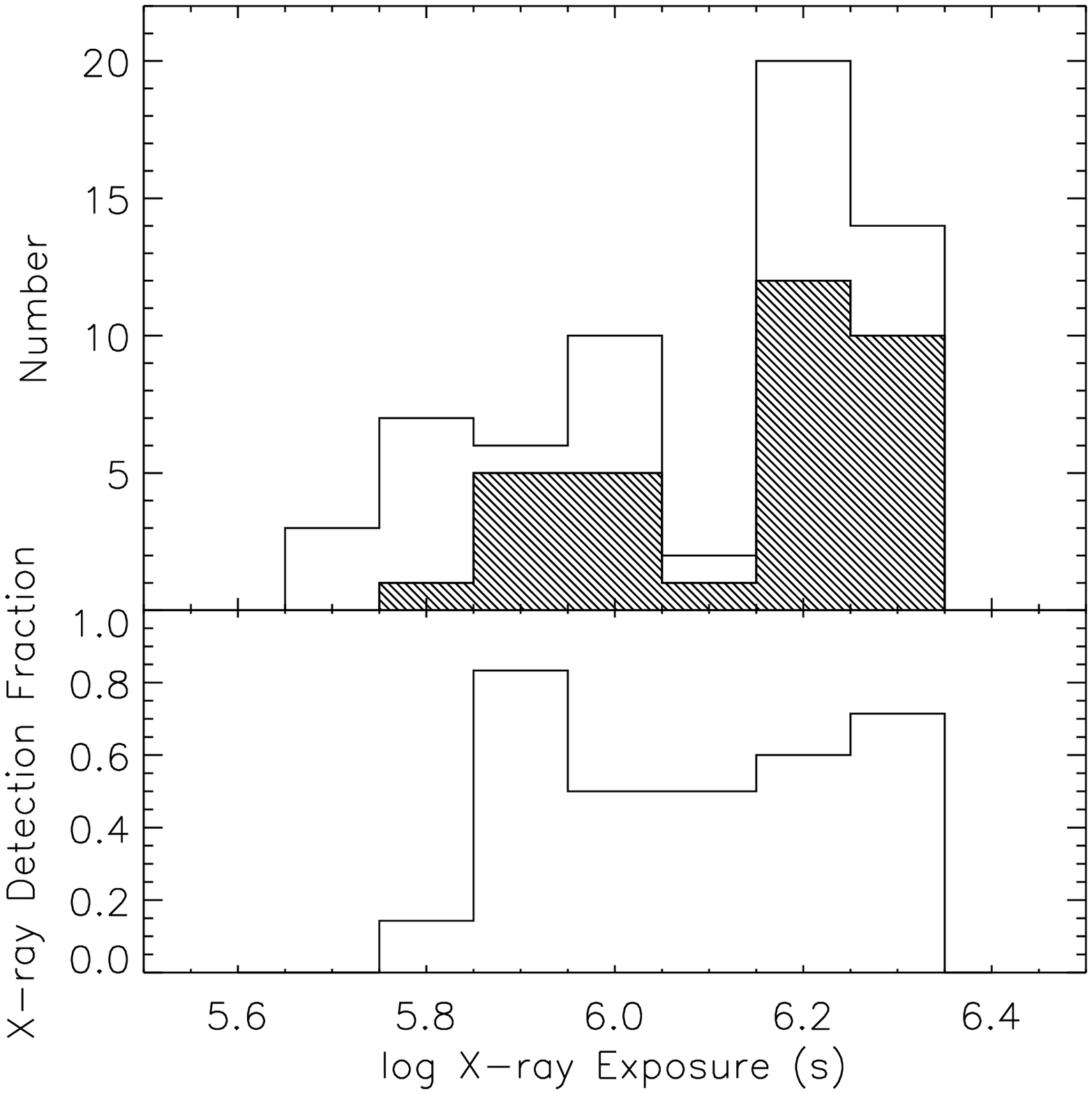} \\
\includegraphics[angle=0,scale=0.35]{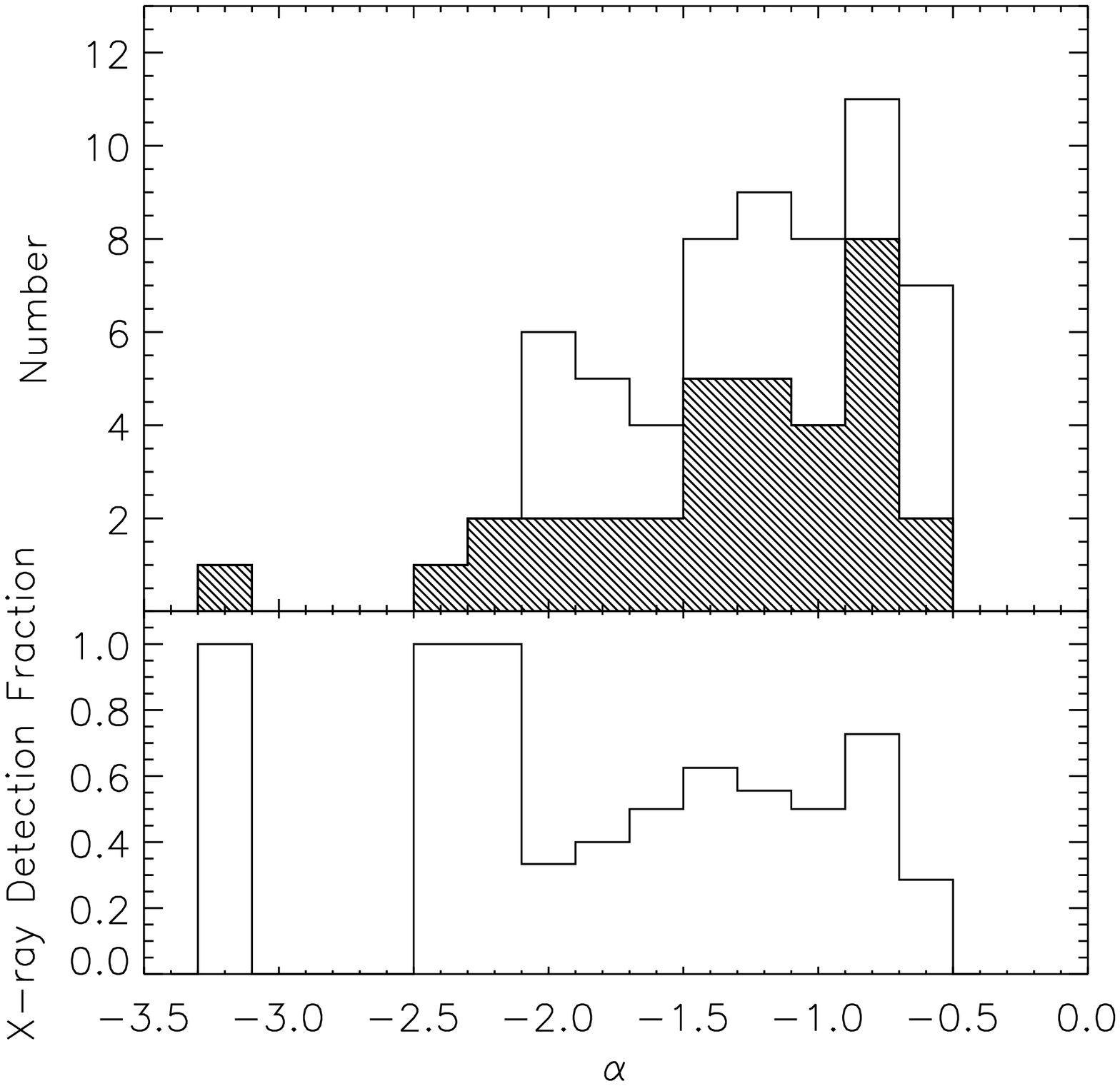} &
\includegraphics[angle=0,scale=0.35]{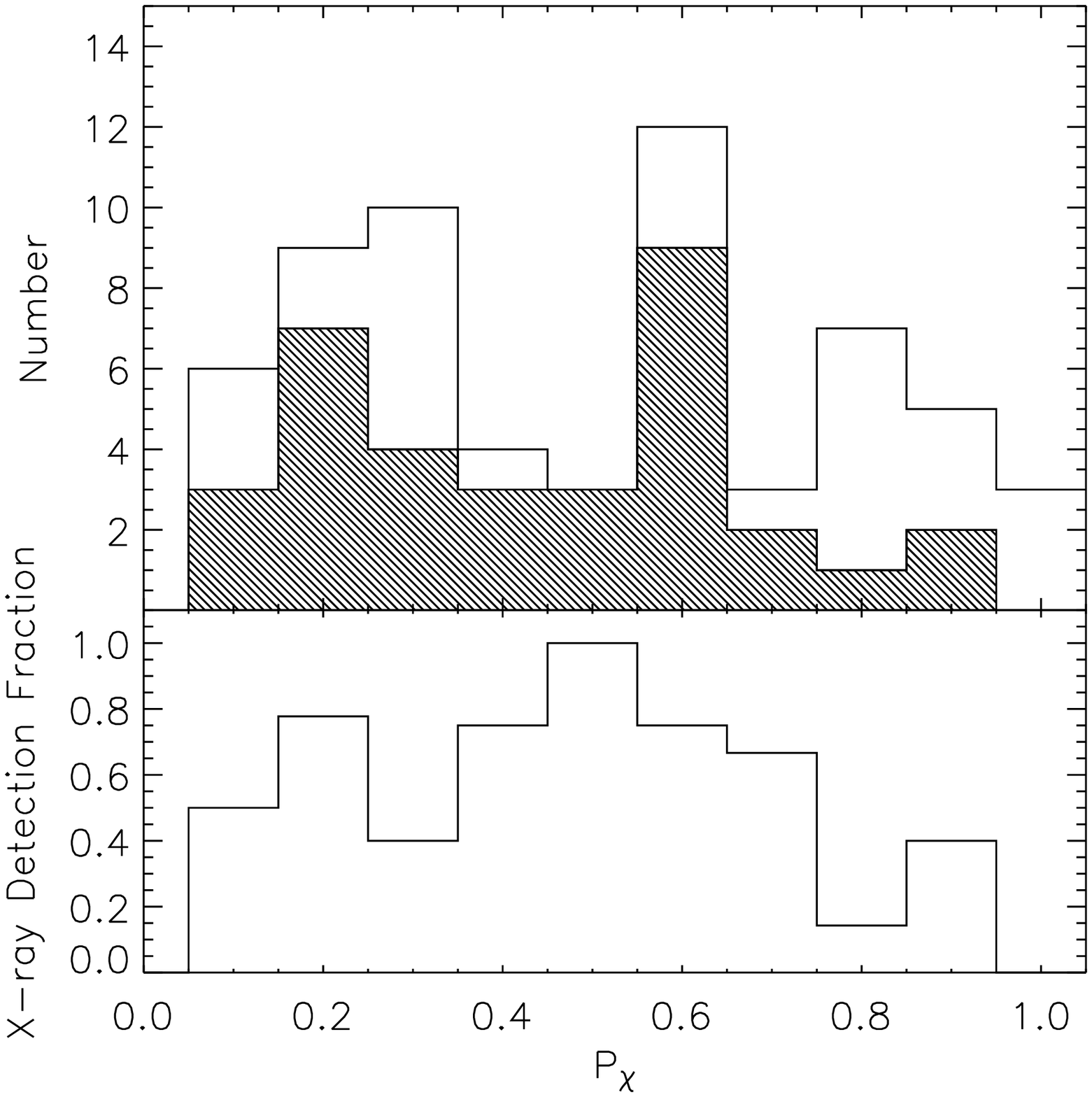} \\
\end{array}$
\end{center}
\caption{X-ray detection fraction as a function of redshift (top left), 
X-ray exposure time (top right), power-law spectral index $\alpha$
(bottom left), and power-law fit probability $P_{\chi}$ (bottom
right). Shaded histograms represent the X-ray--cataloged \plagas;
clear histograms represent the full power-law sample.}
\end{figure*}


\section{Multiwavelength Properties}

In the following sections, we discuss the X-ray, infrared, radio, and
optical properties of the \plagas.  We show that while only 55\% of
them have cataloged X-ray counterparts, as many as 85\% show evidence
for faint X-ray emission.  These X-ray--detected \plagas\ make up a
significant fraction of the X-ray luminous AGN in the comparison
sample.  As such, they are detectable to large distances, explaining
the high redshifts of the sample.

The \plagas\ also have a 24 \micron\ detection fraction of nearly
unity, which we show to be indicative of an intrinsically luminous
X-ray population.  While only 29\% of the sources have radio
counterparts, and all but 2 are radio-quiet, nearly all of the
\plagas\ also have radio luminosities or upper limits consistent with that
of AGN-dominated sources. It is therefore likely that the \plagas\ not
detected in the X-ray are intrinsically luminous AGN hidden behind
high columns of obscuring gas and dust. The optical--MIR SEDs of the
sample, which are flatter than the typical radio-quiet (star-formation
subtracted) AGN SED (Elvis et al. 1994), also suggest increasing
obscuration in the sources not detected in the X-ray.  The optical
detection fraction in the GOODS field is high (85\%), with
approximately 50\% of the X-ray sources having compact optical
counterparts.

\subsection{X-ray Emission}

Of the 62 \plagas, only 34 (55\%) have cataloged X-ray counterparts in
the Alexander et al. (2003) catalog of the 2~Ms CDF-N X-ray
field\footnote{Throughout the paper, we use only the main source
catalog from Alexander et al. (2003), and do not include the
lower-significance X-ray sources detected in the supplemental
optically bright X-ray catalog.  No \plagas\ and only 2 of the
color-selected sources discussed in \S6.2 are detected in the
optically bright catalog.}.  Sources in the Alexander et al. (2003)
catalog were chosen using WAVDETECT (Freeman et al. 2002) with a
false-positive probability threshold of $1 \times 10^{-7}$.  For those
\plagas\ without X-ray counterparts, we searched for faint X-ray
emission following the procedure outlined in Donley et al. (2005).  We
restricted this search to sources with off-axis angles of $\theta < 10
\arcmin$, as sources that lie outside this radius fall in regions of
rapidly changing exposure, and often have sky backgrounds that cannot
be well-fit by Poisson distributions; 84\% (52) of the \plagas\ met
this criterion. Of these, 58\% (30) are detected in the Alexander et
al. catalog, 19\% (10) have at least a 2.5$\sigma$ detection in one
X-ray band and at one encircled-energy radius (EER)\footnote{The 80\%
EER refers to the radius within which 80\% of the energy is found.},
and 13\% (7) remain non-detected.  Coadding the 7 non-detected
\plagas\ does not lead to a detection.  Four additional sources lie
too close to a cataloged X-ray source to search for faint emission,
and 1 source (CDFN~8362) has an irregular sky background, preventing
us from accurately testing for detection.  If we assume that, like
those sources for which we could test for emission, 85\% of the 5
sources with nearby counterparts or irregular sky backgrounds would be
detected (either strongly or weakly), then 85\% of the
\plagas\ show evidence for X-ray emission, while 15\% do not.
Changing the detection threshold to 3, 4, and 5 $\sigma$ changes the
detected fraction to 77\%, 71\%, and 65\%, respectively.

We list the X-ray properties of the weakly-detected \plagas\ in Table
2. If a source was not detected to $>2.5 \sigma$ in one or more of the
three bands, we list in Table 2 a conservative $2.5\sigma$ upper
limit, measured by adding any positive source flux to a 2.5$\sigma$
limit on the local sky background. The 70\% EER was used to calculate
both the source counts and sky background; no aperture correction was
applied. For those \plagas\ without a cataloged or weak X-ray
counterpart, we list in Table 1 5$\sigma$ upper limits, calculated as
described above.

While \aah\ found that their sample of \plagas\ included about 1/3 of
the hard X-ray sources with 24 \micron\ detections, our more
conservative selection detects only 14\% of the X-ray sources in the
Alexander et al. (2003) catalog that meet our exposure time cut of
0.5~Ms and that have 24 \micron\ flux densities of $S_{\rm 24} >
80$\microjy.  (Further restricting the comparison to those X-ray
sources that also meet our IRAC significance cut of $S/N > 6$ only
increases the selection fraction to 17\%). If we consider only those
X-ray and MIR sources in the CDF-N catalog with log~$L_{\rm
x}$(erg~s$^{-1}$)$> 42$ (to rule out sources dominated by
star-formation), the X-ray--detected \plagas\ comprise $\sim 20\%$ of
the X-ray sources (with exposures of $> 0.5$~Ms) detected at 24
\micron\ (to $S_{\rm 24} > 80$\microjy).  Because our selection criteria
were designed to be quite robust, we exclude a number of IRAC
color-selected AGN candidates with power-law like emission (see
\S6). Determining the intrinsic proportion of ``power-law'' AGN will
therefore require a more careful analysis. As mentioned in \S4 and
shown in Figure 4, the \plagas\ comprise a significantly higher
fraction of the high-redshift X-ray sources.

\subsubsection{Dependence on Sample Properties}

We plot in Figure 5 the X-ray detection fraction for the power-law
galaxies as a function of redshift, X-ray exposure time, power-law
slope, and power-law fit probability, where we consider as 'detected'
only those sources with cataloged X-ray counterparts.  For the sources
with redshift estimates, X-ray detection appears to be relatively
insensitive to redshift.  The X-ray detection fraction, however, is
higher for those sources with redshift estimates (70\%) than for the
full power-law sample (55\%). The detection fraction as a function of
X-ray exposure time is lower than average at the lowest exposures, as
expected, but does not increase substantially at the highest
exposures.  We note that in our analysis of the X-ray detection
fraction in \S5.1, we considered only those \plagas\ with off-axis
angles of $\theta < 10 \arcmin$, and therefore excluded 6 of the 10
sources in the lowest two exposure bins. There is significant
variation in the detection fraction as a function of power-law index,
with an increase at the steepest indices and a possible decrease at
the flattest indices. The power-law fit probability, however, appears
to have a minimal effect on the X-ray detection fraction, with only a
slight drop towards the highest values of $P_{\chi}$ (recall that
$P_{\chi} = 0.5$, not $P_{\chi} = 1$, is ideal).  These results
suggest that the X-ray detection fraction of the power-law sample is
relatively insensitive to variations in redshift, X-ray exposure time,
and power-law properties, to the limits imposed by our selection
criteria.

\subsubsection{X-ray Luminosity}

We plot in Figure 6 the observed 0.5-8~keV X-ray luminosities of the
\plagas\ and the members of the comparison sample detected in the 
X-ray catalog. The luminosities of sources selected via the Lacy et
al. selection criteria (see \S6.2) are also plotted.  As was discussed
by \aah, \plagas\ have X-ray luminosities typical of AGN (log~$L_{\rm
x}$(ergs~s$^{-1}$)$\ge 42$), and significantly higher than those of
star-forming galaxies (log~$L_{\rm x}$(~ergs~s$^{-1}$)$ < 41$).

The \plagas\ comprise a significant fraction of the high luminosity
sample. This is not surprising, as the power-law selection requires
the AGN to be energetically dominant, and therefore preferentially
selects the most luminous AGN.  To illustrate the effect of X-ray
luminosity on the AGN contribution to the optical through MIR
continuum, we plot in Figure 7 the median rest- and observed-frame
optical-MIR SEDs of the X-ray--detected members of the comparison
sample, as a function of X-ray luminosity.  We normalize the SEDs to
the 12 \micron\ monochromatic power, which, for the rest-frame
comparison, is representative of bolometric luminosity and independent
of galaxy or AGN type (Spinoglio \& Malkan 1989).  As can be seen,
low-luminosity X-ray sources are dominated by the 1.6 \micron\ stellar
bump in the optical-NIR bands (e.g. Alonso-Herrero et al. 2004).  The
relative strength of this feature decreases with increasing X-ray
luminosity, and disappears almost entirely at luminosities of $L_{\rm
x} > 44.0$, taking on the characteristic power-law shape required by
the selection used here.  The fraction of X-ray sources in the
comparison sample that can be fit by red, AGN-dominated power-laws
($\alpha \le -0.5$) is therefore a strong function of X-ray
luminosity, increasing from 0\% at log~$L_{\rm x}$(ergs~s$^{-1}$)$ =
41-42$ to 3\%, 24\%, and 59\% at luminosity limits of log~$L_{\rm
x}$(ergs~s$^{-1}$)$ = 42-43,43-44,\rm{and}\ 44-45$.  Blue power-laws
($\alpha > -0.5$) in the IRAC bands are indicative of
stellar-photosphere-dominated SEDs in the rest-frame near-infrared
(for $z \le 1.5$). The corresponding percentages of X-ray sources fit
by blue power-laws are 39\%, 46\%, 41\%, and 14\%, showing the reduced
proportion of stellar-dominated objects in the highest luminosity bin.

\begin{figure}
\centering
\epsscale{1.}
\plotone{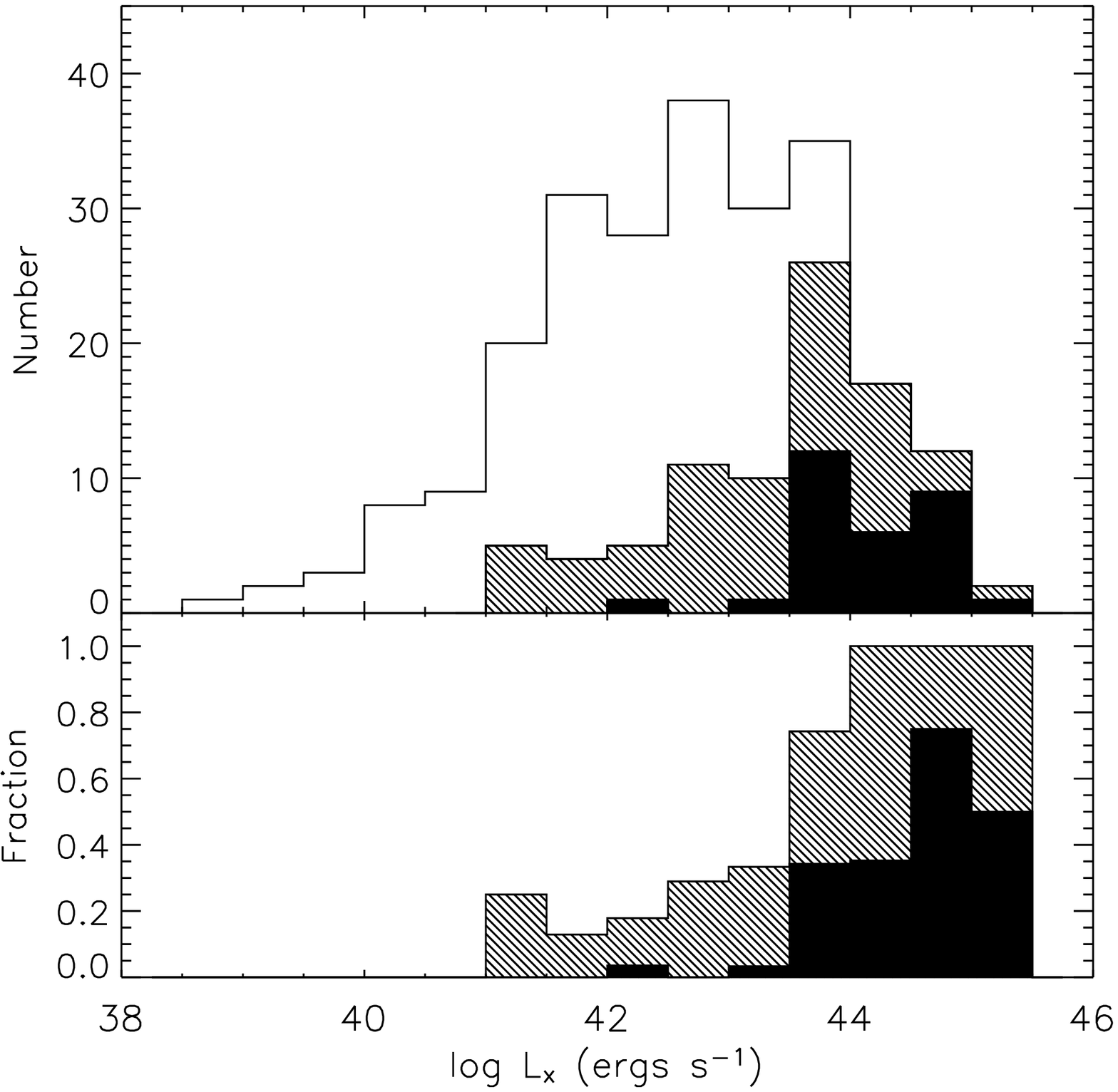}
\caption{Comparison between the observed 0.5-8 keV X-ray luminosity (L$_{\rm x}$)
distributions of the cataloged X-ray sources in the comparison sample
(unshaded histogram), the power-law sample (filled histogram), and the
Lacy et al. sample (lightly shaded histogram; see \S6.2). The lower
panel gives the fraction of the X-ray sources that meet the power-law
or Lacy et al. criteria.}
\end{figure}

\begin{figure*}
\centering
\epsscale{1}
\plottwo{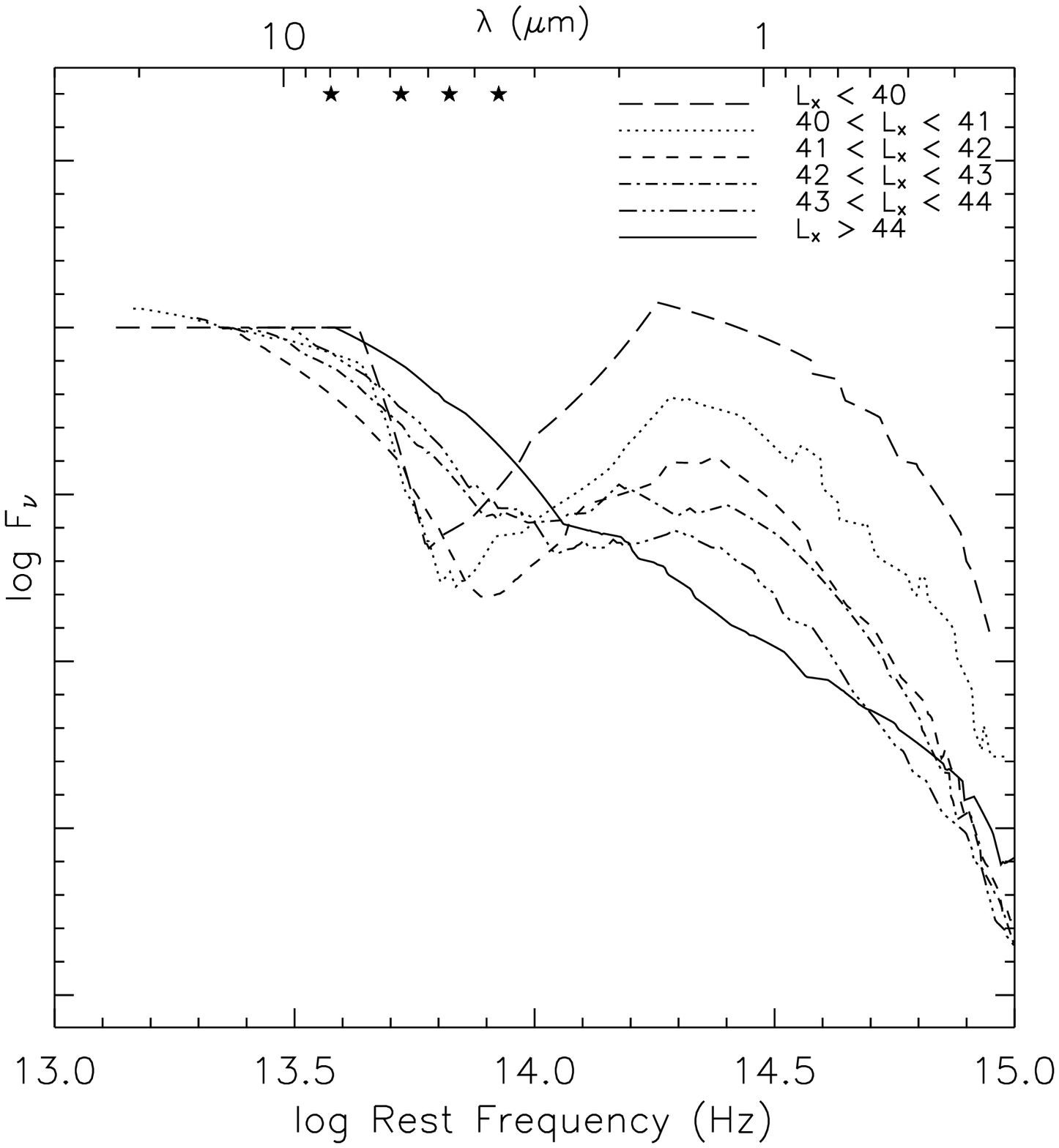}{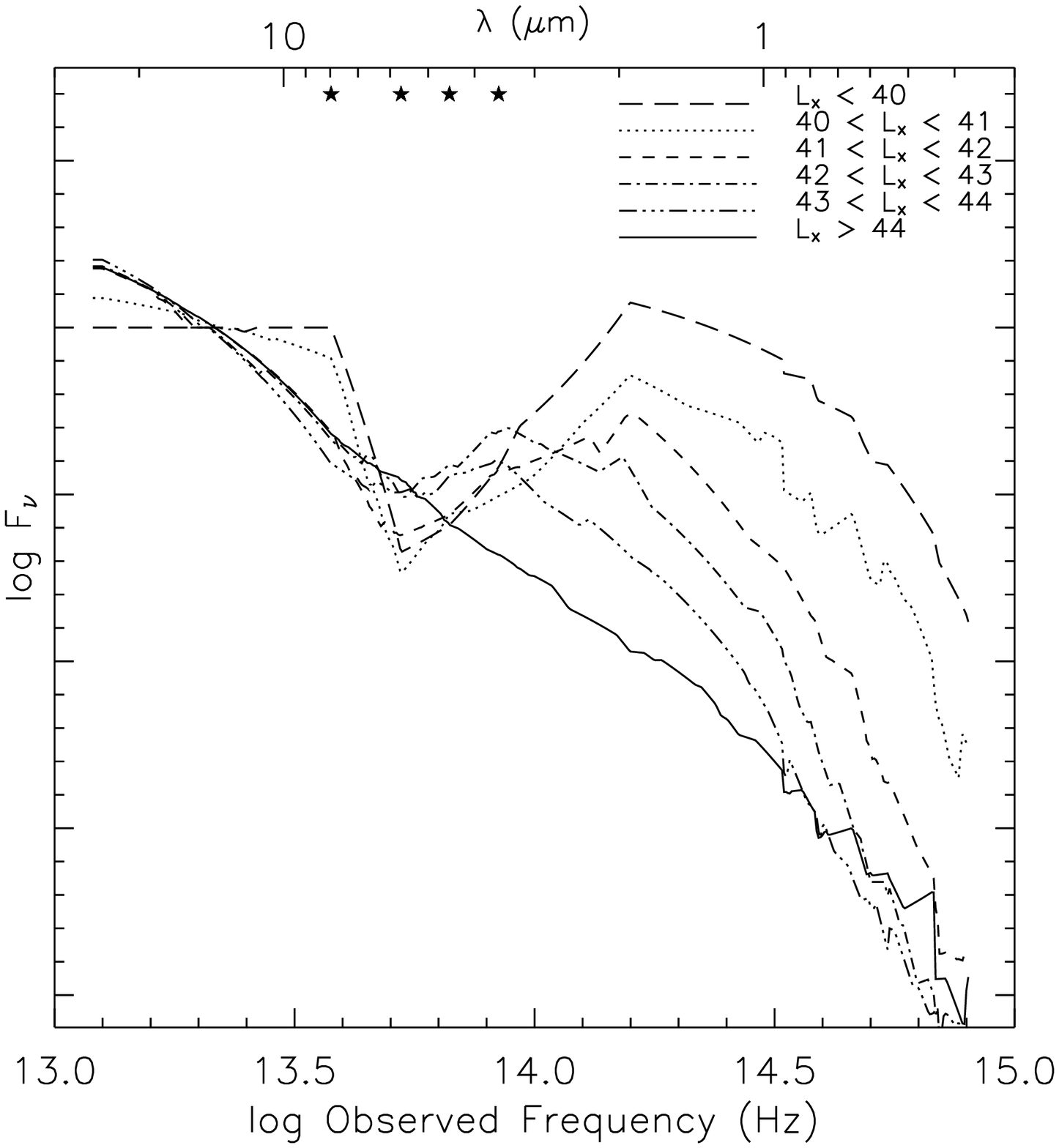}
\caption{Median rest-frame (left) and observed-frame (right) SEDs of the X-ray--detected 
members of the comparison sample, as a function of observed 0.5-8 keV
luminosity (in units of ergs~s$^{-1}$). Stars indicate the wavelengths
of the four IRAC bands.}
\end{figure*}

A similar trend is seen as a function of redshift. In a flux limited
sample such as this, low-luminosity AGN are detected primarily at low
redshift, whereas high-luminosity AGN are detectable over a range of
redshifts (see Figure 8).  To first order, the high redshifts of the
power-law sample can therefore be explained by their preferentially
high luminosities.  In addition, AGN with high X-ray luminosities have
comoving space densities that peak at higher redshifts than
low-luminosity AGN, amplifying the selection effect discussed above
(e.g. Ueda et al. 2003).  The color-redshift relation for AGN may also
play a minor role in this trend.  Richards et al. (2006) show that the
3.6 \micron\ to 5.8 \micron\ color (log($S_{\rm 5.8}/S_{\rm 3.6}$)) of
an AGN with the median radio-quiet SED (Elvis et al. 1994) shifts from
0.2 at $z=0$ to 0.44 at $z=1.5$ before falling to 0.08 at $z=4$.  As
will be shown in \S5.4, however, the \plagas\ tend to have flatter
SEDs than the Elvis et al. (1994) AGN template and as such, they
should have colors less sensitive to changes in redshift.


\subsection{24 \micron\ Emission}

Unlike \aah, we do not require a 24 \micron\ detection for source
selection. However, all but 5 of the \plagas\ have a cataloged 24
\micron\ counterpart within a 2\arcsec\ radius. The 24 \micron\ flux 
densities are listed in Table 1. Of the remaining 5 sources, 3
(CDFN~7514, CDFN~14642, and CDFN~25218) have a cataloged counterpart
within 2-3\arcsec.  For these \plagas, we use the flux density of the
nearby source as an upper limit if it exceeds our nominal limit of 80
\microjy. One additional power-law galaxy, CDFN~45610, lies within the
PSF of a bright 24 \micron\ source, preventing a detailed analysis of
its properties.  Only one source, CDFN~38580, an X-ray--detected
power-law galaxy at $z=2.91$ (Chapman et al. 2005), shows no evidence
for 24 \micron\ emission.  As was found for the power-law X-ray
sources in the EGS (Barmby et al. 2006), there is only weak agreement
between the 24 \micron\ flux densities and those predicted by
extending the IRAC power-law fit, with the X-ray--detected and X-ray
non-detected \plagas\ having observed 24 \micron\ flux densities that
tend to lie within only 53\% and 60\% of the predicted values,
respectively.

The fraction of AGN in MIR samples has been shown to increase with
increasing 24 \micron\ flux density (Treister et al. 2005; Brand et
al. 2006) in general agreement with the models of Pearson (2005).  We
would therefore expect luminous AGN in a flux-limited sample to have a
high 24 \micron\ detection fraction, such as that seen for the
power-law sample.  To test this, we plot in Figure 9 the 24 \micron\
detection fraction for the members of the comparison sample as a
function of X-ray luminosity.  The detection fraction is relatively
high ($\sim 90$\%) at X-ray luminosities typical of starburst galaxies
(log~$L_{\rm x}$(~ergs~s$^{-1}$)$ < 41$, see Alexander et al. 2002).
This is not surprising, as rapidly star-forming galaxies with large
numbers of X-ray binaries will also be luminous infrared sources.  The
detection fraction then drops to $\sim 65$\% at luminosities typical
of starbursts and low-luminosity AGN ($41 < \rm{log}\ L_{\rm
x}$(ergs~s$^{-1}$)$ < 42$) before rising to 100\% at log~$L_{\rm
x}$(ergs~s$^{-1}$)$ > 44-45$. Because the comparison sample is
composed of objects with high significance IRAC MIR detections, the 24
\micron\ detection fraction is higher than that of an unbiased sample.
For comparison, Rigby et al. (2004) find an overall 24 \micron\
detection fraction of only 60\% for hard X-ray sources in the
CDF-S. Nevertheless, for AGN-dominated sources, a high 24 \micron\
detection fraction like that of the \plagas\ accompanies an
X-ray-luminous AGN population.  We note that of the 18 \plagas\ not
detected in X-rays, all but one are detected at 24 \micron, with the
exception being CDFN~45610, the power-law galaxy that lies in the PSF
of a bright 24 \micron\ source.  Therefore, if these sources are
AGN-dominated as expected, it is likely that they, along with the rest
of the power-law sample (\S5.1.2 and Figure 6), are intrinsically
X-ray luminous but heavily obscured.


\subsection{Radio Loudness}

Eighteen of the 62 \plagas\ (29\%) have cataloged radio counterparts
from the Richards (2000) 1.4 GHz VLA survey of the CDF-N.  This survey
has a detection limit of 40 \microjy\ and is 95\% complete at
80~\microjy.  In addition, 3 of these \plagas\ (CDFN~16796,
CDFN~27641, and CDFN~37999) are members of the radio-selected
sub-millimeter SCUBA sample of Chapman et al. (2005), as is one radio
source that fell below the Richards (2000) detection limit
(CDFN~38580). For comparison, Polletta et al. (2006) found that half
of their obscured AGN sample were radio-detected, although their radio
limits were deeper and their IRAC coverage more shallow.

Since all of the radio-detected \plagas\ are also detected at
24~\micron, we can test for radio-loudness via the parameter $q$,
where $q$ = log (f$_{24~\micron}$/f$_{1.4~\rm GHz}$) (see Appleton et
al. 2004). Sources with $q < 0$ have far more radio emission (with
respect to MIR emission) than is expected from the radio/infrared
correlation of radio-quiet AGN and star-forming galaxies, and are
referred to as radio-excess AGN (e.g. Yun, Reddy, \& Condon 2001,
Donley et al. 2005). Kuraszkiewicz et al. (2007) show that $q$ is less
sensitive to reddening and host galaxy dilution than measures of radio
loudness based on radio and optical flux densities. Only 2 of the 18
radio-detected power-law galaxies are radio-excess AGN; the remaining
galaxies are consistent with the radio/infrared
correlation\footnote{We note that the radio-excess AGN CDFN~12939 was
not included in the sample of Donley et al. (2005) because its X-ray
exposure time fell slightly below the more stringent cut of 1~Ms. The
source CDFN~20246 corresponds to VLA~123709+622258 in Donley et
al. (2005).}.  Kuraszkiewicz et al. (2007) find that $\lesssim$15\% of
AGN are radio-intermediate or radio-loud based on a cut of $q < 0$,
consistent with the radio-loud fractions of Kellermann et al. (1989)
and Stern et al. (2000). The fraction of radio-excess AGN in the
power-law sample ($\sim$3\%) may therefore be lower than average.

Both sources classified as radio-excess AGN have a cataloged X-ray
counterpart; 11 of the remaining 16 radio sources are also bright
X-ray sources, while 3 are weakly-detected.  Only two radio sources
(CDFN~40913 and CDFN~51788) are not detected in the X-ray. As such,
selection criteria that include a radio flux density cut may be more
likely to select X-ray bright AGN than selection via MIR properties
alone. If such a cut were based on the current sensitivity of the
CDF-N radio and X-ray surveys, however, only 13 of the 34 \plagas\
detected in the X-ray catalog (38\%) would be selected; the remaining
62\% would be missed.

While the majority of the \plagas\ with radio counterparts appear to
be radio-quiet as defined by the radio/infrared correlation, their
radio luminosities are consistent with those of AGN-powered sources.
In the local universe, galaxies with log~$L_{\rm
radio}$~(W~Hz$^{-1}$)$> 23$ tend to be AGN-dominated (Condon et
al. 2002; Yun et al. 2001).  In the GOODS-North (GOODS-N) field,
Morrison et al. (2006) show that galaxies with log~$L_{\rm
radio}$~(W~Hz$^{-1}$)$> 24$ and $z>1$ have up to an 80\% X-ray
detection rate.  While a source of this luminosity could be a
star-formation powered ULIRG (from the radio/infrared correlation, a
radio luminosity of log~$L_{\rm radio}$~(W~Hz$^{-1}$)$= 24$
corresponds to a total infrared (TIR, 8-1000 \micron) luminosity of $4
\times 10^{12} L_\sun$ (Bell 2003)), the high X-ray detection rate
suggests that the population is dominated by AGN.

\begin{figure}[t]
\centering
\epsscale{1.1}
\plotone{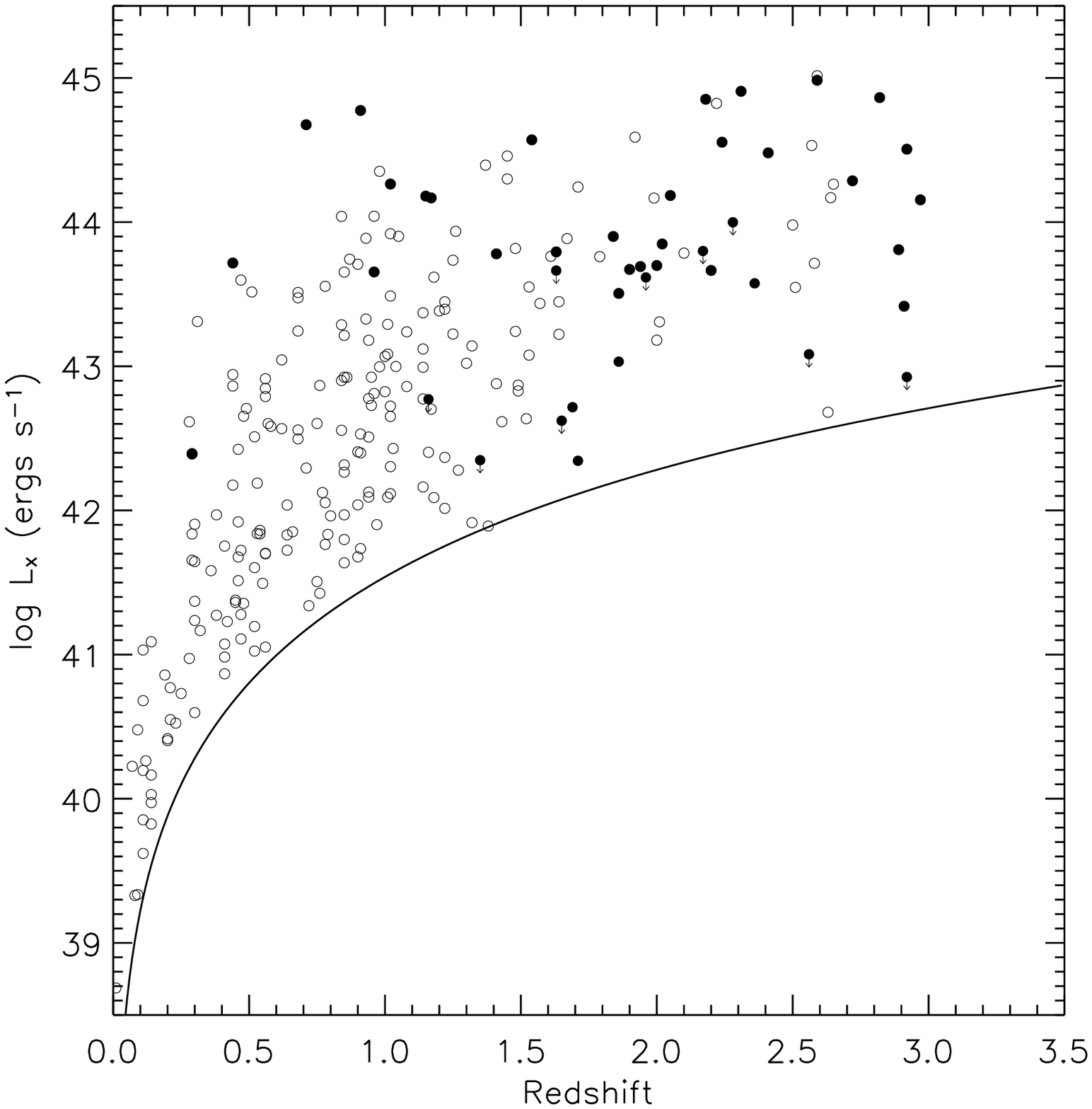}
\caption{X-ray luminosity vs. redshift for the X-ray sources in the comparison 
sample (open circles) and the \plagas\ with available redshift
estimates (filled circles). The on-axis X-ray flux limit of $7 \times
10^{-17}$~erg~s$^{-1}$~cm$^{-2}$ (Alexander et al. 2003) is plotted
for reference.}
\end{figure}

All 16 of the 18 radio-detected \plagas\ with redshifts have observed
1.4 GHz luminosities in excess of log~$L_{1.4
\rm{GHz}}$~(W~Hz$^{-1}$)$>23$, and 14 (88\%) have log~$L_{1.4
\rm{GHz}}$~(W~Hz$^{-1}$)$>24$.  In addition, 26 of the 27 radio
non-detected \plagas\ with redshifts have upper limits greater than
log~$L_{1.4\rm{GHz}}$~(W~Hz$^{-1}$)$ = 23$ (the exception being
CDFN~49937, a low-luminosity X-ray source), and 21 (78\%) have limits
greater than log~$L_{1.4\rm{GHz}}$~(W~Hz$^{-1}$)$ = 24$.  Using a cut
of log~$L_{1.4 \rm{GHz}}$~(W~Hz$^{-1}$)$>24$, Morrison et al.  (2006)
found that 60\% of luminous radio galaxies in the GOODS-N field have
power-law dominant SEDs. Using the same cut, we find that as many as
81\% of power-law selected AGN are radio luminous, though not
necessarily radio-loud.


\subsection{Optical Morphology and Detection Fraction}

While it is difficult to determine optical morphologies for faint
sources, particularly those with only ground-based optical data, at
least 20 of the \plagas\ have point-like optical/NIR counterparts (8
of which were detected in the GOODS \textit{Hubble Space Telescope}
(\textit{HST}) field, and 12 of which have only ground-based data).
Of these, 18 are detected in the X-ray, one is weakly-detected, and
one is non-detected.  As such, at least $50$\% of the X-ray--detected
sources have optical counterparts dominated by the central engine. Of
the 15 sources with BLAGN spectra, 13 are point-like, compared to 2 of
the 4 sources with NLAGN spectra.  Of the \plagas\ with available
GOODS \textit{HST} data, $\sim 30\%$ of the sources with optical
counterparts are point-like.

\begin{figure}[t]
\centering
\epsscale{1.1}
\plotone{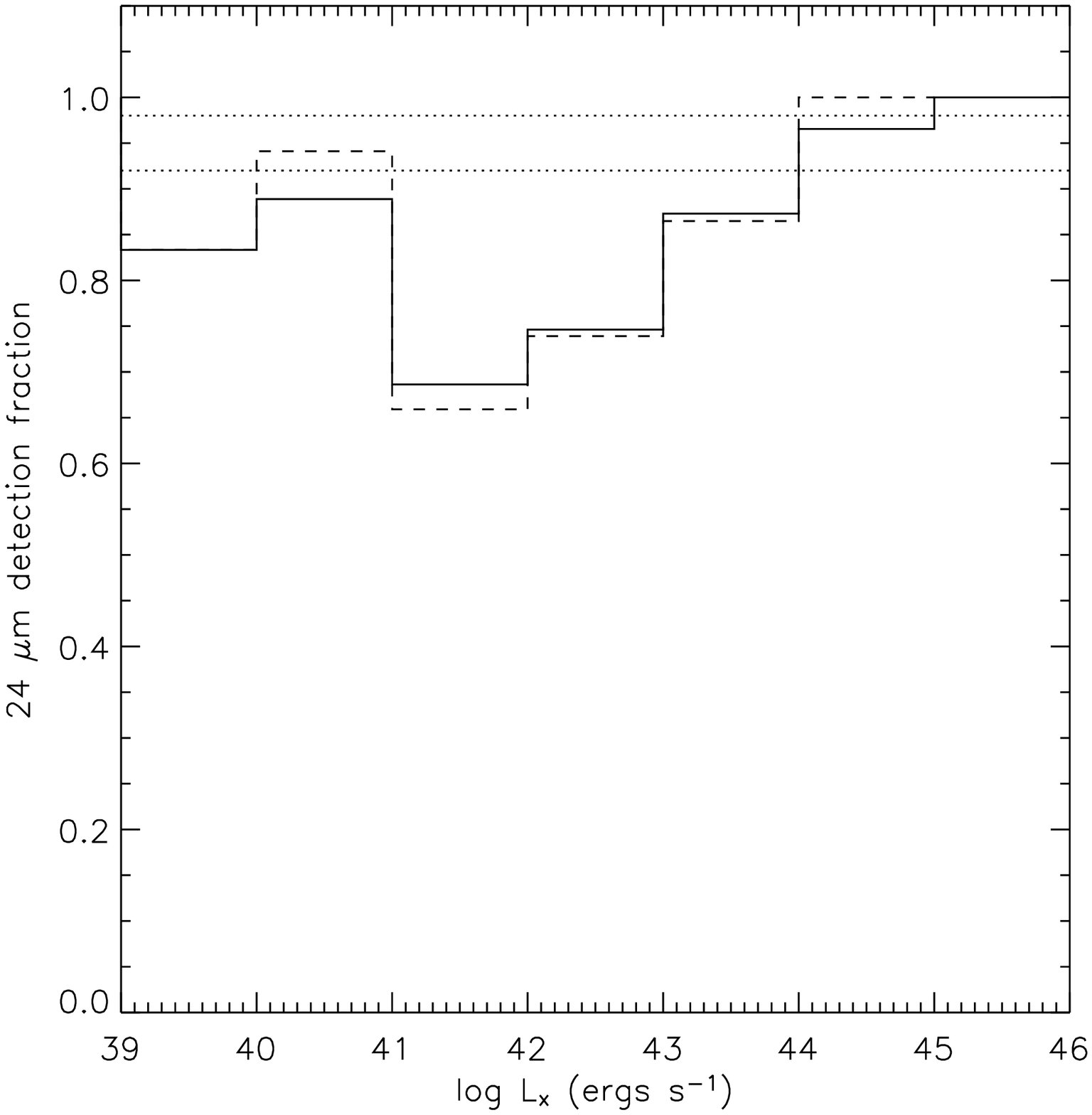}
\caption{24 \micron\ detection fraction as a function of X-ray luminosity
for the members of the comparison sample detected in the X-ray.  The
solid and dashed lines give the detection fraction for all sources
with redshift estimates and for those with spectroscopic redshifts,
respectively.  Dotted lines give the upper and lower limit on the 24
\micron\ detection fraction of the \plagas.}
\end{figure}

\begin{figure*}
\centering
\epsscale{0.8}
\plotone{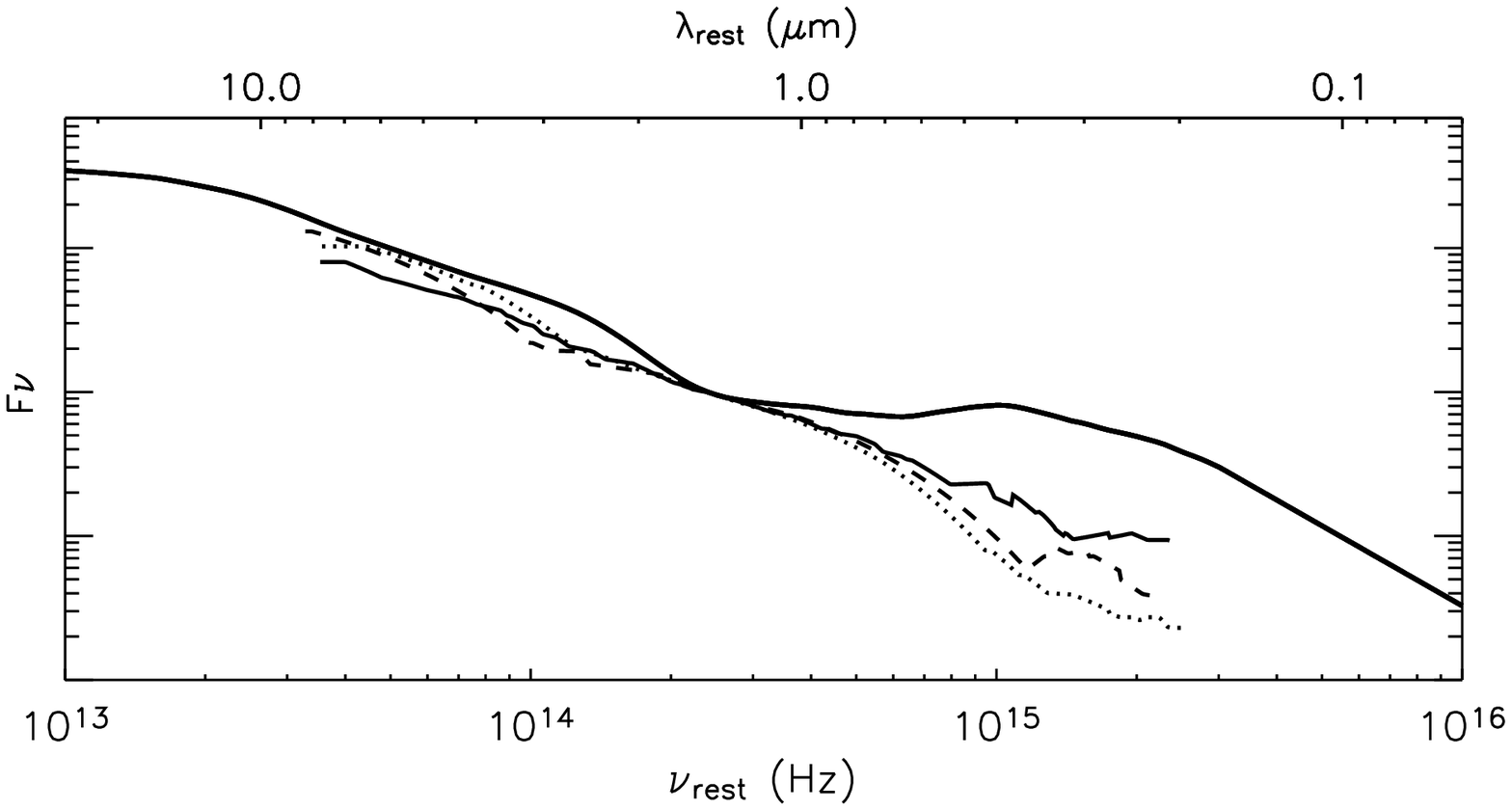}
\caption{Median optical through MIR rest-frame SEDs of the X-ray--detected 
(solid line), weakly-detected (dashed line), and non-detected (dotted
line) \plagas.  The median radio-quiet SED of Elvis et al. (1994) is
shown by a solid line that spans the frequency range of the plot.  The
SEDs have been normalized to their 1.25 \micron\ flux densities.}
\end{figure*}

Worsley et al. (2006) stacked the X-ray emission from optical sources
in the GOODS Fields to test whether these galaxies could account for
the remaining unresolved fraction of the CXRB.  While the stacked
emission of galaxies not individually detected in the X-ray accounts
for the unresolved portion of the CXRB at 0.5-6~keV, the reduced
sensitivity of \chandra\ at higher energies makes a determination of
the resolved fraction more difficult.  They estimate that GOODS
optical sources account for at most 40\% of the unresolved emission
from 6-8 keV, suggesting that some of the missing CXRB sources may
also be missed in the GOODS images.  Of the 27 \plagas\ that lie in
the GOODS region, 23 (85\%) have cataloged GOODS optical counterparts.
Two of the sources missed in the catalog (CDFN~32673 and CDFN~39529)
are obscured X-ray sources similar to the extreme X-ray/optical
sources (EXOs, e.g. Koekemoer et al. 2004), one (CDFN~33052) is not
detected in the X-ray, and one (CDFN~25218) is an obscured,
weakly-detected X-ray source that may have a faint B-band counterpart.
Therefore, at most 7\% (2/27) of the power-law galaxies in the GOODS
region are obscured AGN not detected in either the GOODS optical or
the X-ray catalogs.

\subsection{Optical--MIR SEDs}

We plot in Figure 10 the median optical through MIR rest-frame SEDs of
the cataloged, weakly-detected, and X-ray non-detected \plagas,
normalized by their 1.25 \micron\ monochromatic flux densities. For
comparison, we plot the median radio-quiet, starlight-subtracted AGN
SED of Elvis et al. (1994).  The median SEDs of the \plagas\ are
generally flatter than the Elvis et al. SED, and show no evidence of a
UV bump.  The median SED of the \plagas\ with X-ray detections
resembles that of the BLAGN SED class of \aah; that of the
non-detected \plagas\ drops much more rapidly in the optical,
presumably due to increasing optical obscuration, and resembles that
of the the NLAGN SED class of \aah.  The weakly-detected sources have
optical SEDs that fall between those of the detected and non-detected
sources, suggesting an intermediate level of obscuration. The steepest
IRAC spectral index found in our sample is $\alpha = -3.15$, only
slightly steeper than that found by \aah\ ($\alpha = -2.8)$ and Rigby
et al. (2005) ($\alpha = -2.9)$.

Brand et al. (2006) show that X-ray detected AGN in the 5~ks
XBo\"{o}tes field and MIR-detected sources with clear signs of AGN
activity in their optical spectra tend to have flat MIR spectral
slopes ($\rm{log}\ \nu f_{\rm \nu}(24)/\nu f_{\rm \nu}(8)=0$,
e.g. Elvis et al. 1994), whereas star-forming galaxies typically have
higher values of log$\nu f_{\rm \nu}(24)/\nu f_{\rm \nu}(8) \sim 0.5$.
Power-law galaxies have log$\nu f_{\rm \nu}(24)/\nu f_{\rm \nu}(8) =
-0.24$ to $0.73$, with a mean slope in $\nu f_{\rm \nu}$ of $0.18 \pm
0.22$, suggesting that, when compared to unobscured AGN flat in
log~$\nu f_{\rm \nu}$, either relatively more flux is observed at 24
\micron\ or relatively less flux is observed at 8 \micron. The
\plagas\ in the X-ray catalog tend to have slightly flatter slopes
($0.12 \pm 0.21$) than those not strongly detected in X-rays ($0.26
\pm 0.21$).  This trend can be seen in Figure 10, where the \plagas\
not detected in the X-ray rise more prominently in log~$\nu f_{\rm
\nu}$ with increasing wavelength than do those galaxies detected in
X-rays, suggesting that the X-ray non-detected \plagas\ are more
heavily obscured in the optical.

\section{Comparison with other MIR selection techniques}

In the following section, we compare the power-law selection technique
to a number of other MIR-based AGN selection criteria.  While the
\plagas\ display a large range of infrared-to-optical ratios, and cannot
be selected on this basis, they comprise a significant fraction of the
sources with high infrared-to-optical ratios, suggesting that many
heavily optically-reddened sources have power-law continua in the
NIR/MIR. The power-law selection more closely matches the \spitzer\
color-color selection criteria of Lacy et al. (2004) and Stern et
al. (2005).  We compare the X-ray completeness of these techniques,
and show that, like the power-law selection, these color criteria tend
to select high-redshift X-ray luminous AGN in the deep X-ray fields.
We also discuss the X-ray detection fractions of the color-selected
samples, which are lower than that of the power-law sample.  We argue
that this is due, at least in part, to a larger contamination by
star-forming galaxies.


\subsection{Optical vs 24 \micron\ Emission}

The selection of infrared-bright, optically-faint sources has been
suggested as a means of identifying obscured AGN (Houck et al. 2005;
Weedman et al. 2006; Yan et al. 2004, 2005).  Approximately 40 such
sources have been observed with the \spitzer\ IRS; the majority (75\%)
are dominated by silicate absorption, indicative of heavy obscuration,
or are featureless.  Although no deep X-ray information is available
for those sources observed thus far, their spectral characteristics
suggest that many are likely to be obscured AGN.

Only one member of the power-law sample, CDFN~27360 (an
X-ray--detected AGN with a column density of $N_{\rm H} = 2.5 \times
10^{23}$~cm$^{-2}$ (see Table 3)), meets the strict criterion used by
Houck et al. (2004) and Weedman et al. (2006) to select
infrared-bright, optically-faint sources: $f_{24} > 0.75$~mJy and
$R>24.5$ (Houck et al. 2004), or $f_{24} > 1.0$~mJy and $R>23.9$
(Weedman et al. 2006).  These selection criteria, however, have been
set to ensure that sources can be followed-up effectively with IRS.
We plot in Figure 11 the positions of the comparison sample (discussed
in \S4) on the color-color plot of Yan et al. (2004); with the
exception of the 7 sources that are X-ray non-detected at the
2.5$\sigma$ level (see \S5.1), 6 of which have no R-band counterpart,
only sources detected at 8\micron, 24\micron, and R-band are shown.

\begin{figure}
\centering
\epsscale{1.1}
\plotone{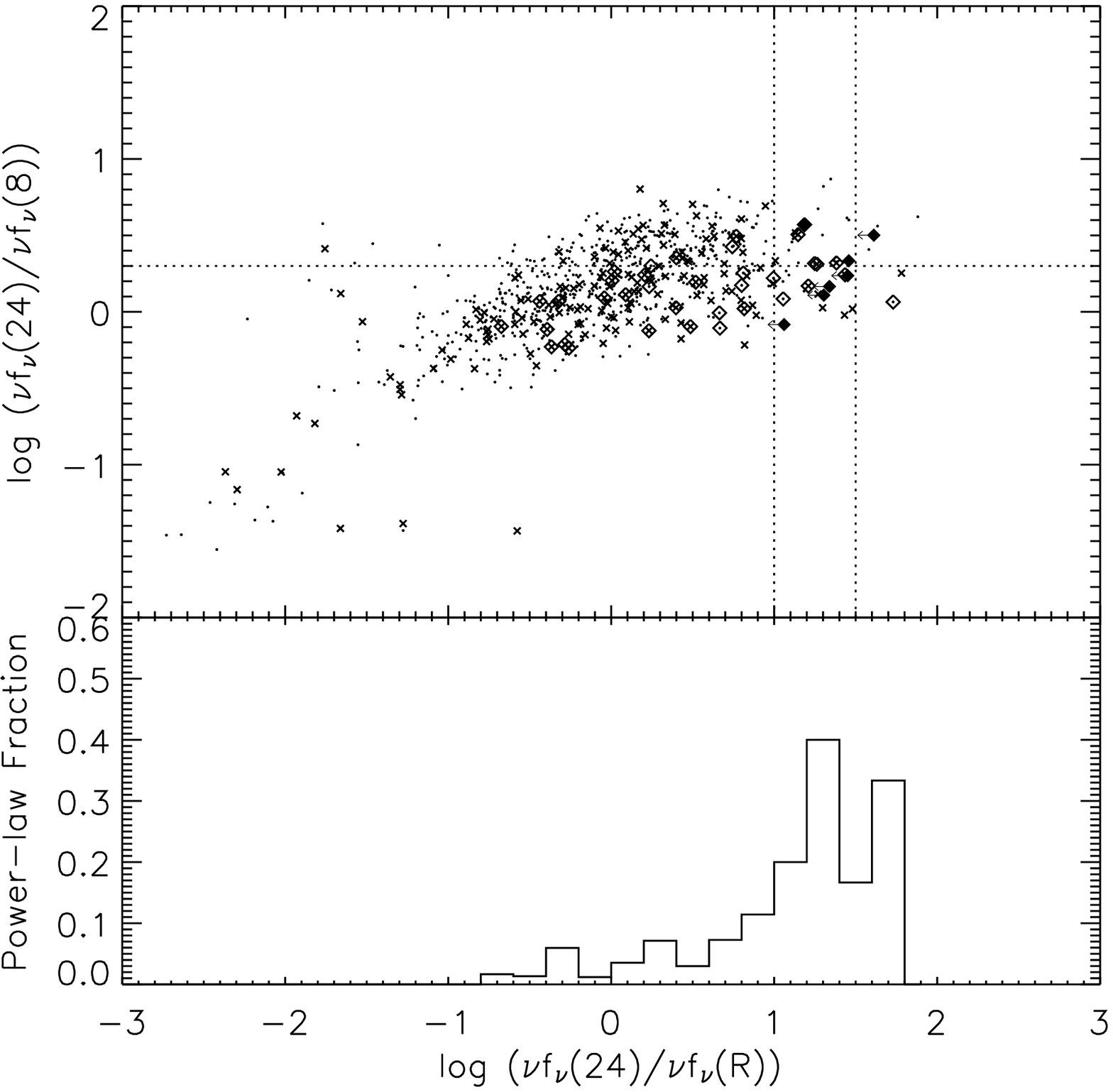}
\caption{Location of the comparison sample on the color-color diagram
of Yan et al. (2004).  All members of the comparison sample are shown
(dots), as are the \plagas\ (diamonds) and the X-ray--detected members
of the comparison sample (crosses).  The 7 \plagas\ that are X-ray
non-detected to the 2.5 $\sigma$ level are indicated by filled
diamonds. Dotted lines show the selection criteria of Yan et
al. (2004), as discussed in \S6.1. The lower panel shows the power-law
fraction of the comparison sample as a function of R(24,0.7). }
\end{figure}

Following Yan et al. (2004, 2005) we define the following quantities:
\begin{equation}
R(24,0.7) = \rm{log} (\nu f_{\nu}(24\micron)/\nu f_{\nu}(R))
\end{equation}
\begin{equation}
R(24,8)   = \rm{log} (\nu f_{\nu}(24 \micron)/\nu f_{\nu}(8 \micron))
\end{equation}

\noindent According to Yan et al. (2004), sources with $R(24,0.7) >
1.5$ and $R(24,8) \sim 0.5$ are likely to be dust-reddened AGN,
although 75\% of the luminous starburst candidates with $R(24,0.7) >
1$ and $R(24,8) > 0.3$ followed-up with IRS by Yan et al. (2005)
appear to be either unobscured AGN or galaxies with both a buried AGN
and a starburst component.  While the \plagas\ (and X-ray--detected
members of the comparison sample) cover a large range of $R(24,0.7)$
and $R(24,8)$, as is expected for a sample of AGN with a variety of
redshifts and obscurations, the \plagas\ comprise an increasingly
significant fraction of the highly optically-reddened members of the
comparison sample with $R(24,0.7) \ge 1.2$.  This suggests that
power-law selection is capable of detecting both optically obscured
and unobscured AGN, as expected, and that a significant fraction
(20-40\%) of the infrared-bright/optically-faint sources in the
comparison sample have power-law SEDs in the NIR/MIR. In addition,
those \plagas\ shown to be X-ray undetected to the 2.5$\sigma$ level
all have upper limits of $R(24,0.7) \ge 1$, consistent with the
expectations for obscured AGN.  In the following section, we compare
the power-law selection to selection criteria that more closely match
those used here.

\subsection{\spitzer\ Color-Color Selection}

MIR AGN color selection criteria have been defined by Ivison et
al. (2004), Lacy et al. (2004), Stern et al. (2005), and
Hatziminaoglou et al. (2005).  The completeness and reliability of the
latter three criteria in selecting X-ray--detected AGN in the Extended
Groth Strip (EGS) is discussed by Barmby et al. (2005). We plot in
Figures 12 and 13 the position of the \plagas\ with respect to the
IRAC color-color cuts of Lacy et al. (2004) and Stern et al. (2005),
both of which have been designed for relatively shallow surveys. In
addition, we overplot the redshifted IRAC colors of the Dale \& Helou
(2002) star-forming template (which is degenerate with their parameter
$\alpha$ for the wavelengths of interest), the median radio-quiet AGN
SED of Elvis et al. (1994), and the SEDs of the ULIRGs Mrk 273,
IRAS~17208-0014, and Arp 220, described in Appendix A.

The selection criteria of Lacy et al. are based on SDSS quasars, and
therefore are not designed to select AGN in which the host galaxy
dominates the MIR energy, as well as AGN that are obscured in the MIR.
As discussed in \aah\ and as shown in Figure 12, IRAC power-law
galaxies fall along a line well within the Lacy et al. selection
region, although these galaxies fill only a small fraction of the
available color space within the defined cut. Only 16\% of the objects
from the comparison sample that satisfy the Lacy et al. color cuts
also meet the power-law criteria.

The Stern et al. (2005) selection criteria provide a slightly closer
match to the power-law selection technique.  Stern et al. define
their color cut using $\sim 9400$ sources from AGES\footnote{the AGN
and Galaxy Evolution Survey, C. S. Kochanek et al., in prep}, 800 of
which are spectroscopically confirmed AGN. Their color criterion
identifies 91\% of the BLAGN and 40\% of the NLAGN in the AGES sample.
17\% of the sources that met their AGN criteria were not classified as
AGN based on their optical spectra, but may be optically normal
(optically dull) AGN. All of the \plagas\ meet the Stern et
al. criteria, and 29\% of the sources from the sample that meet the
Stern et al. criteria also meet the power-law criteria.  The \plagas\
show a larger scatter about the power-law locus in the Stern et
al. diagram (Figure 13) than they do in the Lacy et al. diagram
(Figure 12); this is likely to be due, at least in part, to the
smaller wavelength baseline probed by the Stern et al. colors.

\subsubsection{X-ray completeness of color-selected samples}

In the EGS, with an X-ray exposure time of $\le 200$~ks, the Lacy et
al. and Stern et al. color criteria selected 73\% and 51\% of the
X-ray AGN, respectively (Barmby et al. 2005).  In the deeper CDF-N
field, these criteria select only 21\% and 17\% of the Alexander et
al. (2003) X-ray sample.  However, only 50\% of the X-ray sources in
the CDF-N meet the exposure time and IRAC $S/N$ cuts used to define
the comparison and color-selected samples.  When we consider only
those X-ray sources in the comparison sample, the Lacy et al. and
Stern et al. criteria select 39\% and 33\% of the X-ray sources,
respectively. Further restricting the comparison to sources in the
comparison sample with X-ray luminosities indicative of AGN activity,
log~$L_{\rm x}$~(erg~s$^{-1}$)$ \ge 42$, increases the selection
fractions to 52\% and 47\%, respectively.  For comparison, our strict
power-law criteria recover $\sim 20\%$ of the X-ray luminous AGN in
the comparison sample.

Power-law galaxies, a subsample of both the Lacy et al. and Stern et
al. criteria, tend to preferentially lie at both high luminosity and
high redshift.  The same is true for the full color-selected samples,
as shown in Figures 14 and 15, in which we plot only sources with
spectroscopic redshifts and only the Lacy et al. color criteria.
Including photometric redshifts does not change the results, and the
same trends are seen for both the Lacy et al. and Stern et
al. selection techniques.

\begin{figure}
\centering
\epsscale{1.2}
\plotone{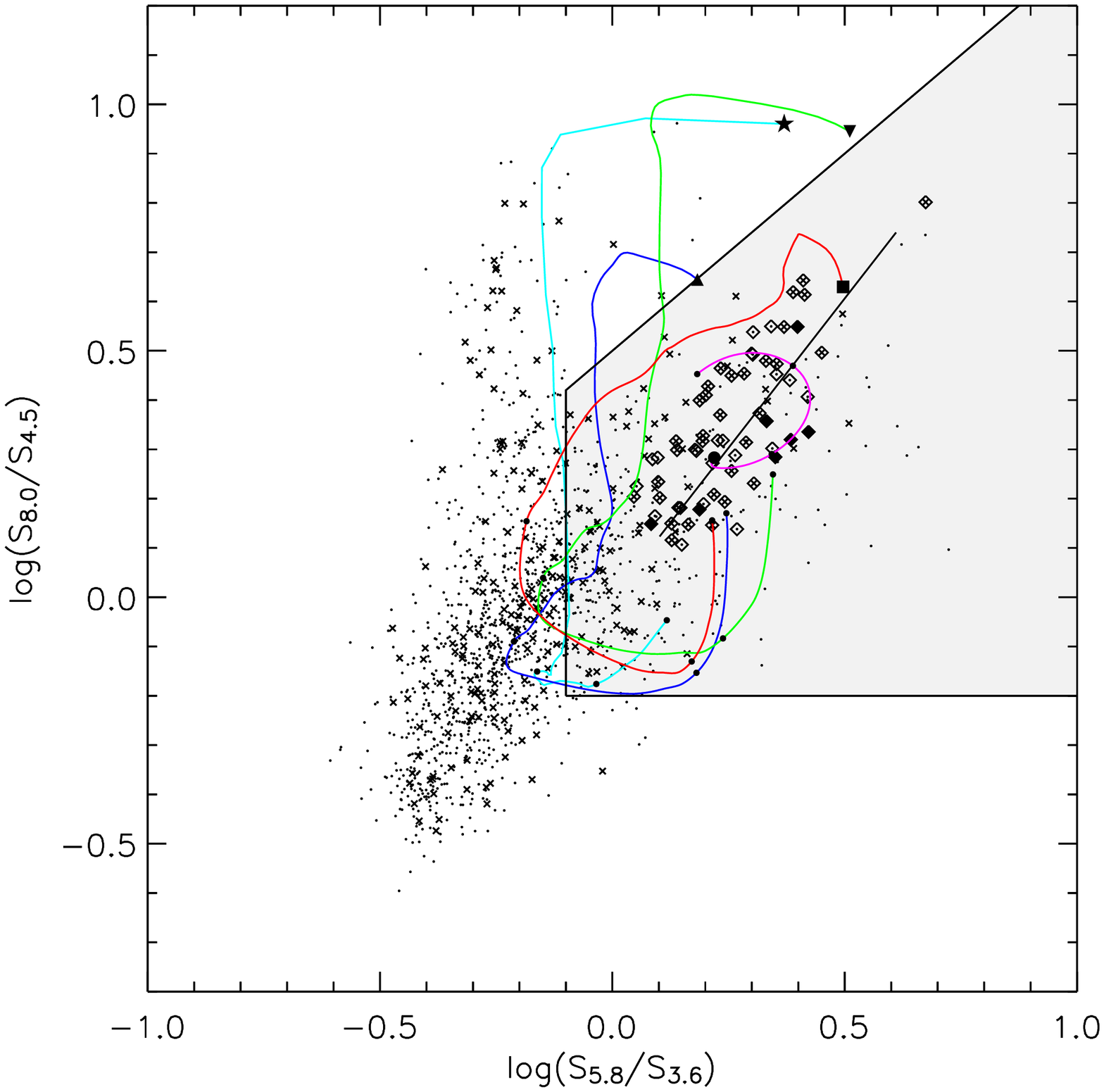}
\caption{Location of the comparison sample on the color-color diagram
of Lacy et al. (2004). Symbols are as described in Figure 11.
Overplotted are the power-law locus from $\alpha = -0.5$ to $\alpha =
-3$ (thin black line) and the redshifted IRAC colors of a typical
star-forming galaxy (Dale \& Helou 2002, cyan, star), the cold
(starburst dominated) ULIRGs Arp 220 (blue, triangle) and IRAS~17208-0014
(green, upside down triangle), the ULIRG/Sey 2 Mrk 273 (red, square)
and the radio-quiet AGN SED from Elvis et al. (1994, magenta, circle),
where the indicated point represents the colors at $z=0$, and small
circles mark the colors at $z=1,2,\rm{ and}\ 3$.}
\end{figure}

At low redshift, only a small fraction of the comparison sample lies
inside the Lacy et al. selection region (see Figure 14).  At redshifts
of $z>1$, however, the vertical branch in color-space populated by
low-redshift ($z<0.3$) aromatic-dominated sources (Sajina, Lacy, \&
Scott 2005) disappears, and at $z>2$, all of the remaining IRAC
sources meet the Lacy et al. criteria.  Similarly, at log~$L_{\rm
x}$~(erg~s$^{-1}$)$ < 42$, only a small fraction of the sources in the
comparison sample fall in the selection region (see Figure 15).  In
contrast, 72\% of the X-ray sources in the comparison sample with
log~$L_{\rm x}$~(erg~s$^{-1}$)$ > 43$ and 100\% of those sources with
log~$L_{\rm x}$~(erg~s$^{-1}$)$ > 44$ meet the Lacy et al. selection
criteria.  These high luminosity sources lie in the same region of
color space as the power-law selected AGN (see Figure 12), although as
is apparent from Figure 6, the Lacy et al. criteria (and the Stern et
al. criteria) select a more complete sample of AGN with high X-ray
luminosities than does the power-law selection. As is also shown in
Figure 6, however, the X-ray luminosities of the color-selected
sources extend below the traditional AGN limit of log~$L_{\rm
x}$~(erg~s$^{-1}$)$ > 42$, suggesting either heavy obscuration of
highly-luminous AGN, or possible contamination by star-forming
galaxies.

\vspace{2cm}
\subsubsection{AGN Reliability}

Of the sources selected via the power-law criteria, 55\% were detected
in the X-ray catalog of Alexander et al. (2003). The X-ray detection
fractions of the sources selected via the Lacy et al. and Stern et
al. criteria are lower: 25\% and 38\%, respectively. We searched for
faint X-ray emission from the color-selected sources using the method
described in \S5.1; 65\% of the Lacy et al. sources and 70\% of the
Stern et al. sources show evidence for X-ray emission (at the
2.5$\sigma$ level or higher), compared to 85\% for the \plagas. Do
these techniques select a larger fraction of AGN not detected in
X-rays, or do they suffer from greater contamination from star-forming
galaxies?

\begin{figure}
\centering 
\epsscale{1.2}
\plotone{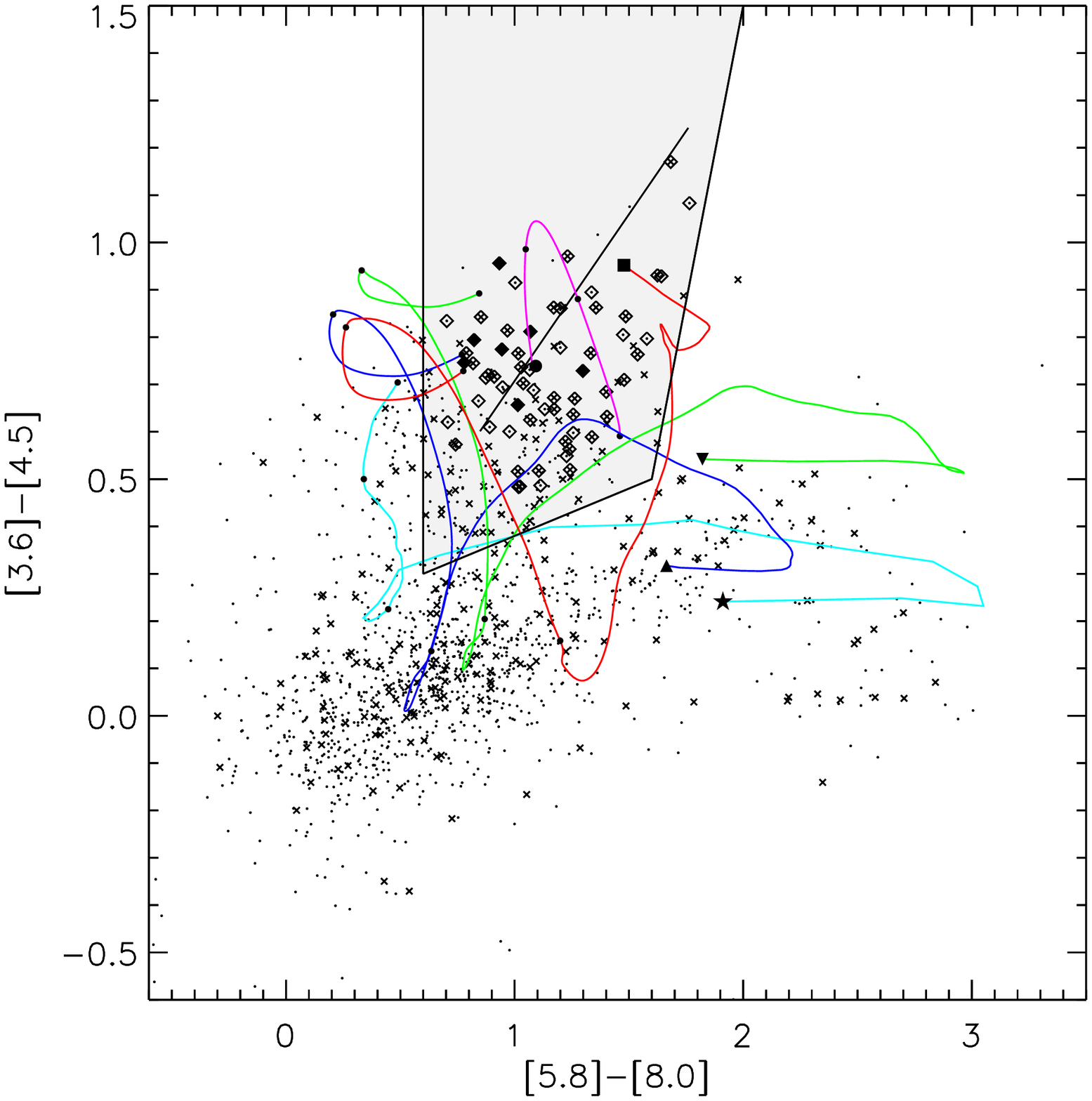}
\caption{Location of the comparison sample on the color-color diagram
of Stern et al. (2005). Symbols are as described in Figure 11 and
templates are as described in Figure 12.}
\end{figure}

\begin{figure*}
\centering
\epsscale{0.75}
\plotone{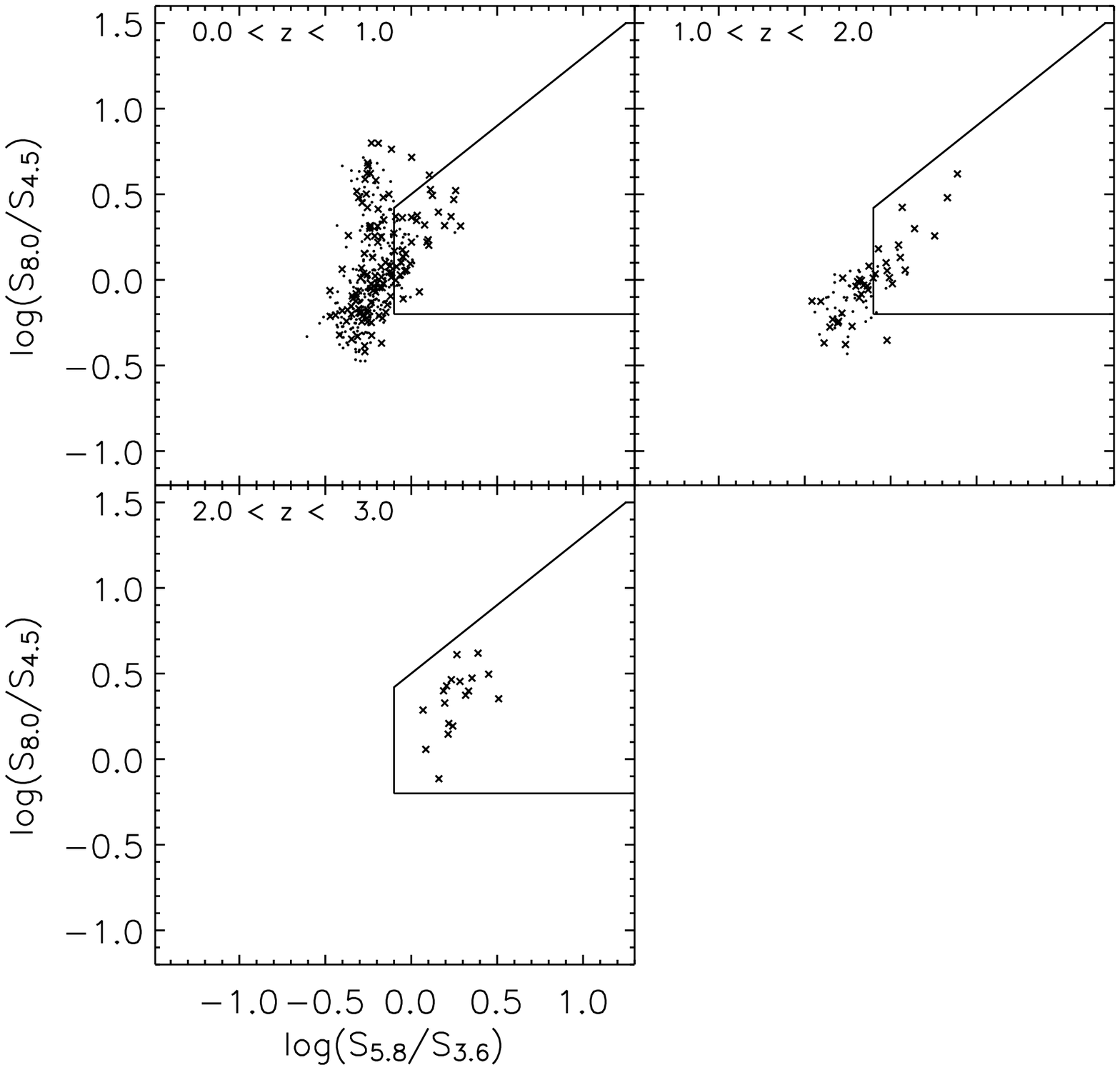}
\caption{Location in the Lacy et al. (2004) color space of the
comparison sample as a function of redshift. Only sources with
spectroscopic redshifts are shown. X-ray sources in the comparison
sample are indicated by crosses. }
\end{figure*}

\begin{figure*}
\centering
\epsscale{0.75}
\plotone{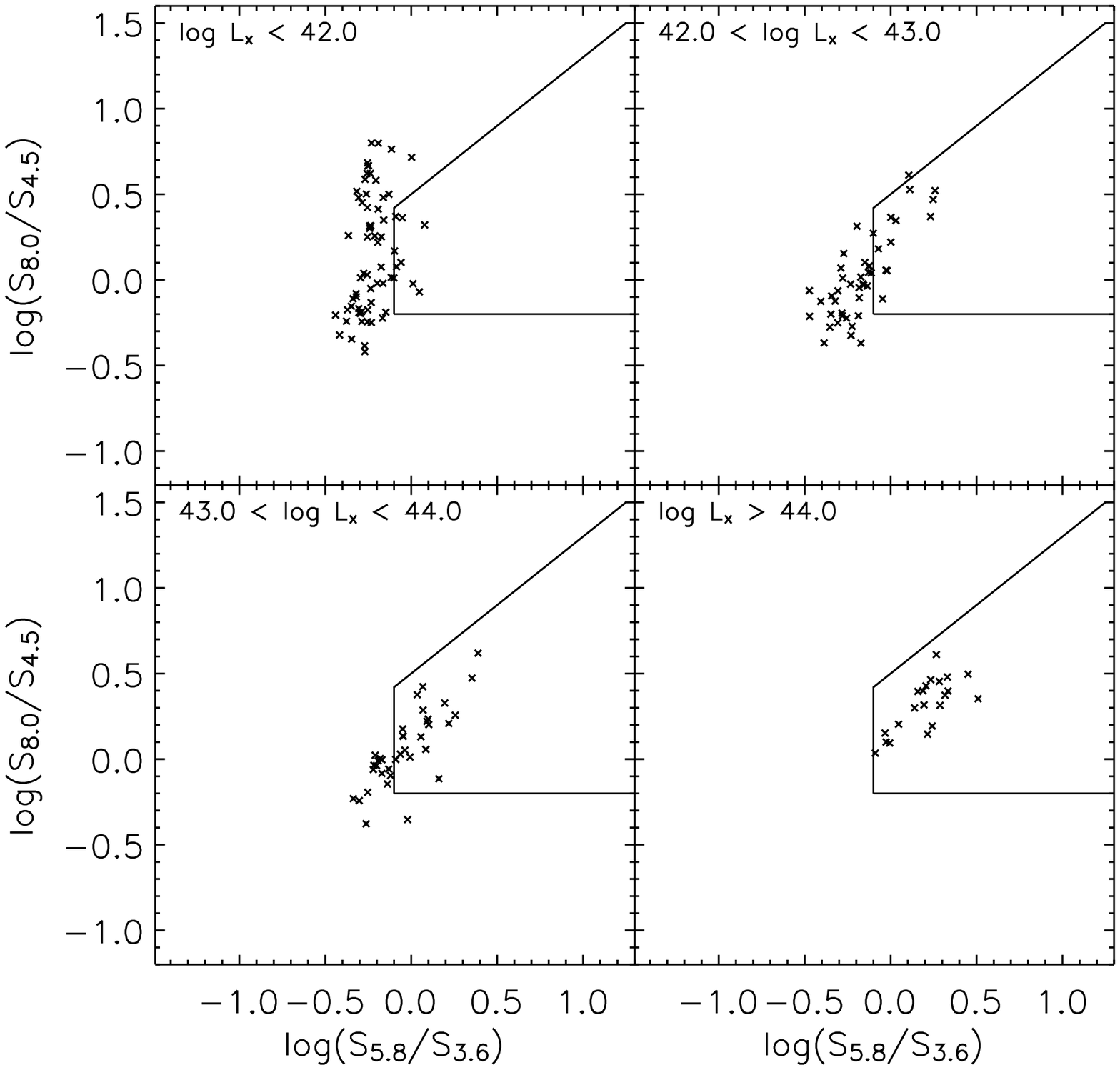}
\caption{Location in the Lacy et al. (2004) color space of the
comparison sample as a function of X-ray luminosity (in units of
ergs~s$^{-1}$). Only sources with spectroscopic redshifts are shown.}
\end{figure*}

\begin{figure}
\centering
\epsscale{1.1}
\plotone{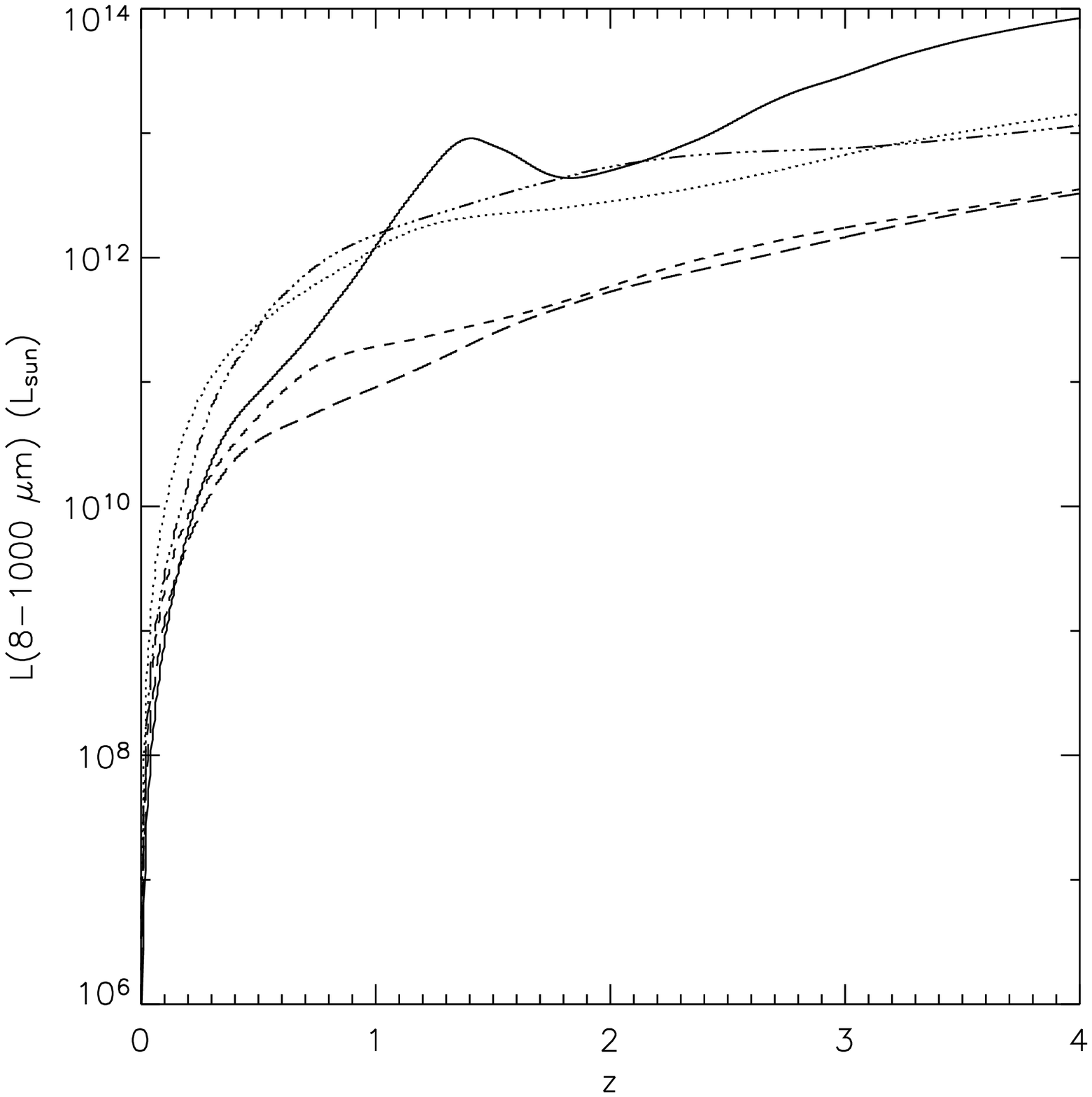}
\caption{Total infrared luminosity (8-1000 \micron) required for Arp~220
to meet the detection limits of our survey (see \S3). The lines
represent the luminosity required for detection in the MIPS 24
\micron\ (solid) and IRAC 3.6 \micron\ (long-dashed), 4.5 \micron\ (short-dashed), 
5.8 \micron\ (dotted), and 8.0 \micron\ (dot-dashed) bands.}
\end{figure}

A spectroscopic follow-up of candidate obscured AGN selected via the
Lacy et al. criteria reveals that most ($\sim 80\%$) have spectral
signatures typical of type-2 AGN (Lacy et al. 2005).  However, the 12
candidate sources for which spectra were taken were chosen to have 8
\micron\ flux densities $> 1$~mJy, and all lie at $z<1.34$. As shown
in Figure 12, the SEDs of the star-formation dominated ULIRG Arp~220
enters the Lacy et al. selection region at low ($z=0.23$~to~0.67) and
moderately high ($z>1.43$) redshift, as does that of IRAS~17208-0014
($z=0.36$~to~0.89 and $z>1.35$).  The Dale \& Helou (2002)
star-forming template shows similar behavior, falling just inside the
AGN selection region at low redshift ($z=0.48$~to~0.58) and
re-entering the region at high redshift ($z>1.66$), suggesting that
pure color-selection is capable of selecting both normal and
highly-luminous star-forming galaxies at low and moderately high
redshift.  The simulations of Sajina et al. (2005) also show a
significant number of aromatic feature-dominated sources near the
outskirts of the Lacy et al. AGN selection region, and suggest that
while the IRAC colors provide an effective means of identifying
obscured AGN up to $z=2$, such criteria are less effective at higher
redshift, where the IRAC bands sample the NIR light from the galaxy
(see also Alonso-Herrero et al. 2006). As expected, the AGN template
(Elvis et al. 1994) lies in the same region of color space as the
\plagas, as does the low-redshift template of Mrk~273, a ULIRG with
AGN signatures in the optical and NIR (Veilleux et al. 1999; Risaliti
et al. 2006). Mrk~273 meets the Lacy et al. criteria at all redshifts
except $z=0.83$~to~1.40.

The Stern et al. criteria also suffer from potential contamination
from cold ULIRGs and star-forming galaxies, although the star-forming
templates tend to fall in the selection region for smaller redshift
intervals than in the Lacy et al. color-space.  The star-forming
template of Dale \& Helou (2002) lies along the edge of the selection
region at low redshift ($z=0.51$~to~0.56) but it does not re-enter the
selection region at the redshifts plotted here ($z<3$). Arp~220 falls
in the selection region twice (from redshifts of $z=0.28$~to~0.55 and
$z=1.17$~to~1.44), as does IRAS~17208-0014 ($z=0.49$~to~0.58 and
$z=1.15$~to~1.60). Once again, both the AGN template (Elvis et
al. 1994) and the low-redshift Mrk~273 template have colors similar to
those of the \plagas. The Mrk~273 template meets the Stern et
al. criteria in three redshift intervals ($z=0$~to~0.06,
$z=0.19$~to~0.39, and $z=1.22$~to~1.59), whereas the AGN template
falls in the selection region at all redshifts plotted here.

The Mrk~273, Arp~220, and IRAS~17208-0014 templates all lie near regions of
color-space populated by \plagas.  The template of Mrk~273, a ULIRG
with AGN signatures, meets our power-law criteria ($\alpha \le -0.5$,
$P>0.1$, monotonically rising) at redshifts of $z=0$~to~0.08,
$z=0.18$~to~0.39 and at $z>2.88$ (assuming 10\% flux errors in the 3.6
and 4.5 \micron\ bands, and 15\% flux errors in the 5.8 and 8.0
\micron\ bands).  The Arp~220 template, however, does not meet these
criteria at redshifts less than $z \sim 2.9$, as expected.  With the
exception of a narrow redshift window of $z=0.51$~to~0.53 in which
IRAS~17208-0014 appears as a power-law galaxy, the same is true for
this star-forming template, which meets our criteria at $z \gtrsim
2.8$. Only five of the sources in our sample with redshift estimates
lie at $z<1$; none of these lie between $z=0.51$~and~0.53, and all are
detected in the X-ray.  We therefore expect little contamination from
low-redshift ULIRGS, although the presence of such sources cannot be
ruled out. As for contamination from high redshift star-forming
galaxies, a ULIRG with an SED like that of Arp~220 that lies at
$z=2.8$ must have a total infrared luminosity (TIR, 8-1000 \micron) of
log~$L_{\rm TIR}$(L$_\sun$)$ > 13.3$ to have a 24 \micron\ flux
density $> 80$ \microjy\ (see Figure 16). The fraction of ULIRGS
dominated by AGN increases with infrared luminosity, with 50\% of
ULIRGs showing Seyfert-like properties at log~$L_{\rm TIR}$(L$_\sun$)$
> 12.3$ (Veilleux et al. 1999).  It is therefore unlikely that
HyperLIRGS with log~$L_{\rm TIR}$(L$_\sun$)$ > 13$ have NIR
luminosities dominated by star-formation.  As such, we also expect
little contamination in the power-law sample from high-redshift
star-forming galaxies, so long as the power-law objects are also
detected at 24 \micron\ (as is the case both for the sample of \aah\
and for the majority of our sample).

As a final check, we plot in Figures 12 and 13 the seven power-law
galaxies in our sample that are not detected in the X-ray at the 2.5
$\sigma$ level. With one or two exceptions, they are well removed from
the colors of Arp~220 and IRAS~17208-0014 for any value of $z<2.8$
(particularly in Figure 12).  We therefore believe they contain AGN
that are heavily obscured in the X-ray. This test illustrates the
utility of the color-color plots in identifying regions of color space
within the color selection regions and near the power-law locus in
which contamination by star-forming galaxies is possible. The current
templates suggest that pure color selection is likely to select
star-forming galaxies and ULIRGs at both low and high
redshifts. Several other lines of evidence also point towards a higher
fraction of star-forming galaxies in the color-selected samples. While
$\ge 92$\% of the \plagas\ are detected at 24 \micron, the 24 \micron\
detection fractions of the Lacy et al. and Stern et al. samples are
lower: 72\% and 83\%, respectively.  In addition, while 88\% of the
radio-detected \plagas\ with redshift estimates have radio
luminosities of $L(1.4\rm{GHz})>24$~W~Hz$^{-1}$, the value typical of
AGN-powered radio galaxies in GOODS-N (Morrison et al. 2006), only
49\% and 54\% of the Lacy et al. and Stern et al. radio sources (with
redshift estimates) have luminosities that exceed this value. Finally,
we plot in Figure 17 the position of the color-selected sources on the
X-ray to optical diagnostic diagram. While the current detections and
upper limits place only two \plagas\ within the transition region and
none near the region populated by quiescent galaxies (see Figure 2),
15\% and 10\% of the Lacy et al. and Stern et al. sources fall in
these regions, and more have upper limits that place them just
outside.

\begin{figure*}
\centering
\epsscale{1}
\plottwo{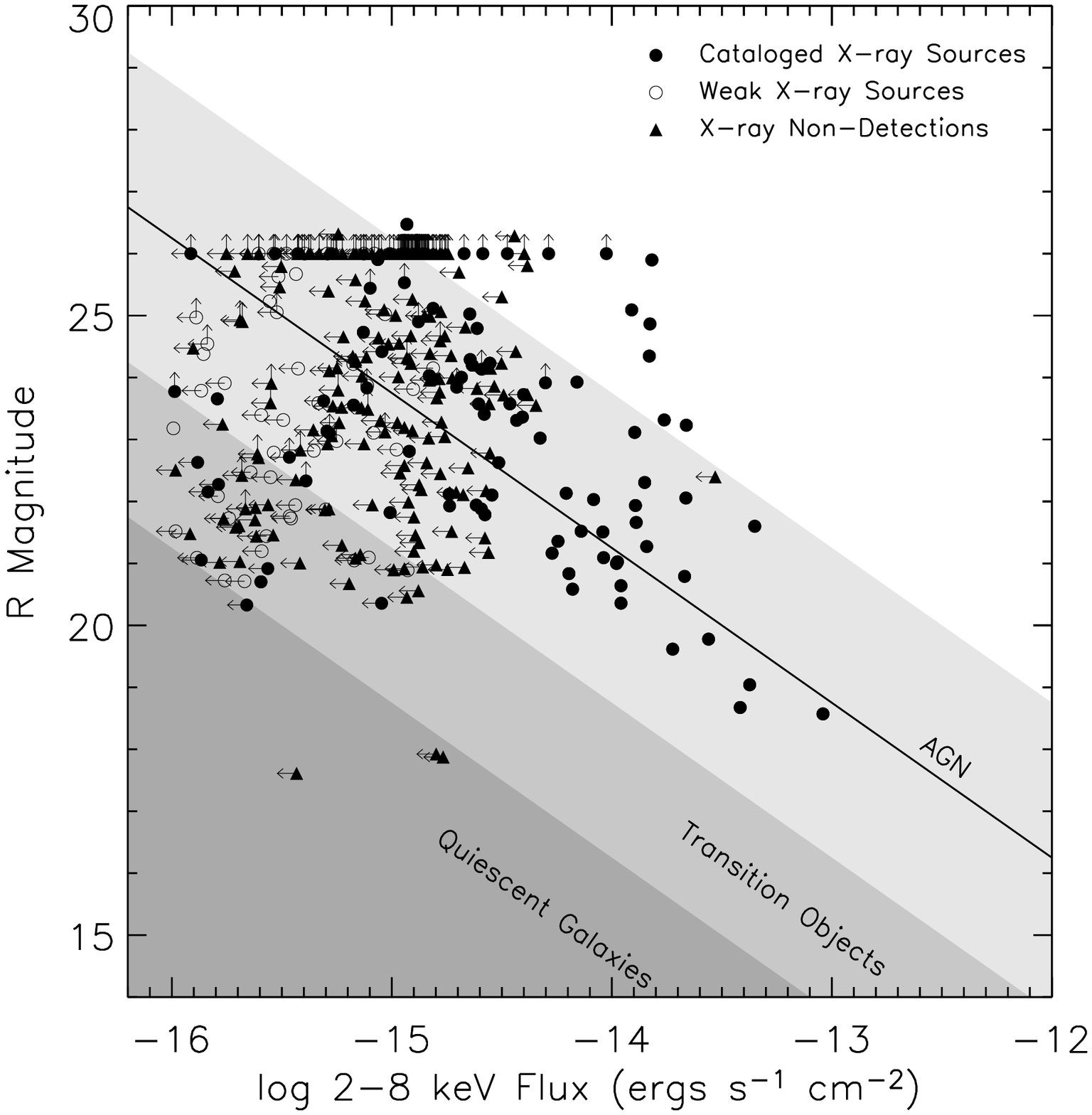}{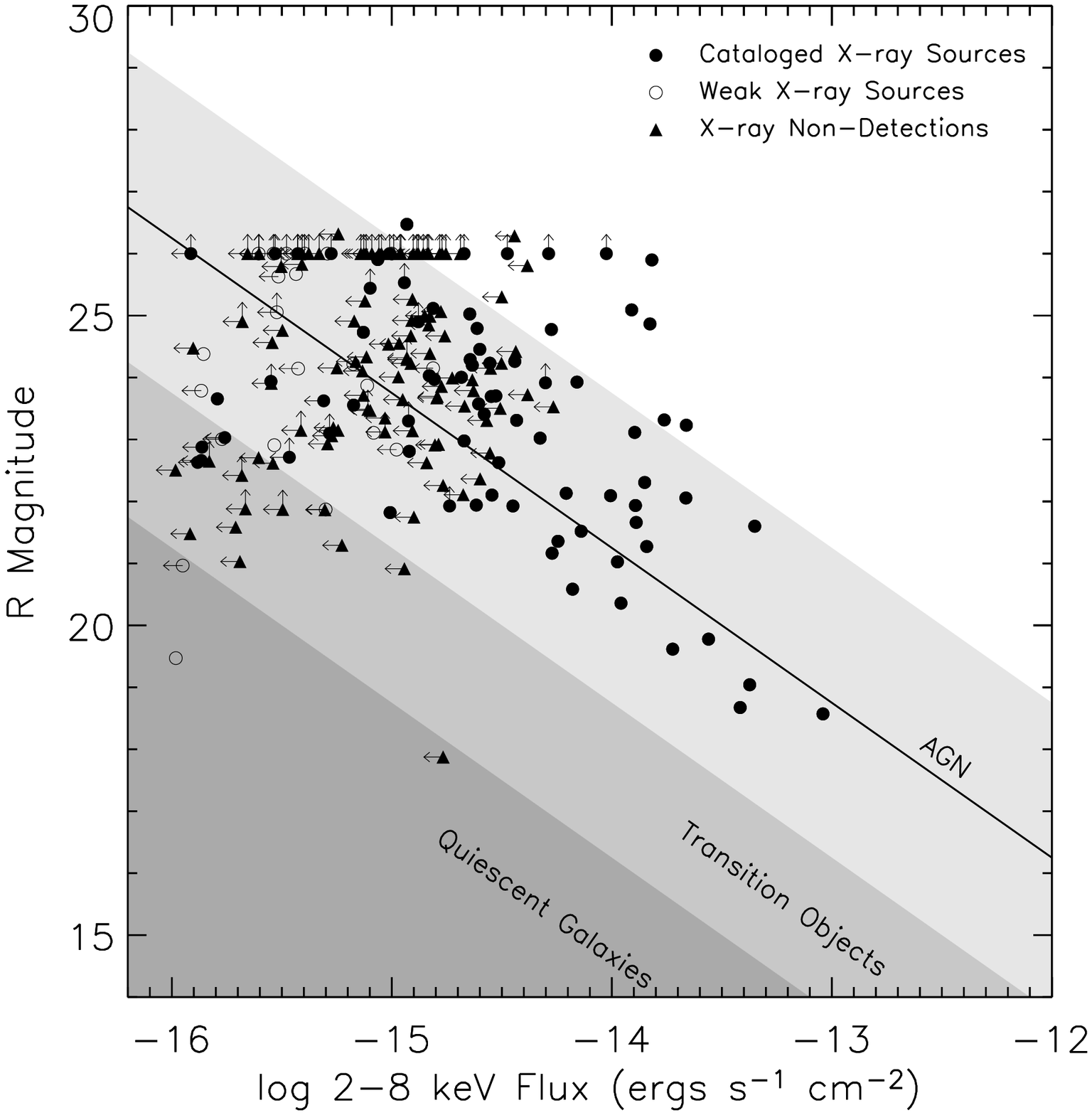}
\caption{Relationship between the observed R-band magnitude and hard
(2--8 keV) X-ray fluxes for the Lacy et al. (2004) sources (left) and
the Stern et al. (2005) sources (right).  Symbols, lines, and shading
are as described in Figure 2.}
\end{figure*}

\subsubsection{SEDs of color-selected sources}

How do the MIR SEDs of the Lacy et al. and Stern et al. selected
sources compare to those of the \plagas?  Of the Lacy et al. sources,
20\% can be fit by red, AGN-dominated power-laws ($\alpha < -0.5$) and
30\% can be fit by blue, stellar-dominated power-laws ($\alpha >
-0.5$). A higher fraction of the Stern et al. sources have good
power-law fits: 37\% are red power-laws and 34\% are blue
power-laws\footnote{We note we have applied only a power-law fit
probability to derive the above numbers, and have excluded the other
criteria discussed in \S3 (i.e. no turnovers in flux density, no
apparent stellar bump). The number of galaxies identified with red
power-laws using just these two criteria is therefore 79, not 62.}.
The comparison sample (see Section 4) shows the behavior of objects
selected without regard to the presence of an AGN; 6\% and 39\% of the
full comparison sample can be fit by red and blue power-laws, as can
16\% and 35\% of the cataloged X-ray sources in the comparison sample.

\section{Obscuration}

In the local universe, Seyfert 2's are four times more numerous than
Seyfert 1's (Maiolino \& Rieke 1995) and at least half of all Seyfert
2's are Compton-thick (Maiolino et al. 1998; Risaliti et al. 1999).
In the more distant universe, X-ray background and luminosity function
synthesis models predict global obscured (log~$N_{\rm H}$(cm$^{-2}$)$
\ge 22$) ratios of 3:1 to 4:1 (Comastri et al. 2001; Ueda et al. 2003;
Gilli 2004; Treister et al. 2005; Tozzi et al. 2006), significantly
higher than the observed type 2 to type 1 ratio of spectroscopically
identified X-ray sources in the deep fields, 2:1 (e.g., Barger et
al. 2003; Szokoly et al. 2004; Treister et al. 2005), but slightly
lower than the 6:1 ratio observed for high-redshift X-ray--detected
SCUBA galaxies (Alexander et al. 2005).

In the following discussion, we estimate the X-ray column densities of
the power-law sample and compare the resulting obscured fraction to
predictions from the XRB.  If we consider only those \plagas\ detected
in the X-ray, we find an obscured ratio of 2:1.  Including the
\plagas\ both weakly- and non-detected in the X-ray results in an obscured
ratio of $\lesssim 4:1$. We use a Monte Carlo code to measure the
dispersion in this ratio, and investigate the change in the obscured
fraction with X-ray luminosity and redshift. By comparing the space
density of obscured AGN in the power-law sample to the predictions of
Treister et al. (2006), we estimate that at most 20-30\% of obscured,
MIR-detected AGN have SEDs that meet our robust power-law criteria.
We also discuss the effect of large obscuring columns on the NIR/MIR
continuum, and show that while heavily obscured AGN can have NIR
emission dominated by the AGN, the power-law criteria may be biased
against the most heavily obscured (Compton-thick) AGN.

\vspace{1cm}
\subsection{Column Densities}

We estimated the intrinsic column density, $N_{\rm H}$, of each
X-ray--detected power-law and color-selected galaxy by comparing the
observed hard to soft-band X-ray flux ratio to that expected for a
typical AGN at a given redshift, after correcting the flux ratios for
a Galactic column density of $N_{\rm H} = 1.6 \times
10^{20}$~cm$^{-2}$.  When no redshift was available, we estimated the
column densities at redshifts of $z=0.5$ and $z=3.0$, the approximate
limits of the power-law sample.  We assumed an intrinsic photon index
of $\Gamma = 1.81$ (Tozzi et al. 2006), and did not include a Compton
reflection component, which tends to decrease the necessary column by
$\sim 20$\% (see Donley et al. 2005).  The estimated column densities
of the \plagas\ are given in Tables 2 and 3 and plotted in Figure 18.

\subsubsection{Obscured fraction}

A comparison between the X-ray-cataloged and weakly-detected \plagas\
is complicated, because only one of the latter has both a hard and
soft-band detection. We can therefore place only upper limits on the
columns of 6 of the weakly-detected AGN, lower limits on two, and no
limit on the remaining one.  However, all weakly-detected \plagas\ for
which we can estimate $N_{\rm H}$ are consistent with being obscured
($N_{\rm H} > 10^{22}$~cm$^{-2}$), but not Compton-thick ($N_{\rm H} >
10^{24}$~cm$^{-2}$).  Of the X-ray--detected \plagas, $\sim$~68\% are
obscured, in agreement with obscured fractions found by Ueda et
al. (2003) for AGN at similar redshifts.  If we consider only those
\plagas\ in the X-ray catalog, we therefore derive an obscured to
unobscured ratio of $\sim$~2:1.

Several lines of evidence point towards high obscuration in the
sources not detected in X-rays.  First, the power-law selection
criteria (\S5.1.2) and high 24 \micron\ detection fraction (\S5.2)
appear to be indicative of a population of intrinsically X-ray
luminous sources (see Fig. 10), sources that, if unobscured, should be
detectable out to high redshift (see Fig. 9).  Secondly, the median
optical--MIR SED of the sources not detected in X-rays falls below
that of the weakly-detected sources at optical wavelengths, which in
turn falls below that of the strongly-detected sources, suggesting a
continuously increasing optical obscuration for a given 1.25 \micron\
flux density (see Fig. 11).  Adding the 10 weakly-detected \plagas,
all of which are likely to be obscured (see Table 2), we calculate an
obscured fraction of $\sim$~75\%.  If we futher include the \plagas\
not detected in X-rays, assuming that all are obscured, the maximum
obscured fraction of \plagas\ rises to 82\% (4:1-5:1). (Considering
only those \plagas\ with $\theta < 10 \arcmin$ gives an obscured
fraction of 81\%).  This upper limit on the obscured fraction is
consistent with the fraction of NLAGN SEDs found by \aah\ for
power-law sources in the CDF-S ($\sim 75$\%).

To place a lower limit on the obscured fraction, we consider two
additional scenarios. If we assume that all of the X-ray non-detected
\plagas\ are obscured, as before, but assume instead that the 6 
weakly-detected \plagas\ with upper limits on their column densities are
\textit{unobscured}, the obscured fraction drops to $\sim$ 71\%.
If we further consider only those \plagas\ with column density
estimates (i.e. those sources strongly or weakly detected in the
X-ray) and assume again that all of the weakly-detected \plagas\ with
upper limits on their column densities are unobscured, the minimum
obscured fraction drops to $\sim 60\%$.  This lower limit, however,
falls below the obscured fraction of the X-ray--detected \plagas\
(68\%).  As it is unlikely that the \plagas\ not detected in the X-ray
are significantly less obscured than those detected in the X-ray, the
obscured fraction of the X-ray--detected \plagas\ can be taken as the
lower limit on the obscured fraction.

To investigate the dispersion in the observed ratio of obscured to
unobscured AGN, we ran a Monte Carlo simulation in which we varied
both the assumed intrinsic photon index, $\Gamma$, and the photometric
redshifts, and then recalculated the column densities of the
\plagas. The intrinsic photon index was drawn from the distribution of
Tozzi et al (2006), as measured for the brightest 30 sources in their
sample: $\Gamma = 1.81 \pm 0.20$.  We allowed the photometric
redshifts of the \plagas\ to vary about their mean by $\Delta (z) =
0.1 (1+z)$ (see \S4).  For the simulation, we assigned the 6 cataloged
and weakly-detected sources with hard-to-soft flux ratios but no
redshift estimate the median redshift of our sample, $z=1.9$. The
results are shown in Figure 19. The maximum obscured fraction varies
from $\sim 75$\% to $85$\% (3:1-6:1), with a mean of $81\%$ (4:1).

\subsubsection{Redshift and luminosity dependence of obscured fraction}

\begin{figure}
\centering
\epsscale{1.1}
\plotone{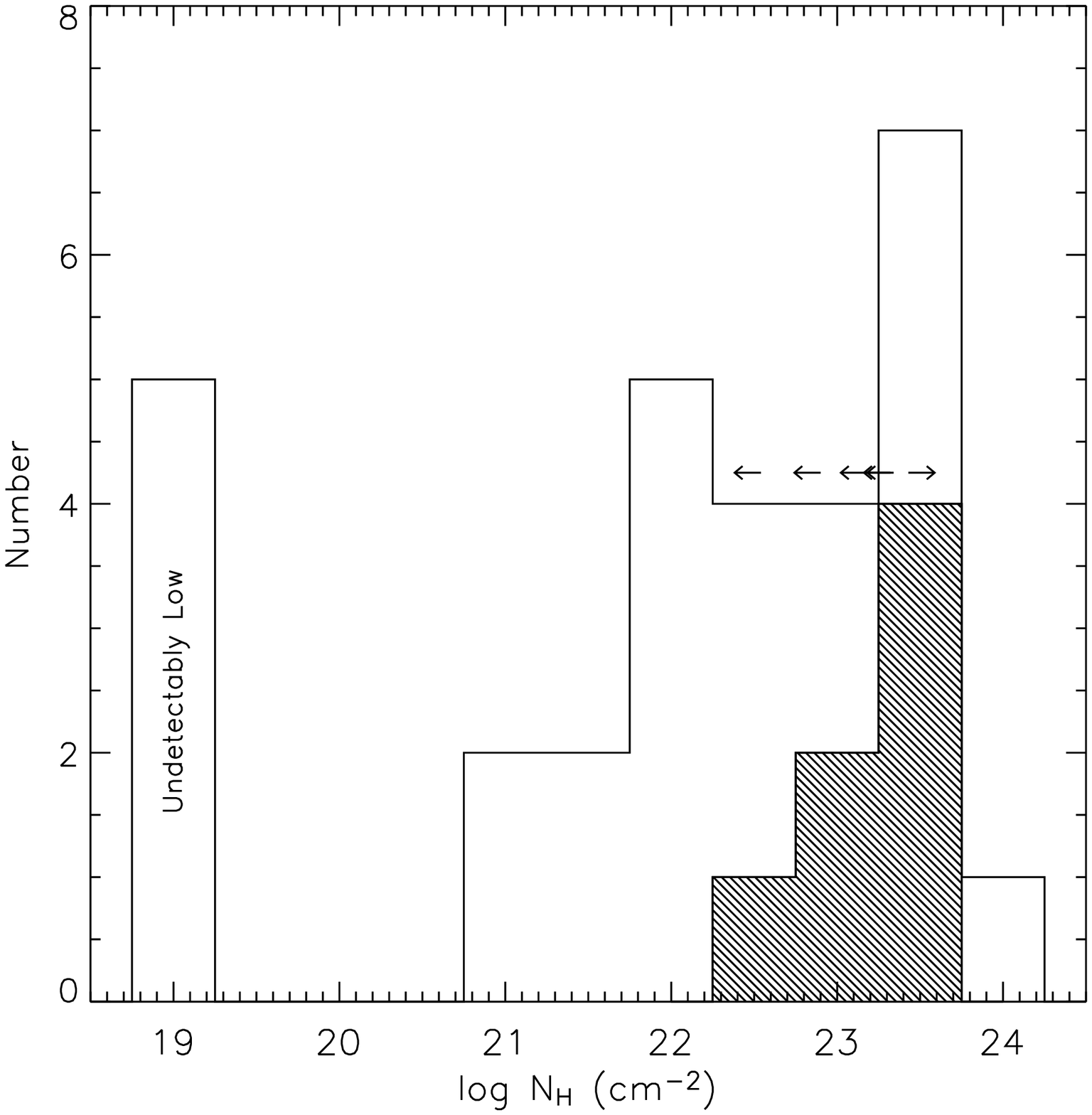}
\caption{Distribution of X-ray column densities $N_{\rm H}$(cm$^{-2}$) for 
the \plagas\ in the X-ray catalog (unshaded) and those only
weakly-detected (shaded). We place only upper or lower limits on the
column densities of all but one of the weakly-detected sources; the
limits of sources with redshift estimates are shown by small arrows.
Sources with undetectably low column densities were assigned columns
of log~$N_{\rm H}$(cm$^{-2}$)$ = 19$ for plotting.}
\end{figure}

We plot in the right panel of Figure 19 the obscured fraction as a
function of redshift and observed X-ray luminosity.  Solid and dotted
lines represent \plagas\ with low ($\rm{log}\ L_{\rm
x}$~(erg~s$^{-1}$)$ < 44$) and high ($\rm{log}\ L_{\rm
x}$~(erg~s$^{-1}$)$ > 44$) absorption-corrected X-ray luminosities,
and thin and thick lines represent \plagas\ with low ($z<2$) and high
($z>2$) redshifts, respectively.  We include all X-ray sources with
redshifts for which column densities are available, and exclude those
sources with only upper limits on the X-ray luminosity. In addition to
having a relatively high obscured to unobscured ratio, \plagas\ appear
to be more heavily obscured at low X-ray luminosities and at high
redshifts.  The upper limit on the obscured fraction varies from 0.5,
for low-redshift, high-luminosity \plagas, to 1, for high-redshift,
low-luminosity sources. As we can plot only those \plagas\ detected in
the X-ray for which redshift estimates are available (37/63),
systematic effects are likely to be present (e.g. Treister et
al. 2005).

The trends we find are consistent with previous results. A number of
authors have predicted an increase in the obscured fraction of AGN
with increasing redshift (e.g., Gilli et al. 2001, 2003; La Franca et
al. 2005; Ballantyne et al. 2006; Tozzi et al. 2006), although a
constant obscured fraction also appears to be consistent with the
current data (Ueda et al. 2003; Gilli 2004; Treister et al. 2004;
Treister \& Urry 2005). Akylas et al. (2006) suggest that the observed
increase in the obscured fraction with redshift may be due to
statistical fluctuations in the small number of counts in the low
energy bands, which can cause the column densities of high redshift
sources to be overestimated.  Stronger evidence exists for a decrease
in the obscured fraction with increasing X-ray luminosity (Ueda et
al. 2003; Steffen et al. 2003; Hasinger 2004; Szokoly et al. 2004;
Barger et al. 2005; Treister et al. 2005; Ballantyne et al. 2006), as
expected from the receding torus model (Lawrence 1991). For instance,
Szoloky et al. (2004) find X-ray Type 2 fractions of 75\%, 44\%, and
33\% for AGN with X-ray luminosities of log~$L_{\rm
x}$~(erg~s$^{-1}$)$ = 42-43, 43-44, \rm{and}\ 44-45$, respectively,
and Ballantyne et al. (2006) find that the model that simultaneously
best describes local AGN data, the CXRB spectrum, and X-ray number
counts requires an obscured ratio of 4:1 for low luminosity AGN and
1:1 to 2:1 for AGN with log~$L_{\rm x}$~(erg~s$^{-1}$)$ > 44$.

\begin{figure*}
\centering
\epsscale{1}
\plottwo{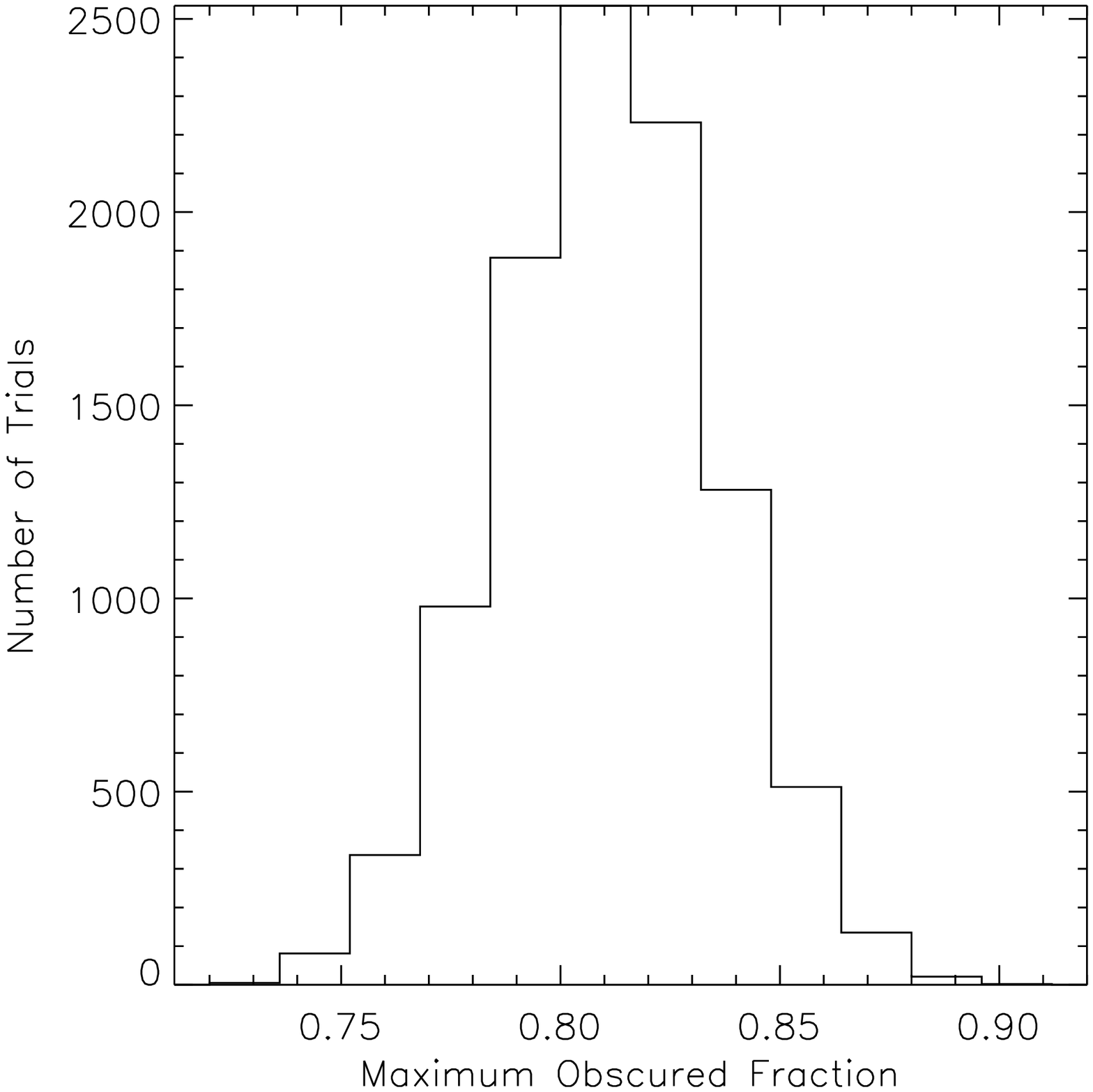}{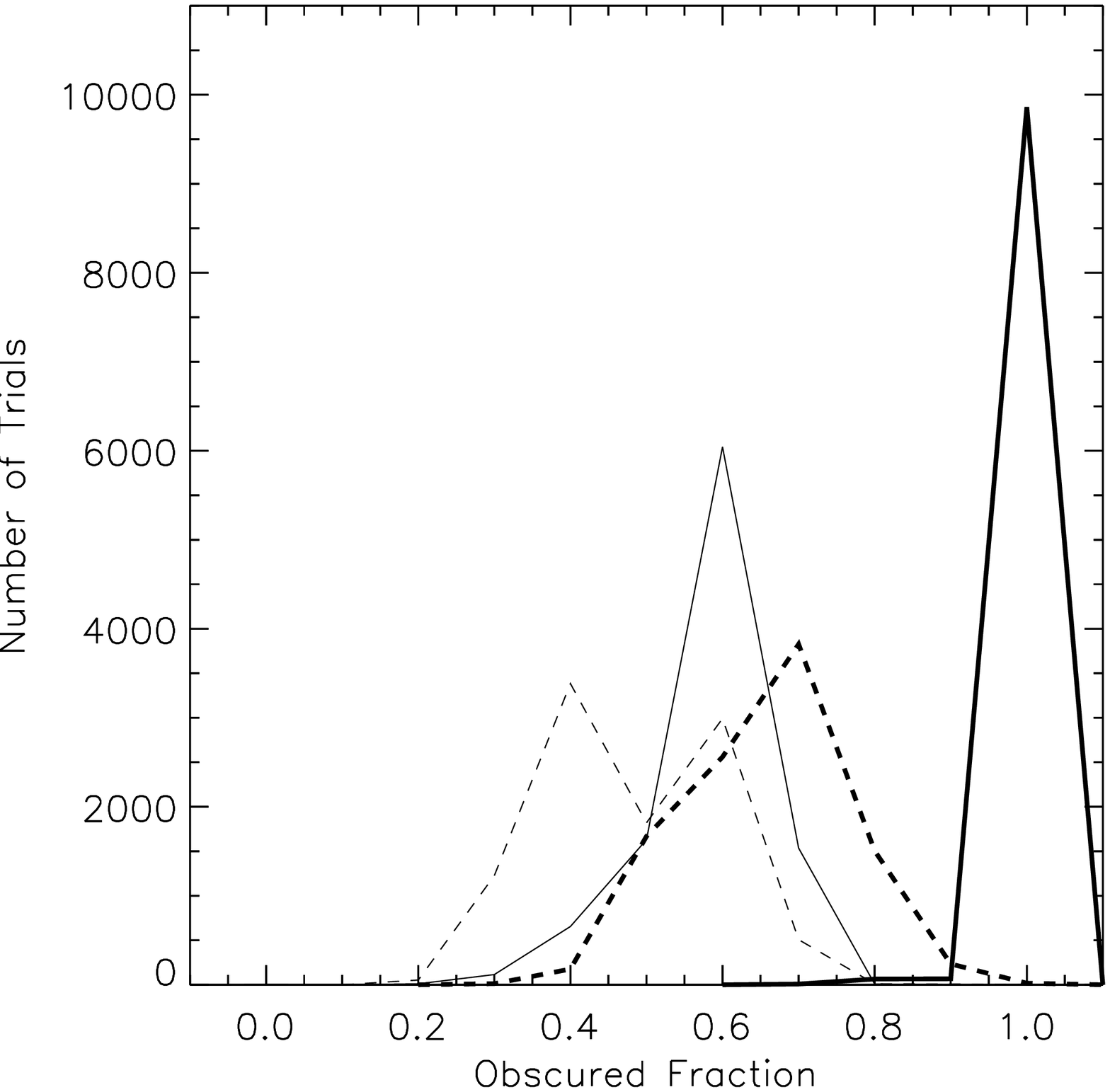}
\caption{Left: Upper limit on the fraction of obscured 
(log~$N_{\rm H}$(cm$^{-2}$)$ \ge 22$) \plagas. All \plagas\ not
strongly or weakly detected in X-rays were assumed to be
obscured. Right: Upper limit on the obscured fraction as a function of
redshift and X-ray luminosity. Sources with absorption-corrected
$\rm{log}\ L_{\rm x}$~(erg~s$^{-1}$)$ < 44$ are shown by solid lines;
dotted lines represent sources with $\rm{log}\ L_{\rm
x}$~(erg~s$^{-1}$)$ > 44$.  Thin and thick lines represent sources
with redshifts of $z<2$ and $z>2$, respectively.  For viewing clarity,
we do not plot the histograms, but instead connect their points.}
\end{figure*}

\subsubsection{Obscured fraction of color-selected samples}

If we assume that all of the color-selected sources are AGN, and that
all X-ray--non-detected sources are obscured, the Lacy et al. and
Stern et al. selection criteria give upper limits on the obscured
fraction of 92\% (12:1) and 88\% (7:1).  As discussed in \S6.2.2,
however, it is likely that these samples contain star-forming galaxies
as well as AGN; these estimates of the obscured fractions are
therefore almost certainly high.  By comparing the numbers of IR
color-selected and optically-selected AGN, Stern et al. (2004)
estimated an obscured AGN fraction of 76\%, similar to that found for
the \plagas, although Richards et al. (2006) use a different
infrared-to-optical flux ratio to revise this estimate to $\ge 41$\%.
Lacy et al. (2004) similarly found an obscured ratio of 46\% for the
brightest members of their sample.  It is important to note, however,
that both the color- and power-law criteria select luminous AGN whose
NIR/MIR emission is dominated by the central engine.  Low-luminosity
and heavily-obscured AGN whose NIR/MIR emission is dominated by the
host galaxy will tend to fall outside of the color-selection and
power-law regions, increasing the total obscured fraction of AGN.

\subsection{Space Densities of Obscured AGN}

If we assume that all of the \plagas\ not detected in X-rays are
obscured, the power-law sample contains at most $\sim$51 obscured AGN
and at least $\sim$11 unobscured AGN (assuming that CDFN~50281, with a
column density of $N_{\rm H} \le 10^{22}$~cm$^{-2}$ is unobscured).
How do these numbers compare to the predicted space densities of Type
1 and Type 2 AGN? 

Treister et al. (2006, private comm.) predict 199 AGN in a
350~arcmin$^{2}$ area (the area of our survey) down to a 24 \micron\
limiting flux density of $80$~\microjy, $\sim 105-110$ of which should
be detected in the hard X-ray band and $\sim 174$ of which are
obscured (Type II). For comparison, the X-ray catalog of Alexander et
al. (2003) contains 147-149 hard X-ray sources that meet our exposure
cut (and therefore lie in the survey area) and which have $S_{\rm 24}
> 80$~\microjy\ and AGN-like X-ray luminosities of log~$L_{\rm
x}$(ergs~s$^{-1}$)$ > 42$.  At a flux density of 80~\microjy, the
predicted number counts (for X-ray detected AGN) therefore lie within
a factor of $\sim 1.4$ of the observed values.

As discussed in \S5.2, all but 5 of the power-law galaxies have 24
\micron\ counterparts and all but one of the MIPS-detected sources
have a 24 \micron\ flux density $>80$~\microjy.  Of the 5 sources
without 24 \micron\ counterparts, however, only one source is a clear
non-detection. The power-law sample therefore contains 56-60 galaxies
with 24 \micron\ fluxes $>80$~\microjy.  Based on the Treister et
al. (2006) predictions, these sources should account for $\sim
28-30\%$ of the MIR-detected AGN (or $\sim 20-22\%$ if we boost the
Treister et al. counts by a factor of 1.4).  Likewise, the 45-49
obscured \plagas\ with 24 \micron\ fluxes $>80$~\microjy\ should
account for at most $\sim 26-28\%$ ($\sim 18-20\%$) of the obscured,
Type II, MIR-detected AGN.

Our results therefore suggest that at most $\sim 20-30$\% of the
MIPS-detected AGN and MIPS-detected obscured AGN have high $S/N$
power-law IRAC continua; the remainder could have SEDs dominated by or
strongly affected by the host galaxy, red power-law SEDs that fall
below our IRAC detection limit, or have been rejected due to noise
that caused them to fail our relatively stringent signal-to-noise
criteria.  It is not surprising that we find a fairly small portion of
power-law AGN, as our selection criteria require the AGN to be both of
high luminosity and energetically dominant over the other source
components in the near- to mid-IR.  Franceschini et al. (2005) find
that of a sample of \chandra-detected AGN in the SWIRE survey, only
62\% have optical-IR SEDs typical of Type 1 or Type 2 AGN, and 40-60\%
of X-ray--selected AGN are known to be optically-dull (Hornschemeier
et al. 2001; Barger et al. 2001; Giacconi et al. 2001).  The fraction
of AGN with optical--IR SEDs dominated by star-formation is likely to
be even higher for heavily obscured AGN samples, such as that
identified here. Polletta et al. (2006) select obscured AGN candidates
on the basis of red non-stellar SEDs, similar to the power-law
criterion used here; only 40\% of their obscured AGN candidates show
AGN signatures in the optical/NIR SEDs.

\subsection{Effect of Reddening on Obscured AGN}

Is it reasonable to assume that heavily obscured AGN at high redshift
could have NIR/MIR SEDs still dominated by the central engine?
Observations of QSOs indicate that AGN can be up to 2--3 magnitudes
brighter than their host galaxies in the NIR (e.g. McLeod \& Rieke
1994; Percival et al. 2001; Marble et al. 2003), although the
contribution of the non-stellar continuum varies from source to source
and as a function of wavelength. The total H-band contribution of
luminous AGN, for example, ranges from $\sim$35 to 90\% (McLeod \&
Rieke 1994), and while the non-stellar continuum of NGC~1068 accounts
for $>$80\% of the flux at 2.3 \micron\ in the inner $4\farcs4$, the
contribution drops to 30\% at 1.6
\micron\ (Origlia et al. 1993).

In the Milky Way, a column density of log~$N_{\rm H}$(cm$^{-2}$)$ =
23$ corresponds to an $A_{\rm V}$ of $\sim 50$~mag (Bohlin, Savage, \&
Drake 1978).  The $A_{\rm V}$ to $N_{\rm H}$ ratio of AGN, however,
appears to be an order of magnitude lower than that of the Milky Way
(e.g. Maccacaro, Perola, \& Elvis 1982; Reichert et al. 1985; Granato
et al. 1997; Maiolino et al. 2001a), due either to a higher
gas-to-dust ratio, or to the formation of large grains in the dense
AGN environment (Maiolino et al. 2001a, 2001b).  Following
Mart\'{\i}nez-Sansigre (2006), we therefore adopt an AGN $N_{\rm H}$
to $A_{\rm V}$ conversion of $2.0 \times 10^{-23}$, for which column
densities of log~$N_{\rm H}$(cm$^{-2}$)$ = 22,\ 23,\ \rm{and}\ 24$
correspond to $A_{\rm V} = 0.2,\ 2,\ \rm{and}\ 20$~mag, respectively.

Assuming the IR extinction law of Rieke \& Lebofsky (1985), we
estimate the observed IRAC extinctions for an obscured AGN at $z=2$,
where IRAC samples the $1-3$ \micron\ rest-frame emission.  At a
column density of log~$N_{\rm H}$(cm$^{-2}$)$ = 22$, the extinctions
at the center of the IRAC channels are negligible (0.02-0.06~mag), but
they rise with column density to 0.2-0.6~mag for log~$N_{\rm
H}$(cm$^{-2}$)$ = 23$ and 1.8-6.2~mag for log~$N_{\rm H}$(cm$^{-2}$)$
= 24$.  It is therefore reasonable to assume that luminous obscured
AGN could dominate the emission of their host galaxies at all but the
highest column densities.  

If the interstellar extinction law is applicable to AGN, differential
extinction across the IRAC bands will redden the observed spectral
slope, potentially increasing the number of heavily obscured luminous
AGN in the power-law sample.  At $z=2$, a column density of
log~$N_{\rm H}$(cm$^{-2}$)$ = 22$ has only a minor effect on the
measured slope: an observed spectral index of $\alpha = -0.5$
corresponds to an unreddened intrinsic slope of $\alpha = -0.45$.  At
higher column densities of log~$N_{\rm H}$(cm$^{-2}$)$ = 23$ and
log~$N_{\rm H}$(cm$^{-2}$)$ = 24$, however, the best-fit intrinsic
slopes at $z=2$ increase to $\alpha= 0.0$ and $\alpha = 4.5$.  The
unrealistic predicted intrinsic slope of the high-redshift
Compton-thick model suggests that highly obscured AGN at high redshift
are likely to have steep observed spectral slopes, if their IR and
X-ray emission are subject to the same obscuring material.  For
example, the spectral slope of CDFN~44836, $\alpha = -3.2$, could be
caused by the extinction of an $\alpha = 0$ source by a column density
of log~$N_{\rm H}$(cm$^{-2}$)$ = 23.8$ at $z=2$.  At such high column
densities, however, the de-reddened source begins to steepen at the
bluest bands, no longer resembling a power-law source.

While the above results therefore suggest that power-law selection may
be biased against Compton-thick AGN, there exist a population of AGN
whose X-rays are more heavily obscured than their optical/NIR emission
(e.g. Akiyama et al. 2003; Brusa et al. 2003; Page et al. 2003; Wilkes
et al. 2005).  For instance, 10\% of the BLAGN sample of Perola et
al. (2004) have X-ray columns of log~$N_{\rm H}$(cm$^{-2}$)$ \ge 22$.
Shi et al. (2006) also find several Compton-thick AGN that do not
follow an observed correlation between X-ray column density and the
strength of the silicate emission or absorption feature. Instead,
their silicate absorption is weaker than expected, suggesting a
Compton-thick absorber that obscures the X-ray emission, but not the
IR emission.  These outliers could also be explained, however, by a
Compton-thick absorber that obscures the mid-IR so strongly that the
output in this range is dominated by star formation in the host
galaxies.


\section{Conclusions}

We define a sample of 62 \plagas\ in the CDF-N.  Sources were required
to be detected to $S/N > 6$ in each of the IRAC bands, have IRAC
slopes of $\alpha < -0.5$ (where $f_{\nu} \propto \nu^{\alpha}$), and
to lie in regions with both deep X-ray ($>0.5$~Ms) and radio coverage.
We studied the multiwavelength properties of the \plagas\ and compared
the power-law selection technique to other MIR-based AGN selection
criteria.  We then measured the intrinsic obscuring column densities
of the \plagas\ to estimate the obscured fraction of the sample, which
should be less affected by obscuration than optical and/or X-ray
selected samples.  The main results of this paper are as follows:

\begin{itemize}

\item
Power-law selection requires the AGN to be energetically-dominant in
the near/mid infrared.  On average, AGN with X-ray luminosities of
log~$L_{\rm x}$~(erg~s$^{-1}$)$ < 44$ have SEDs dominated by or
strongly affected by the stellar continuum.  Therefore, power-law
galaxies tend to have high X-ray luminosities, and make up a
significant fraction of the X-ray luminous AGN population.

\item
Power-law galaxies lie at significantly higher redshifts than the
typical IRAC-- or X-ray--detected source, primarily due to their high
luminosities. While the power-law sample defined here accounts for only
$\sim 20\%$ of the X-ray and MIR-detected AGN in the comparison
sample, approximately 50\% of the high-redshift X-ray sample are
\plagas.

\item  
45\% of the \plagas\ are not detected in the X-ray catalog of
Alexander et al. (2003) at exposures of $>0.5$~Ms.  A search for faint
emission reveals that 15\% remain undetected at the $2.5\sigma$
detection level. X-ray detection is not a strong function of redshift,
X-ray exposure time (at $>1$~Ms), power-law slope, or power-law fit
probability.

\item
Almost all (93-98\%) of the IRAC-selected \plagas\ are detected at 24
\micron.  The 24 \micron\ detection fraction of AGN in the comparison
sample increases with X-ray luminosity. Because we require all
\plagas\ to be detected in each of the four IRAC bands, the 24
\micron\ detection fraction is almost certainly higher than would be
expected for an unbiased sample of AGN.  Nevertheless, comparison with
a sample suffering from the same bias (the comparison sample) shows
that the high detection fraction is representative of an intrinsically
X-ray luminous AGN population.

\item
30\% of the \plagas\ have radio counterparts from the Richards et
al. (2000) VLA survey of the CDF-N.  Only two of the 18 radio-detected
sources (11\%) have radio emission in excess of that predicted by the
radio-infrared correlation, suggesting an overall radio-loud fraction
of only 3\%.  In comparison, almost all of the radio-excess AGN of
Donley et al. (2005) have optical/MIR SEDs dominated by the stellar
bump.  Further exploration of a possible systematic difference in the
radio properties of the two samples would be interesting. While there
is very little overlap between the radio-excess and power-law samples,
both have similarly low X-ray detection fractions, suggesting that
this may be a common feature of AGN selected independently of their
X-ray and optical properties, regardless of redshift or luminosity.

\item
The optical--MIR SEDs of the \plagas\ are flatter than the median
radio-quiet SED of Elvis et al. (1994).  Sources not detected in
X-rays have SEDs that drop off more rapidly in the optical than those
of the \plagas\ detected in X-rays, as discussed in \aah, presumably
due to increasing optical obscuration.  Weakly-detected
\plagas\ have intermediate optical SEDs.

\item
At least half of the \plagas\ detected in the X-ray catalog have
compact optical counterparts, suggesting that the optical light is
dominated by the AGN, as expected.  Only 2 non-- or weakly-detected
sources have compact optical counterparts. 15\% of the \plagas\ in the
GOODS field do not have optical counterparts brighter than the GOODS
limiting magnitudes.
   
\item
Power-law galaxies comprise a subset of the MIR sources selected via
color criteria (e.g. Lacy et al. 2004; Stern et al. 2005).  While
color-selected AGN samples include a higher fraction of
high-luminosity AGN than does the power-law selected sample, the color
criteria select larger fractions of sources not detected in X-rays,
due at least in part to a higher degree of contamination by
star-forming galaxies. Combining color-selection with additional tests
designed to rule out emission from star-forming galaxies is likely to
produce more reliable samples of AGN.

\item
68\% (2:1) of the X-ray--detected \plagas\ are obscured (log~$N_{\rm
H}$(cm$^{-2}$)$ > 22$), and all of the weakly-detected \plagas\ are
consistent with being obscured, but not Compton-thick (log~$N_{\rm
H}$(cm$^{-2}$)$ > 24$).  If we assume that all of the X-ray
non-detected \plagas\ are obscured, we derive a maximum obscured
fraction of 81\% (4:1).  Power-law galaxies also appear to be more
heavily obscured at low X-ray luminosities and at high redshift.

\item
Power-law galaxies detected to high $S/N$ in the IRAC bands account
for $\sim 20-30$\% of both the MIR-detected AGN and the MIR-detected
obscured AGN predicted by the X-ray luminosity function synthesis
models of Treister et al. (2006) down to 24 \micron\ flux densities of
$80$ \microjy.  This percentage is best interpreted as a lower limit
since our conservative selection criteria may exclude some power-law
objects.

\item
At all but the highest column densities, NIR/MIR extinction should
have only a minor effect on the power-law emission of luminous AGN.
IRAC power-law selection, however, is likely to be biased against
high-redshift Compton-thick AGN if the covering fractions of the X-ray
and NIR-emitting regions are the same.

\end{itemize}


\acknowledgments

This work was supported by an NSF Graduate Research Fellowship and by
NASA through contract 1255094 issued by JPL/California Institute of
Technology.  We thank the referee for helpful comments that improved
the paper.

\vspace{4cm}
\begin{appendix}

To understand the performance of our power-law AGN selection at high
redshift and high luminosity, it is necessary to test it with template
spectral energy distributions of appropriately luminous infrared
galaxies. It has recently become possible to construct accurate
templates in the critical 0.8 to 10 \micron\ range using a combination
of Spitzer data, 2MASS total galaxy measurements, and ground-based
spectroscopy in the 3 - 4 \micron\ range. We used these data sources
to build templates for 3 ultra-luminous infrared galaxies. Arp 220 and
IRAS 17208-0014 are both strongly dominated by star formation, whereas
Mrk 273 appears to be powered by a mixture of star formation and an
obscured AGN (e.g., Ptak et al. 2003; Farrah et al. 2003).

The references in Table A4 provide total-galaxy flux densities in the
optical, near infrared, and at 3.6 and 4.5 \micron. We derived the
near infrared stellar continuum starting from the stellar spectra of
Strecker et al. (1979). They have the advantage of being consistently
calibrated, are not affected by terrestrial atmospheric absorption,
and they extend to 5.5 \micron. Specifically, we used the spectrum of
$\beta$ And for the near infrared stellar continuum, since its CO
absorption approximately matches that of actively star forming
galaxies. At wavelengths shortward of 1.2 \micron, our starting point
was the spectral energy distribution of Arp 220 from Silva et
al. (1998). For each galaxy, we joined these two spectra at 1.2
\micron\ and then adjusted the normalization and reddening to give a
good overall fit through the photometry (UBVR as available plus
JHK). The resulting spectrum was taken as the stellar photospheric
part of the template (since we need a template valid up to z = 3 in
the IRAC 3.6 \micron\ band, the behavior at wavelengths shortward of
0.8 \micron\ is unimportant).  From Goldader et al. (1995), all three
galaxies have strong CO stellar absorption features (bandhead at rest
2.3 \micron), so the photospheric spectrum is not significantly
diluted by other sources at least to 2.4 \micron. Therefore, we use it
to define the template from 0.8 to 2.4 \micron.

For wavelengths longer than about 6 \micron, the templates are based
on the spectra listed in Table A4. Because the infrared activity is
generally concentrated in the nuclei, and the beams used for these
spectra are relatively large, we made no corrections for extended
emission. To bridge from 2.4 to 6 \micron, we proceeded as
follows. The IRAC photometry at 3.6 and 4.5 \micron\ matches the
apertures for the 2MASS total galaxy measurements, so it defines the
overall output across the bridge region. The spectrum of M82 (St\"urm
et al. 2000) indicates that we should expect no strong spectral
features between 4 and 6 \micron\ (other than Br $\alpha$, which
should have small equivalent width).  To fold in the 3 - 4 \micron\
region, we used the spectra indicated in Table A4. We estimated the
photospheric contribution to these spectra by applying corrections to
our stellar template from small aperture photometry in the JHK region,
or from small aperture photometry at L, or from both. This
contribution was subtracted from the spectrum (using the $\beta$ And
template) and the remainder, representing the excess emission, was
added to the total stellar template. This excess spectrum was scaled
until the total template matched the photometric point at 3.6 \micron\
(the scaling factors required were small, between 1 and 1.5,
indicating that the excess emission was nearly all within the $\sim$
1" slits used for the spectroscopy).  We joined the longest wavelength
point from these spectra to the shortest wavelength point in the 6 -
15 \micron\ spectra with a scaled and adjusted spectrum from M82
(Sturm et al. 2000). We found that adjusting the slope as a power law
in wavelength and then normalizing gave a smooth connection that also
was compatible with the photometry at 4.5 \micron. The templates are
shown in Figure 20 and are given in Table A5.

\newpage
\begin{figure}
\centering
\figurenum{20}
\includegraphics[angle=90,scale=0.5]{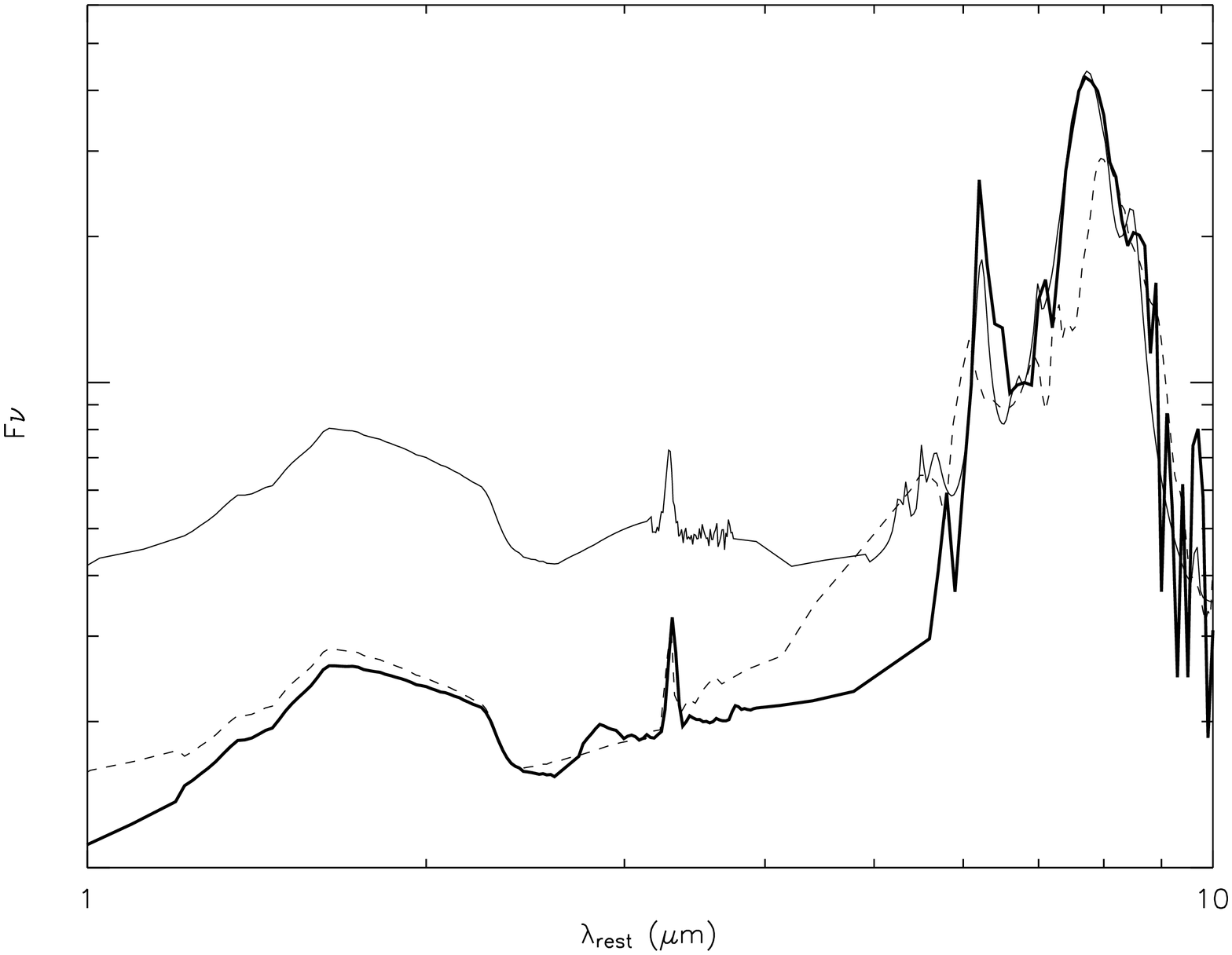}
\caption{SED templates for Arp 220 (light solid line), IRAS
17208-0014 (heavy solid line) and Mrk 273 (light dashed line) between
1 and 10 \micron. The templates have been normalized to the same value
near 7 \micron. The influence of the AGN in Mrk 273 is apparent both
in the reduced equivalent widths of the aromatic features and in the
filling-in of the SED near 5 \micron.}
\end{figure}

\end{appendix}


\clearpage
\begin{landscape}
\begin{centering}
\begin{deluxetable}{lllllcrrrrccr}
\pagestyle{empty}
\tablenum{1}
\tabletypesize{\tiny}
\tablewidth{0pt}
\tablecaption {Power-law Galaxy Sample}
\tablehead{
\colhead{ID}                     &
\colhead{$\alpha_{\rm 2000}$}    &
\colhead{$\delta_{\rm 2000}$}    &
\colhead{$P_{\chi}$\tablenotemark{a}}   &
\colhead{$\alpha$\tablenotemark{b}} &
\colhead{$z$}                    &
\colhead{24 \micron}             &
\colhead{1.4 GHz}                &
\colhead{0.5-8 keV}              &
\colhead{log L$_{\rm x}$\tablenotemark{c}}        &
\colhead{X-ray}                  &
\colhead{$\theta$\tablenotemark{d}}               &
\colhead{$q$\tablenotemark{e}}                    \\
\colhead{}                     &
\colhead{}                     &
\colhead{}                     &
\colhead{}                     &
\colhead{}                     &
\colhead{}                     &
\colhead{Flux Density}         &
\colhead{Flux Density}         &
\colhead{X-ray Flux}           &
\colhead{(ergs~s$^{-1}$)}      &
\colhead{Exposure}             &
\colhead{($\arcmin$)}          &
\colhead{}                     \\
\colhead{}                     &
\colhead{}                     &
\colhead{}                     &
\colhead{}                     &
\colhead{}                     &
\colhead{}                     &
\colhead{(\microjy)}          &
\colhead{(\microjy)}          &
\colhead{(ergs~s$^{-1}$~cm$^{-2}$)} &
\colhead{}                     &
\colhead{(s)}                  &
\colhead{}                     &
\colhead{}                     
}
\startdata                                                          
CDFN 11348  &  12 35 25.58  &  62 16 09.41  &  0.86  & -0.86  &              \nodata  &      204.0  &  $<$  80.0  &  $<$  8.10E-16  &   \nodata  & 9.59E+05  &   9.6  &  $>$   0.41 \\
CDFN 17315  &  12 35 29.37  &  62 12 56.53  &  0.22  & -1.94  &                 2.41  &      251.9  &      135.0  &       7.02E-15  &      44.5  & 7.43E+05  &   8.9  &        0.27 \\
CDFN 11516  &  12 35 37.11  &  62 17 23.40  &  0.49  & -0.89  &                 2.05  &      234.9  &  $<$  80.0  &       5.27E-15  &      44.2  & 9.00E+05  &   8.7  &  $>$   0.47 \\
CDFN 07514  &  12 35 38.08  &  62 19 40.56  &  0.86  & -0.79  &      2.97 $\pm$ 0.40  &  $<$ 157.2  &  $<$  80.0  &       2.00E-15  &      44.2  & 9.86E+05  &   9.7  &     \nodata \\
CDFN 12939  &  12 35 38.52  &  62 16 42.91  &  0.69  & -1.15  &                 0.71  &     1413.8  &     3550.0  &       2.23E-13  &      44.7  & 9.60E+05  &   8.3  &       -0.40 \\
CDFN 14642  &  12 35 43.35  &  62 16 17.15  &  0.84  & -1.17  &      2.36 $\pm$ 0.34  &  $<$ 117.6  &  $<$  80.0  &       9.19E-16  &      43.6  & 1.25E+06  &   7.6  &     \nodata \\
CDFN 08362  &  12 35 46.64  &  62 20 13.23  &  0.57  & -1.39  &              \nodata  &      118.5  &  $<$  80.0  &  $<$  7.85E-16  &   \nodata  & 9.66E+05  &   9.3  &  $>$   0.17 \\
CDFN 10934  &  12 35 48.89  &  62 19 04.46  &  0.26  & -1.58  &              \nodata  &      115.4  &  $<$  80.0  &  $<$  4.92E-16  &   \nodata  & 9.84E+05  &   8.4  &  $>$   0.16 \\
CDFN 16796  &  12 35 49.43  &  62 15 36.52  &  0.22  & -0.79  &                 2.20  &      585.6  &       74.6  &       1.34E-15  &      43.7  & 1.58E+06  &   6.8  &        0.89 \\
CDFN 25218  &  12 35 50.36  &  62 10 41.70  &  0.11  & -2.04  &              \nodata  &  $<$ 108.5  &  $<$  80.0  &       4.21E-16  &   \nodata  & 7.04E+05  &   7.2  &     \nodata \\
CDFN 20981  &  12 35 54.30  &  62 13 43.37  &  0.95  & -0.69  &              \nodata  &      131.7  &  $<$  80.0  &  $<$  2.04E-16  &   \nodata  & 1.74E+06  &   6.0  &  $>$   0.22 \\
CDFN 25248  &  12 36 06.37  &  62 12 32.41  &  0.81  & -0.60  &      2.89 $\pm$ 0.39  &      139.2  &  $<$  80.0  &       9.64E-16  &      43.8  & 1.81E+06  &   4.8  &  $>$   0.24 \\
CDFN 33052  &  12 36 08.87  &  62 08 03.56  &  0.10  & -1.37  &              \nodata  &      151.8  &  $<$  80.0  &  $<$  4.67E-16  &   \nodata  & 5.43E+05  &   7.3  &  $>$   0.28 \\
CDFN 27463  &  12 36 11.95  &  62 11 47.18  &  0.77  & -0.53  &      2.92 $\pm$ 0.39  &      180.3  &  $<$  80.0  &  $<$  1.23E-16  &   $<$42.9  & 1.68E+06  &   4.5  &  $>$   0.35 \\
CDFN 31054  &  12 36 14.04  &  62 09 47.93  &  0.78  & -0.59  &      1.65 $\pm$ 0.27  &       60.8  &  $<$  80.0  &  $<$  2.46E-16  &   $<$42.6  & 1.79E+06  &   5.6  &  $>$  -0.12 \\
CDFN 23661  &  12 36 22.98  &  62 15 26.28  &  0.16  & -1.40  &                 2.59  &      454.5  &  $<$  80.0  &       1.88E-14  &      45.0  & 1.85E+06  &   3.0  &  $>$   0.75 \\
CDFN 39358  &  12 36 23.37  &  62 06 05.23  &  0.21  & -1.92  &      1.86 $\pm$ 0.29  &      479.5  &       71.7  &       4.71E-16  &      43.0  & 5.42E+05  &   8.3  &        0.83 \\
CDFN 37928  &  12 36 29.21  &  62 07 37.61  &  0.91  & -0.59  &      1.35 $\pm$ 0.23  &      161.6  &  $<$  80.0  &  $<$  2.15E-16  &   $<$42.3  & 1.77E+06  &   6.6  &  $>$   0.31 \\
CDFN 37999  &  12 36 32.59  &  62 07 58.95  &  0.11  & -2.14  &                 2.00  &      711.8  &       90.6  &       1.83E-15  &      43.7  & 1.78E+06  &   6.2  &        0.90 \\
CDFN 27641  &  12 36 35.60  &  62 14 23.51  &  0.86  & -1.82  &                 2.02  &     1345.2  &       87.8  &       2.52E-15  &      43.8  & 1.92E+06  &   1.2  &        1.19 \\
CDFN 28773  &  12 36 36.66  &  62 13 46.46  &  0.63  & -0.74  &                 0.96  &      459.6  &  $<$  80.0  &       1.01E-14  &      43.7  & 1.91E+06  &   1.1  &  $>$   0.76 \\
CDFN 14667  &  12 36 36.90  &  62 22 27.27  &  0.57  & -2.03  &              \nodata  &      292.9  &  $<$  80.0  &       3.45E-15  &   \nodata  & 1.06E+06  &   8.6  &  $>$   0.56 \\
CDFN 37334  &  12 36 37.08  &  62 08 51.90  &  0.32  & -0.90  &      1.71 $\pm$ 0.27  &      301.7  &       71.1  &       1.19E-16  &      42.3  & 1.78E+06  &   5.2  &        0.63 \\
CDFN 44902  &  12 36 40.59  &  62 04 51.34  &  0.27  & -0.87  &      1.16 $\pm$ 0.22  &      166.6  &  $<$  80.0  &  $<$  8.28E-16  &   $<$42.8  & 6.96E+05  &   9.1  &  $>$   0.32 \\
CDFN 24409  &  12 36 42.22  &  62 17 10.89  &  0.65  & -0.80  &                 2.72  &       91.7  &  $<$  80.0  &       3.35E-15  &      44.3  & 1.65E+06  &   3.2  &  $>$   0.06 \\
CDFN 11965  &  12 36 42.23  &  62 24 39.01  &  0.59  & -0.71  &      1.90 $\pm$ 0.29  &      480.4  &       82.1  &       1.95E-15  &      43.7  & 9.62E+05  &  10.7  &        0.77 \\
CDFN 14360  &  12 36 47.27  &  62 23 49.06  &  0.29  & -1.28  &              \nodata  &      216.0  &  $<$  80.0  &  $<$  7.05E-16  &   \nodata  & 9.75E+05  &   9.9  &  $>$   0.43 \\
CDFN 18906  &  12 36 48.30  &  62 21 06.89  &  0.76  & -1.45  &              \nodata  &      149.4  &  $<$  80.0  &  $<$  2.94E-16  &   \nodata  & 1.68E+06  &   7.2  &  $>$   0.27 \\
CDFN 41981  &  12 36 49.66  &  62 07 37.84  &  0.50  & -2.30  &      1.54 $\pm$ 0.25  &     1239.3  &      307.0  &       2.59E-14  &      44.6  & 1.74E+06  &   6.3  &        0.61 \\
CDFN 49937  &  12 36 58.76  &  62 04 01.80  &  0.12  & -1.17  &                 0.29  &      138.7  &  $<$  80.0  &       9.75E-15  &      42.4  & 7.33E+05  &  10.1  &  $>$   0.24 \\
CDFN 19080  &  12 36 58.96  &  62 22 14.89  &  0.28  & -1.94  &              \nodata  &      106.6  &  $<$  80.0  &  $<$  4.41E-16  &   \nodata  & 1.62E+06  &   8.4  &  $>$   0.12 \\
CDFN 14046  &  12 36 59.07  &  62 25 21.81  &  0.31  & -1.04  &      2.82 $\pm$ 0.38  &      164.3  &  $<$  80.0  &       1.16E-14  &      44.9  & 8.68E+05  &  11.5  &  $>$   0.31 \\
CDFN 44836  &  12 37 01.69  &  62 07 20.21  &  0.56  & -3.15  &              \nodata  &      323.6  &  $<$  80.0  &       8.56E-16  &   \nodata  & 1.74E+06  &   6.9  &  $>$   0.61 \\
CDFN 53291  &  12 37 01.70  &  62 02 22.05  &  0.10  & -1.59  &      2.17 $\pm$ 0.32  &       99.6  &  $<$  80.0  &  $<$  1.89E-15  &   $<$43.8  & 6.41E+05  &  11.8  &  $>$   0.10 \\
CDFN 45610  &  12 37 06.59  &  62 07 26.83  &  0.79  & -1.05  &              \nodata  &       -1.0  &  $<$  80.0  &  $<$  7.08E-16  &   \nodata  & 1.76E+06  &   7.0  &     \nodata \\
CDFN 29120  &  12 37 06.90  &  62 17 02.37  &  0.29  & -1.06  &                 1.02  &      766.3  &  $<$  80.0  &       3.54E-14  &      44.3  & 1.87E+06  &   3.9  &  $>$   0.98 \\
CDFN 16936  &  12 37 07.01  &  62 24 27.81  &  0.92  & -1.25  &              \nodata  &      161.3  &  $<$  80.0  &  $<$  1.00E-15  &   \nodata  & 1.08E+06  &  10.8  &  $>$   0.30 \\
CDFN 45284  &  12 37 09.85  &  62 08 00.82  &  0.39  & -1.40  &                 2.18  &      128.6  &  $<$  80.0  &       2.11E-14  &      44.9  & 1.71E+06  &   6.6  &  $>$   0.21 \\
CDFN 20246  &  12 37 10.00  &  62 22 58.92  &  0.20  & -1.42  &      1.41 $\pm$ 0.24  &      318.6  &      708.0  &       5.21E-15  &      43.8  & 1.60E+06  &   9.4  &       -0.35 \\
CDFN 38580  &  12 37 12.07  &  62 12 11.61  &  0.70  & -1.14  &                 2.91  &  $<$  80.0  &  $<$  80.0  &       3.84E-16  &      43.4  & 1.72E+06  &   3.5  &     \nodata \\
CDFN 32673  &  12 37 13.72  &  62 15 45.14  &  0.52  & -1.14  &              \nodata  &      129.8  &  $<$  80.0  &       4.46E-16  &   \nodata  & 1.90E+06  &   3.7  &  $>$   0.21 \\
CDFN 43967  &  12 37 14.06  &  62 09 16.80  &  0.38  & -0.96  &      1.84 $\pm$ 0.28  &      105.3  &  $<$  80.0  &       3.57E-15  &      43.9  & 1.83E+06  &   5.7  &  $>$   0.12 \\
CDFN 30147  &  12 37 16.69  &  62 17 33.18  &  0.56  & -1.75  &                 1.15  &     1015.7  &      346.0  &       2.17E-14  &      44.2  & 1.84E+06  &   5.1  &        0.47 \\
CDFN 50281  &  12 37 16.99  &  62 05 53.05  &  0.59  & -1.05  &                 1.94  &      398.5  &  $<$  80.0  &       1.94E-15  &      43.7  & 7.26E+05  &   8.9  &  $>$   0.70 \\
CDFN 28149  &  12 37 17.90  &  62 18 55.56  &  0.17  & -1.24  &                 2.24  &      195.2  &  $<$  80.0  &       9.95E-15  &      44.6  & 1.78E+06  &   6.2  &  $>$   0.39 \\
CDFN 46227  &  12 37 21.72  &  62 08 50.41  &  0.84  & -0.80  &      2.56 $\pm$ 0.36  &      292.1  &  $<$  80.0  &  $<$  2.43E-16  &   $<$43.1  & 1.77E+06  &   6.6  &  $>$   0.56 \\
CDFN 51788  &  12 37 23.03  &  62 05 39.45  &  0.99  & -1.08  &      2.28 $\pm$ 0.33  &      146.6  &       78.5  &  $<$  2.65E-15  &   $<$44.0  & 7.43E+05  &   9.4  &        0.27 \\
CDFN 27360  &  12 37 26.58  &  62 20 26.46  &  0.14  & -2.26  &      1.86 $\pm$ 0.29  &      777.7  &      102.0  &       1.40E-15  &      43.5  & 1.40E+06  &   8.0  &        0.88 \\
CDFN 38126  &  12 37 28.65  &  62 14 22.56  &  0.58  & -1.80  &              \nodata  &      226.1  &  $<$  80.0  &       1.20E-16  &   \nodata  & 1.82E+06  &   5.0  &  $>$   0.45 \\
CDFN 40913  &  12 37 30.77  &  62 12 58.50  &  0.97  & -1.78  &              \nodata  &      139.3  &      107.0  &  $<$  2.08E-16  &   \nodata  & 1.79E+06  &   5.3  &        0.11 \\
\enddata

\end{deluxetable}
\end{centering}
\clearpage
\end{landscape}

\clearpage
\begin{landscape}
\begin{centering}
\begin{deluxetable}{lllllcrrrrccr}
\pagestyle{empty}
\tablenum{1}
\tabletypesize{\tiny}
\tablewidth{0pt}
\tablecaption {--- Continued}
\tablehead{
\colhead{ID}                     &
\colhead{$\alpha_{\rm 2000}$}    &
\colhead{$\delta_{\rm 2000}$}    &
\colhead{$P_{\chi}$\tablenotemark{a}}   &
\colhead{$\alpha$\tablenotemark{b}} &
\colhead{$z$}                    &
\colhead{24 \micron}             &
\colhead{1.4 GHz}                &
\colhead{0.5-8 keV}              &
\colhead{log L$_{\rm x}$\tablenotemark{c}}        &
\colhead{X-ray}                  &
\colhead{$\theta$\tablenotemark{d}}               &
\colhead{$q$\tablenotemark{e}}                    \\
\colhead{}                     &
\colhead{}                     &
\colhead{}                     &
\colhead{}                     &
\colhead{}                     &
\colhead{}                     &
\colhead{Flux Density}         &
\colhead{Flux Density}         &
\colhead{X-ray Flux}           &
\colhead{(ergs~s$^{-1}$)}      &
\colhead{Exposure}             &
\colhead{($\arcmin$)}          &
\colhead{}                     \\
\colhead{}                     &
\colhead{}                     &
\colhead{}                     &
\colhead{}                     &
\colhead{}                     &
\colhead{}                     &
\colhead{(\microjy)}          &
\colhead{(\microjy)}          &
\colhead{(ergs~s$^{-1}$~cm$^{-2}$)} &
\colhead{}                     &
\colhead{(s)}                  &
\colhead{}                     &
\colhead{}                     
}
\startdata                                                          
CDFN 39529  &  12 37 36.83  &  62 14 28.72  &  0.20  & -1.44  &              \nodata  &      118.9  &       57.8  &       6.21E-15  &   \nodata  & 1.74E+06  &   6.0  &        0.31 \\
CDFN 26243  &  12 37 39.78  &  62 22 39.67  &  0.38  & -2.08  &              \nodata  &      218.2  &  $<$  80.0  &  $<$  3.33E-15  &   \nodata  & 6.12E+05  &  10.7  &  $>$   0.44 \\
CDFN 44598  &  12 37 41.00  &  62 12 00.29  &  0.26  & -0.63  &                 1.17  &      558.2  &  $<$  80.0  &       2.02E-14  &      44.2  & 1.76E+06  &   6.7  &  $>$   0.84 \\
CDFN 34143  &  12 37 42.58  &  62 18 11.67  &  0.16  & -1.55  &                 2.31  &      149.6  &  $<$  80.0  &       2.08E-14  &      44.9  & 1.69E+06  &   7.8  &  $>$   0.27 \\
CDFN 44751  &  12 37 44.67  &  62 12 18.71  &  0.21  & -0.94  &      1.69 $\pm$ 0.27  &      416.1  &       67.3  &       2.88E-16  &      42.7  & 1.76E+06  &   7.1  &        0.79 \\
CDFN 40112  &  12 37 57.32  &  62 16 27.57  &  0.28  & -1.55  &                 2.92  &      122.3  &  $<$  80.0  &       4.68E-15  &      44.5  & 1.70E+06  &   8.7  &  $>$   0.18 \\
CDFN 54722  &  12 37 57.55  &  62 08 00.51  &  0.65  & -1.13  &              \nodata  &      174.9  &  $<$  80.0  &  $<$  1.01E-15  &   \nodata  & 6.41E+05  &  10.3  &  $>$   0.34 \\
CDFN 49940  &  12 37 59.62  &  62 11 02.09  &  0.58  & -1.42  &                 0.91  &     2132.0  &       85.0  &       1.52E-13  &      44.8  & 7.33E+05  &   9.1  &        1.40 \\
CDFN 45792  &  12 38 00.94  &  62 13 35.90  &  0.43  & -0.72  &                 0.44  &     4092.5  &      190.0  &       7.78E-14  &      43.7  & 1.44E+06  &   8.8  &        1.33 \\
CDFN 53792  &  12 38 31.18  &  62 12 21.98  &  0.59  & -0.77  &      1.63 $\pm$ 0.26  &      297.0  &  $<$  80.0  &       3.76E-15  &      43.8  & 6.62E+05  &  12.4  &  $>$   0.57 \\
CDFN 53753  &  12 38 32.19  &  62 12 30.60  &  0.25  & -0.66  &      1.63 $\pm$ 0.26  &      198.4  &  $<$  80.0  &  $<$  2.79E-15  &   $<$43.7  & 6.15E+05  &  12.5  &  $>$   0.39 \\
CDFN 54888  &  12 38 34.09  &  62 12 05.88  &  0.67  & -1.75  &      1.96 $\pm$ 0.30  &      476.4  &  $<$  80.0  &  $<$  1.59E-15  &   $<$43.6  & 5.57E+05  &  12.8  &  $>$   0.77 \\
\enddata
\tablenotetext{a}{Chi-squared probability (see \S3)}
\tablenotetext{b}{IRAC spectra index; $f_{\nu} \propto \nu^{\alpha}$}
\tablenotetext{c}{0.5-8 keV}
\tablenotetext{d}{X-ray off-axis angle}
\tablenotetext{e}{$q$ = log (f$_{24~\micron}$/f$_{1.4~\rm GHz}$)}

\end{deluxetable}
\end{centering}
\clearpage
\end{landscape}


\begin{centering}
\begin{deluxetable}{lcclrrrrrcc}
\centering
\tablenum{2}
\tabletypesize{\tiny}
\tablewidth{0pt}
\tablecaption{X-ray Properties of X-ray Weakly-Detected Power-law Galaxies}
\tablehead{
\colhead{Source}                 &
\colhead{z}                      &
\colhead{X-ray}                  &
\colhead{EER}                    &
\colhead{Detection }             &
\colhead{Aperture}               &
\colhead{$\Gamma$}               &
\colhead{Flux}                   &
\colhead{H/S\tablenotemark{b}}   &
\colhead{log N$_{\rm H}$}            \\
\colhead{}                       &
\colhead{}                       &
\colhead{Band\tablenotemark{a}}  &
\colhead{(\%)}                   &
\colhead{$\sigma$}               &
\colhead{Corrected}              &
\colhead{}                       &
\colhead{(erg s$^{-1}$ cm$^{-2}$)} &
\colhead{}                       &
\colhead{(cm$^{-2}$)}            \\    
\colhead{}                       &
\colhead{}                       &
\colhead{}                       &
\colhead{}                       &
\colhead{}                       &
\colhead{Count Rate (s${^{-1}}$)} &
\colhead{}                       &
\colhead{}                       &
\colhead{}                       &
\colhead{}                       
}
\startdata         
CDFN14642 &    2.36 &  Full &  80 &        4.6 &        4.44E-05 &            0.28 &        9.19E-16 &         11.17 &                                                       23.7 $^{+  0.1}_{-  0.1}$  \\
          &         &  Hard &  70 &        3.3 &        2.33E-05 &                 &        6.98E-16 &               &                                                                                  \\
          &         &  Soft &  60 &        3.3 &        1.27E-05 &                 &        6.08E-17 &               &                                                                                  \\
CDFN20981 & \nodata &  Full &  80 & $\le$  2.5 & $\le$  1.44E-05 & $\ge$      0.41 & $\le$  1.68E-16 & $\le$    6.96 &          $\le$    22.5 $^{+  0.1}_{-  0.1}$ -$\le$    23.7 $^{+  0.1}_{-  0.1}$  \\
          &         &  Hard &  80 & $\le$  2.5 & $\le$  1.00E-05 &                 & $\le$  2.31E-16 &               &                                                                                  \\
          &         &  Soft &  60 &        3.3 &        6.35E-06 &                 &        3.18E-17 &               &                                                                                  \\
CDFN25218 & \nodata &  Full &  80 &        3.5 &        3.61E-05 & $\le$      0.61 &        4.21E-16 & $\ge$    5.58 &          $\ge$    22.4 $^{+  0.1}_{-  0.1}$ -$\ge$    23.6 $^{+  0.1}_{-  0.1}$  \\
          &         &  Hard &  70 &        2.8 &        2.35E-05 &                 &        5.42E-16 &               &                                                                                  \\
          &         &  Soft &  80 & $\le$  2.5 & $\le$  1.86E-05 &                 & $\le$  9.33E-17 &               &                                                                                  \\
CDFN37334 &    1.71 &  Full &  80 &        2.6 &        1.02E-05 & $\ge$      0.20 &        1.19E-16 & $\le$    8.83 &                                              $\le$    23.3 $^{+  0.1}_{-  0.1}$  \\
          &         &  Hard &  80 & $\le$  2.5 & $\le$  1.17E-05 &                 & $\le$  2.70E-16 &               &                                                                                  \\
          &         &  Soft &  70 &        3.2 &        5.85E-06 &                 &        2.93E-17 &               &                                                                                  \\
CDFN37928 &    1.35 &  Full &  70 & $\le$  2.5 & $\le$  9.22E-06 & $\ge$      0.69 & $\le$  1.07E-16 & $\le$    5.09 &                                              $\le$    22.9 $^{+  0.1}_{-  0.1}$  \\
          &         &  Hard &  70 & $\le$  2.5 & $\le$  8.10E-06 &                 & $\le$  1.87E-16 &               &                                                                                  \\
          &         &  Soft &  60 &        2.8 &        7.03E-06 &                 &        3.52E-17 &               &                                                                                  \\
CDFN38126 & \nodata &  Full &  80 &        2.7 &        1.03E-05 &         \nodata &        1.20E-16 &       \nodata &                                                                         \nodata  \\
          &         &  Hard &  80 & $\le$  2.5 & $\le$  1.03E-05 &                 & $\le$  2.38E-16 &               &                                                                                  \\
          &         &  Soft &  80 & $\le$  2.5 & $\le$  6.97E-06 &                 & $\le$  3.49E-17 &               &                                                                                  \\
CDFN39358 &    1.86 &  Full &  80 &        2.8 &        4.04E-05 & $\le$      0.14 &        4.71E-16 & $\ge$    9.42 &                                              $\ge$    23.4 $^{+  0.1}_{-  0.1}$  \\
          &         &  Hard &  80 &        2.6 &        3.39E-05 &                 &        7.82E-16 &               &                                                                                  \\
          &         &  Soft &  70 & $\le$  2.5 & $\le$  1.59E-05 &                 & $\le$  7.97E-17 &               &                                                                                  \\
CDFN44751 &    1.69 &  Full &  80 &        3.7 &        2.47E-05 & $\ge$      0.52 &        2.88E-16 & $\le$    6.21 &                                              $\le$    23.2 $^{+  0.1}_{-  0.1}$  \\
          &         &  Hard &  80 & $\le$  2.5 & $\le$  2.01E-05 &                 & $\le$  4.64E-16 &               &                                                                                  \\
          &         &  Soft &  80 &        4.2 &        1.43E-05 &                 &        7.17E-17 &               &                                                                                  \\
CDFN44902 &    1.16 &  Full &  70 & $\le$  2.5 & $\le$  4.54E-05 & $\ge$      1.05 & $\le$  5.29E-16 & $\le$    3.41 &                                              $\le$    22.5 $^{+  0.1}_{-  0.2}$  \\
          &         &  Hard &  80 & $\le$  2.5 & $\le$  2.78E-05 &                 & $\le$  6.41E-16 &               &                                                                                  \\
          &         &  Soft &  60 &        5.0 &        3.60E-05 &                 &        1.80E-16 &               &                                                                                  \\
CDFN46227 &    2.56 &  Full &  60 & $\le$  2.5 & $\le$  1.04E-05 & $\ge$      0.83 & $\le$  1.21E-16 & $\le$    4.37 &                                              $\le$    23.3 $^{+  0.1}_{-  0.2}$  \\
          &         &  Hard &  60 & $\le$  2.5 & $\le$  6.83E-06 &                 & $\le$  1.58E-16 &               &                                                                                  \\
          &         &  Soft &  60 &        2.7 &        6.91E-06 &                 &        3.46E-17 &               &                                                                                  \\
\enddata
\tablenotetext{a}{Full band = 0.5-8 keV; Hard band = 2-8 keV; Soft band = 0.5-2 keV}
\tablenotetext{b}{Flux ratio corrected for Galactic absorption}
\end{deluxetable}
\end{centering}
\clearpage


\begin{centering}
\begin{deluxetable}{lccccrrc}
\tablenum{3}
\centering
\tabletypesize{\tiny}
\tablewidth{0pt}
\tablecaption {X-ray Properties of X-ray-Detected Power-law AGN}
\tablehead{
\colhead{Source}                 &
\colhead{z}                      &
\colhead{0.5-8 keV}              &
\colhead{2-8 keV}                &
\colhead{0.5-2 keV}              &
\colhead{$\Gamma$\tablenotemark{a}}               &
\colhead{H/S\tablenotemark{b}}                    &
\colhead{log N$_{\rm H}$\tablenotemark{c}}            \\
\colhead{}                       &
\colhead{}                       &
\colhead{Flux\tablenotemark{a}}                   &
\colhead{Flux\tablenotemark{a}}                   &
\colhead{Flux\tablenotemark{a}}                   &
\colhead{}                       &
\colhead{}                       &
\colhead{(cm$^{-2}$)}            \\     
\colhead{}                       &
\colhead{}                       &
\colhead{(erg s$^{-1}$ cm$^{-2}$)} &
\colhead{(erg s$^{-1}$ cm$^{-2}$)} &
\colhead{(erg s$^{-1}$ cm$^{-2}$)} &
\colhead{}                       &
\colhead{}                       &
\colhead{} 
}
\startdata 
CDFN07514 &    2.97 &          2.00E-15 &          1.54E-15 &          4.19E-16 &        1.1 &          3.54 &                                                       23.3 $^{+  0.1}_{-  0.2}$ \\
CDFN11516 &    2.05 &          5.27E-15 &          2.85E-15 &          2.04E-15 &        1.8 &          1.33 &                                                    21.2 $^{+  1.0  }_{-\cdots}$ \\
CDFN11965 &    1.90 &          1.95E-15 &          1.56E-15 &          3.44E-16 &        0.9 &          4.38 &                                                       23.1 $^{+  0.1}_{-  0.2}$ \\
CDFN12939 &    0.71 &          2.23E-13 &          1.36E-13 &          8.05E-14 &        1.6 &          1.62 &                                                    21.4 $^{+  0.4  }_{-\cdots}$ \\
CDFN14046 &    2.82 &          1.16E-14 &          6.63E-15 &          4.32E-15 &        1.7 &          1.47 &                                                    22.0 $^{+  0.6  }_{-\cdots}$ \\
CDFN14667 & \nodata &          3.45E-15 &          2.13E-15 &          1.13E-15 &        1.6 &          1.81 &                   21.5 $^{+  0.3}_{-  1.3}$ -         22.6 $^{+  0.3}_{-  0.9}$ \\
CDFN16796 &    2.20 &          1.34E-15 &          1.20E-15 &          1.20E-16 &        0.4 &          9.72 &                                                       23.6 $^{+  0.1}_{-  0.1}$ \\
CDFN17315 &    2.41 &          7.02E-15 &          4.72E-15 &          2.18E-15 &        1.5 &          2.08 &                                                       22.6 $^{+  0.3}_{-  0.4}$ \\
CDFN20246 &    1.41 &          5.21E-15 &          4.99E-15 &          1.35E-16 &       -0.6 &         36.19 &                                                       23.6 $^{+  0.0}_{-  0.0}$ \\
CDFN23661 &    2.59 &          1.88E-14 &          1.10E-14 &          7.43E-15 &        1.8 &          1.41 &                                                    21.8 $^{+  0.7  }_{-\cdots}$ \\
CDFN24409 &    2.72 &          3.35E-15 &          2.32E-15 &          9.48E-16 &        1.4 &          2.35 &                                                       22.8 $^{+  0.2}_{-  0.3}$ \\
CDFN25248 &    2.89 &          9.64E-16 &          8.63E-16 &          1.08E-16 &        0.5 &          7.75 &                                                       23.7 $^{+  0.1}_{-  0.1}$ \\
CDFN27360 &    1.86 &          1.40E-15 &          1.32E-15 &          1.52E-16 &        0.5 &          8.43 &                                                       23.4 $^{+  0.1}_{-  0.1}$ \\
CDFN27641 &    2.02 &          2.52E-15 &          2.48E-15 &          2.07E-16 &        0.2 &         11.66 &                                                       23.6 $^{+  0.1}_{-  0.1}$ \\
CDFN28149 &    2.24 &          9.95E-15 &          6.20E-15 &          3.43E-15 &        1.6 &          1.73 &                                                       22.3 $^{+  0.3}_{-  1.2}$ \\
CDFN28773 &    0.96 &          1.01E-14 &          5.35E-15 &          4.57E-15 &        1.9 &          1.12 &                                                                   $\le$    21.2 \\
CDFN29120 &    1.02 &          3.54E-14 &          1.89E-14 &          1.57E-14 &        1.9 &          1.15 &                                                                   $\le$    21.4 \\
CDFN30147 &    1.15 &          2.17E-14 &          1.73E-14 &          4.26E-15 &        1.0 &          3.92 &                                                       22.6 $^{+  0.1}_{-  0.2}$ \\
CDFN32673 & \nodata &          4.46E-16 &          5.30E-16 & $\le$    2.55E-17 & $\le$  1.4 & $\ge$   19.96 &          $\ge$    22.9 $^{+  0.0}_{-  0.1}$ -$\ge$    24.0 $^{+  0.0}_{-  0.1}$ \\
CDFN34143 &    2.31 &          2.08E-14 &          1.29E-14 &          7.30E-15 &        1.6 &          1.69 &                                                    22.2 $^{+  0.4  }_{-\cdots}$ \\
CDFN37999 &    2.00 &          1.83E-15 &          1.83E-15 &          1.17E-16 &        0.1 &         15.24 &                                                       23.7 $^{+  0.1}_{-  0.1}$ \\
CDFN38580 &    2.91 &          3.84E-16 &          3.74E-16 &          2.93E-17 &        0.2 &         12.42 &                                                       23.9 $^{+  0.1}_{-  0.1}$ \\
CDFN39529 & \nodata &          6.21E-15 &          5.16E-15 &          1.09E-15 &        0.9 &          4.57 &                   22.3 $^{+  0.1}_{-  0.1}$ -         23.5 $^{+  0.1}_{-  0.1}$ \\
CDFN40112 &    2.92 &          4.68E-15 &          2.42E-15 &          2.27E-15 &        2.0 &          1.02 &                                                                   $\le$    20.7 \\
CDFN41981 &    1.54 &          2.59E-14 &          2.18E-14 &          4.34E-15 &        0.9 &          4.85 &                                                       23.0 $^{+  0.1}_{-  0.1}$ \\
CDFN43967 &    1.84 &          3.57E-15 &          2.44E-15 &          1.02E-15 &        1.4 &          2.30 &                                                       22.5 $^{+  0.2}_{-  0.3}$ \\
CDFN44598 &    1.17 &          2.02E-14 &          1.28E-14 &          6.96E-15 &        1.6 &          1.76 &                                                       21.8 $^{+  0.3}_{-  2.2}$ \\
CDFN44836 & \nodata &          8.56E-16 &          9.77E-16 & $\le$    5.71E-17 & $\le$  1.4 & $\ge$   16.43 &          $\ge$    22.8 $^{+  0.1}_{-  0.1}$ -$\ge$    24.0 $^{+  0.1}_{-  0.1}$ \\
CDFN45284 &    2.18 &          2.11E-14 &          1.27E-14 &          7.82E-15 &        1.7 &          1.55 &                                                    22.1 $^{+  0.4  }_{-\cdots}$ \\
CDFN45792 &    0.44 &          7.78E-14 &          3.83E-14 &          3.87E-14 &        2.0 &          0.94 &                                                                         \nodata \\
CDFN49937 &    0.29 &          9.75E-15 &          5.69E-15 &          3.65E-15 &        1.7 &          1.49 &                                                    21.0 $^{+  0.4  }_{-\cdots}$ \\
CDFN49940 &    0.91 &          1.52E-13 &          9.11E-14 &          5.64E-14 &        1.7 &          1.54 &                                                    21.4 $^{+  0.5  }_{-\cdots}$ \\
CDFN50281 &    1.94 &          1.94E-15 &          9.82E-16 &          7.57E-16 &        1.8 &          1.24 &                                                                   $\le$    22.0 \\
CDFN53792 &    1.63 &          3.76E-15 &          3.06E-15 &          6.42E-16 &        0.9 &          4.60 &                                                       23.0 $^{+  0.1}_{-  0.1}$ \\
\enddata
\tablenotetext{a}{Alexander et al. (2003)}
\tablenotetext{b}{Flux ratio corrected for Galactic absorption}
\tablenotetext{c}{Sources with no lower error have an immeasurably low lower limit.}
\end{deluxetable}
\end{centering}

\begin{deluxetable}{lcccccc}
\tabletypesize{\scriptsize}
\tablecaption{A4: Input Data for Templates}
\tablewidth{0pt}
\tablenum{4}
\tablehead{
\colhead{Galaxy} &
\colhead{UBVR} &
\colhead{JHK}&
\colhead{$3-4$ \micron} &
\colhead{3.6, 4.5 \micron} &
\colhead{$6-15$ \micron} &
\colhead{Other}\\
\colhead{}&
\colhead{}&
\colhead{}&
\colhead{spectrum}&
\colhead{}&
\colhead{spectrum}&
\colhead{} \\
}
\startdata
Arp 220 & 1 & 4 & 5 & 7; 54.2, 46.6 mJy & 9 & 12, 13 \\
IRAS 17208-0014 & 2 & 4 & 6 & 8; 17.9, 17.3 mJy & 10 & 14 \\
Mrk 273 & 3 & 4 & 5 & 7; 27.6, 38.2 mJy & 11 & 15 \\
\enddata

\tablerefs{(1) Frueh et al. 1996; (2) Duc, Mirabel, \& Maza 1997; (3) Surace, Sanders, \& Evans 2000;
(4) 2MASS via NED; (5) Imanishi, Dudley, \& Maloney 2006; (6) Imanishi 2006; (7) Spitzer PID 32, 
extraction aperture diameter of 90" for Arp 220 and 50" for Mrk 273; (8) Spitzer PID 3672, 
extraction aperture 30" in diameter; (9) Armus et al. 2006; (10) Rigopoulou et al. 1999;
(11) Higdon et al. 2006; (12) Klaas et al. 1997; (13) Rieke et al. 1985; (14) Scoville et al. 2000;
(15) Rieke 1978}

\end{deluxetable}

\begin{deluxetable}{lcccccc}[t]
\tabletypesize{\scriptsize}
\tablecaption{A5: Templates}
\tablewidth{0pt}
\tablenum{5}
\tablehead{
\multicolumn{2}{c}{Arp 220}            &
\multicolumn{2}{c}{IRAS 17208-0014}    &
\multicolumn{2}{c}{Mrk 273}            \\
\colhead{$\lambda$(\micron)}     &
\colhead{$f_{\rm \nu}$(Jy)}      &
\colhead{$\lambda$(\micron)}     &
\colhead{$f_{\rm \nu}$(Jy)}      &
\colhead{$\lambda$(\micron)}     &
\colhead{$f_{\rm \nu}$(Jy)}      \\
}
\startdata
   1.6572E-02  &   1.2100E-07  &   1.6177E-02  &   1.8200E-09  &   1.6256E-02  &   4.7100E-08 \\
   1.8124E-02  &   1.7100E-07  &   1.7693E-02  &   2.7100E-09  &   1.7778E-02  &   6.6600E-08 \\
   1.9823E-02  &   2.4300E-07  &   1.9352E-02  &   4.0800E-09  &   1.9445E-02  &   9.4900E-08 \\
   2.1690E-02  &   3.3200E-07  &   2.1174E-02  &   5.8800E-09  &   2.1276E-02  &   1.3000E-07 \\
   2.3723E-02  &   4.4200E-07  &   2.3159E-02  &   8.2500E-09  &   2.3271E-02  &   1.7200E-07 \\
\enddata
\tablecomments{The complete version of this table is in the electronic edition of the
Journal.  The printed edition contains only a sample.}

\end{deluxetable}

\end{document}